\documentclass[10pt,journal,compsoc]{IEEEtran}
\usepackage{amsmath}  
\usepackage{graphicx}
\usepackage{caption}
\usepackage{algorithm}
\usepackage{comment}
\usepackage{algpseudocode}
\usepackage{booktabs}
\usepackage{caption}
\usepackage{multirow}
\usepackage{xcolor}
\ifCLASSOPTIONcompsoc
  \usepackage[nocompress]{cite}
\else
  \usepackage{cite}
\fi
\ifCLASSINFOpdf
\else
\fi
\begin{document}
%
\title{Enhancing Worker Recruitment in Collaborative Mobile Crowdsourcing: A Graph Neural Network Trust Evaluation Approach}

\author{Zhongwei Zhan,
        Yingjie Wang*,~\IEEEmembership{Member,~IEEE,}
        Peiyong Duan,
        AKSHITA MARADAPU VERA VENKATA SAI,
        Zhaowei Liu,~\IEEEmembership{Member,~IEEE,}
        Chaocan Xiang,~\IEEEmembership{Member,~IEEE,}
        Xiangrong Tong,
        Weilong Wang,
        and Zhipeng Cai,~\IEEEmembership{Fellow,~IEEE,}
\IEEEcompsocitemizethanks{
\IEEEcompsocthanksitem This work was supported in part by the National Natural Science Foundation of China under Grant 62272405, Grant 62073201, Grant 62072392, Grant 62172351, the Youth Innovation Science and Technology Support Program  of Shandong Provincial under Grant 2021KJ080, the Natural Science Foundation of Shandong Province under Grant ZR2022MF238, the Key projects of Shandong Natural Science Foundation under Grant ZR2020KF019, the Major scientific and technological innovation projects of Shandong Province under Grant 2019JZZY020131, Yantai Science and Technology Innovation Development Plan Project under Grant 2022XDRH023, the Open Foundation of State key Laboratory of Networking and Switching Technology (Beijing University of Posts and Telecommunications) under Grant SKLNST-2022-1-12, the Graduate Innovation Foundation of Yantai University under Grant GIFYTU.
\IEEEcompsocthanksitem Zhongwei Zhan, Yingjie Wang (Corresponding author), Zhaowei Liu, Xiangrong Tong are with the School of Computer and Control Engineering, Yantai University, Yantai 264005, China. E-mail: towangyingjie@163.com.
\IEEEcompsocthanksitem Peiyong Duan is with the department of Electrical and Control, Qilu University of Technology(Shandong Academy of Sciences), Jinan 250000, China.  E-mail: duanpeiyong@sdnu.edu.cn.
\IEEEcompsocthanksitem AKSHITA MARADAPU VERA VENKATA SAI is with the department of Computer Science, Georgia State University, Atlanta, GA 30303.  E-mail: amaradapuveravenkat1@student.gsu.edu
\IEEEcompsocthanksitem Chaocan Xiang is with the College of Computer Science, Chongqing University, Chongqing 400044, China. E-mail: xiangchaocan@cqu.edu.cn.
\IEEEcompsocthanksitem Weilong Wang is with the Department of Computer Science and Engineering, Southeast University, Nanjing 211189, China. E-mail: wang\_wl@seu.edu.cn.
\IEEEcompsocthanksitem Zhipeng Cai is with the department of Computer Science, Georgia State
University, Atlanta, GA 30303. E-mail: zcai@gsu.edu.
}
\thanks{Manuscript received *; revised *}}
%
%

\markboth{Journal of \LaTeX\ Class Files,~Vol.~14, No.~8, August~2015}%
{Shell \MakeLowercase{\textit{et al.}}: Bare Demo of IEEEtran.cls for Computer Society Journals}
%



\IEEEtitleabstractindextext{%
\begin{abstract}
Collaborative Mobile Crowdsourcing (CMCS) allows platforms to recruit worker teams to collaboratively execute complex sensing tasks. The efficiency of such collaborations could be influenced by trust relationships among workers. To obtain the asymmetric trust values among all workers in the social network, the Trust Reinforcement Evaluation Framework (TREF) based on Graph Convolutional Neural Networks (GCNs) is proposed in this paper. The task completion effect is comprehensively calculated by considering the workers' ability benefits, distance benefits, and trust benefits in this paper. The worker recruitment problem is modeled as an Undirected Complete Recruitment Graph (UCRG), for which a specific Tabu Search Recruitment (TSR) algorithm solution is proposed. An optimal execution team is recruited for each task by the TSR algorithm, and the collaboration team for the task is obtained under the constraint of privacy loss. To enhance the efficiency of the recruitment algorithm on a large scale and scope, the Mini-Batch K-Means clustering algorithm and edge computing technology are introduced, enabling distributed worker recruitment. Lastly, extensive experiments conducted on five real datasets validate that the recruitment algorithm proposed in this paper outperforms other baselines. Additionally, TREF proposed herein surpasses the performance of state-of-the-art trust evaluation methods in the literature.
\end{abstract}

\begin{IEEEkeywords}
Mobile Crowdsourcing, Recruitment, Trust Evaluation, Graph Neural Network, Collaboration
\end{IEEEkeywords}}

\maketitle

\IEEEdisplaynontitleabstractindextext

%
\IEEEpeerreviewmaketitle






\section{Introduction}
Mobile Crowdsourcing (MCS) is a paradigm that harnesses the collective intelligence and participation of a diverse group of mobile users within the Internet of Things (IoT) ecosystem \cite{r6}. Supported by technologies such as IoT detection\cite{r26, r72}, blockchain security\cite{r19,r70, r73}, and drone sensing\cite{r28}, MCS continues to play a crucial role in domains like urban sensing, environmental monitoring, and traffic management. Task allocation\cite{r16,r21,r66,r68}, incentive mechanism\cite{r18,r23,r67,r69}, privacy preservation\cite{r13,r11,r14,r71} and quality control\cite{r21, r22, r23} are primarily the fields around which the current research in MCS revolves. In the field of Collaborative Mobile Crowdsourcing(CMCS) research, a key issue is recruiting an appropriate worker team to complete tasks\cite{r8}. Lim et al.\cite{r52} proposed that the task completion effect by worker teams is closely tied to team formation. Efficiency, quality and  coverage\cite{r23, r16} are often prioritized as driving factors in existing recruitment strategies, with collaborative relationships among workers being neglected. This could impact the rationality of task completion effect evaluations\cite{r1}.\\
Online Social Platforms (OSP) such as WeChat, Twitter, Instagram, and Facebook have been rapidly evolving, leading to the attraction of a vast global user base and the emergence of large-scale and intricate social networks\cite{r59,r60}. Zhao et al. \cite{r34} proposed that the positive network externality of social networks has promoted user information sharing, interaction, and dissemination. Additionally, Mesmer-Magnus et al.\cite{r36} found that information sharing significantly enhances the task completion effect of teams. Therefore, in the context of CMCS, the task completion effect could be improved by constructing stable team social networks. It should be noted that the key to building a stable social network is a well-established trust relationship. Two CMCS scenarios are listed in this paper to illustrate how the task completion effect is affected by information sharing through social networks:\\
\begin{itemize}
\item Example 1: Consider a task that necessitates a team to capture, identify, and annotate images within a specified area. Real-time traffic and location information is shared during task execution, enabling congested routes to be avoided and redundant sensing at the same location to be prevented. Furthermore, efficient image annotation could be achieved by the collaboration team members through real-time sharing of sensory data. Ultimately, the task completion effect will be greatly enhanced by information sharing.
\item Example 2: Consider a task that requires real-time target tracking.  As the targets are being tracked, target information needs to be shared among team members for real-time monitoring and tracking, leading to an improvement in the accuracy of target location and status.  Furthermore, through real-time information sharing, team members can swiftly obtain crucial information about the targets and road conditions, allowing for quick decision-making and enhancing the efficiency of target tracking.  Lastly, sharing validation data among team members helps eliminate erroneous data or false information, thereby enhancing the reliability of target tracking.
\end{itemize}
In summary, an enhancement in the task completion effect could be achieved by establishing stable team social networks to facilitate information sharing. The key factors in ensuring the stability of team social networks are identified as strong trust relationships among team members.\\
In the CMCS scenarios, the privacy information of team members, including sensing data, temporal and spatial location information, and social relationships, is generated and stored on the platform. Harbers et al. \cite{r37} found the trade-off between privacy loss and team performance. Privacy information leakage could be a threat to team members' security. Consequently, less information tends to be shared by team members when tasks are executed to prevent compromising their own privacy, resulting in a decrease in the task completion effect. How to improve the task completion effect while reducing privacy loss has been the focus of recent research. However, the focus of most research on privacy issues has been on specific technical improvements in different scenarios, such as pseudonym methods\cite{r15}, differential methods\cite{r12}, and cryptographic methods\cite{r20}. While privacy protection in certain situations could be offered by these methods, they face limitations like information distortion, privacy-utility trade-offs, and static strategies. In contrast, these limitations could be avoided by trust relationships, which provide a more dynamic, precise, and adaptable approach to privacy protection. Greenstadt et al. \cite{r38} studied the relationship between trust relationships and privacy loss. The prevention of each other's privacy information disclosure is facilitated by strong trust relationships among team members, thereby enabling the privacy protection of team social networks \cite{r10}.\\
In conclusion, the enhancement of the task completion effect and the reduction of privacy loss could be achieved by recruiting teams with strong trust relationships. However, the lack of direct trust relationships is often prevalent among most users on any OSP, which makes the evaluation of trust relationships between any two users a challenging task. The following frameworks for trust evaluation have been proposed by some scholars: Hybrid Trust Frameworks\cite{r41, r31, r43}, which incorporate various trust sources. However, challenges are faced in weight allocation and fusion method selection, making it difficult to determine the importance of different trust sources and affecting the accuracy of the final evaluation. Data-Driven Trust Frameworks\cite{r44}, which display good adaptability and comprehensiveness. However, these frameworks exhibit instability when dealing with incomplete or noisy data. Trust frameworks based on Graph Neural Networks(GNNs) possess several advantages in social network trust evaluation\cite{r40, r5}, including better modeling of information propagation, node feature learning, and nonlinear modeling abilities. However, the capture of complex interactions between nodes when dealing with sparse trust relationships is a struggle for traditional GNNs methods.\\
Scholars have previously developed recruitment algorithms that operated on a single server within the platform\cite{r46, r32}. However, when confronted with recruiting tasks of an extensive scale and range, this approach presented several limitations, including performance bottlenecks, network bandwidth constraints, and scalability issues. In the context of CMCS, a promising distributed paradigm has been proven to be edge computing \cite{r15, r27, r45}. Real-time analysis and low-latency processing of IoT devices and sensors are enabled by pushing data processing to the network's edge, thereby reducing latency and conserving bandwidth.\\
To address the aforementioned problems, in this paper, the recruitment algorithm for CMCS scenarios is proposed, and the Trust Reinforcement Evaluation Framework (TREF) is designed to resolve the following three challenges:
\begin{itemize}
\item How to design an efficient and precise trust evaluation framework to obtain trust relationships among all workers.
\item How to reasonably calculate the task completion effect of the team and design an effective recruitment algorithm to address the worker recruitment problem while reducing privacy loss.
\item How to handle worker recruitment on an extremely large scale and implement measures to enhance recruitment efficiency.
\end{itemize}
\begin{figure*}
    \centering
    \includegraphics[height=7cm]{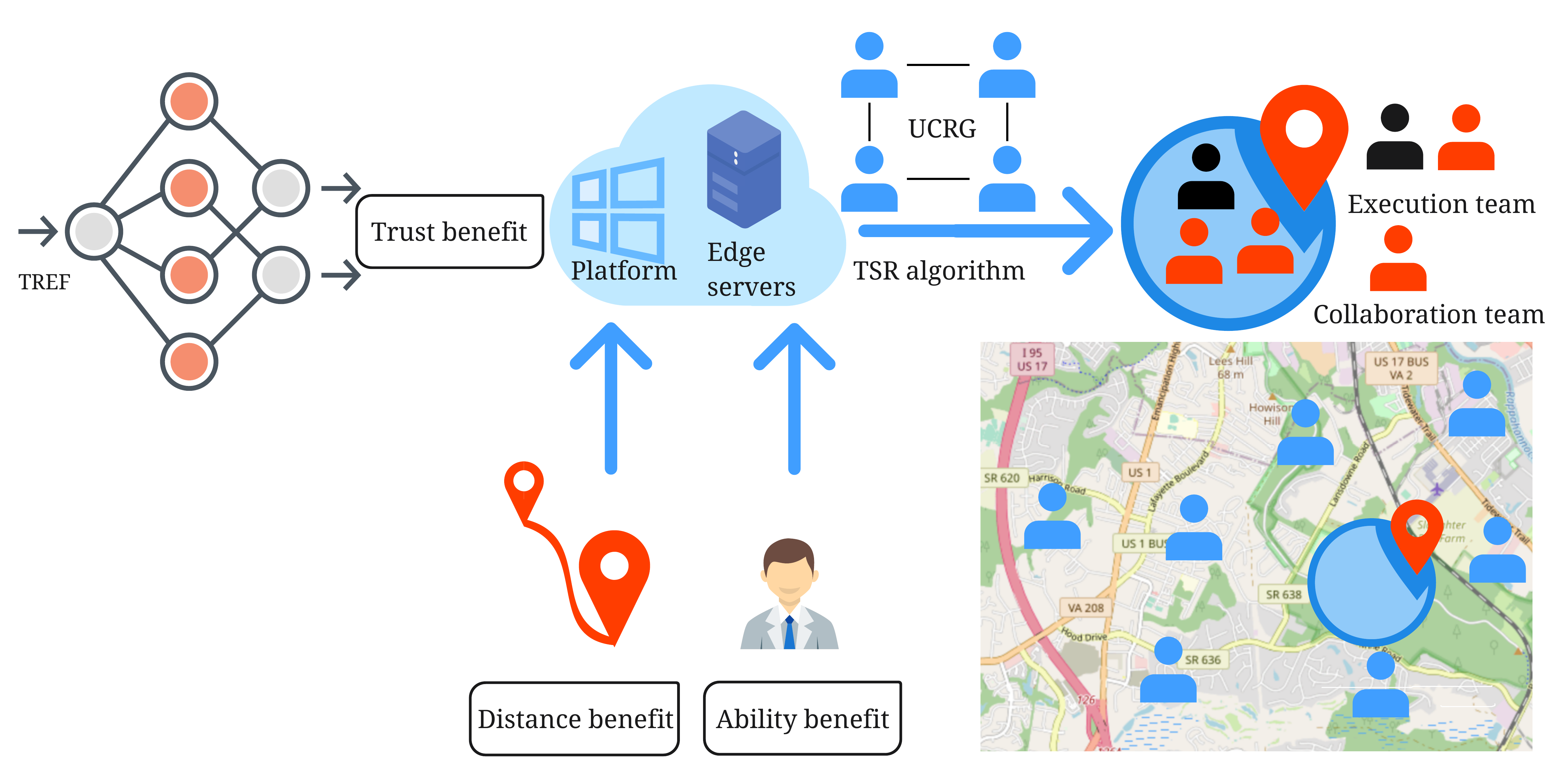} 
    \captionsetup{justification=centering}
    \caption{The architecture of worker recruitment in CMCS scenarios.}
    \label{fig:figure1}
\end{figure*}
The architecture for worker recruitment proposed in this paper is demonstrated in Fig. 1. Firstly, the TREF is designed in this paper based on GCNs to evaluate the asymmetric trust relationships among workers. In TREF, trust propagation is reinforced by expert knowledge that satisfies to trust properties. Subsequently, the Mini-Batch K-Means algorithm is utilized to partition the task publishing regions and deploy edge servers in different regions, enabling the realization of distributed worker recruitment in the form of edge computing. Once the task is published, the worker team is recruited by the edge server associated with the task. Specifically, when workers are within the maximum recruitment range of the task, the platform calculates the trust benefit of workers based on the asymmetric trust values evaluated from TREF. When calculating the ability benefit of workers, the platform takes into account various ability indicators of workers as well as the ability requirements of task publishers. Subsequently, the platform comprehensively considers the trust benefit, ability benefit, and distance benefit to calculate the task completion effect of the team and constructs the Undirected Complete Recruitment Graph (UCRG). In this graph, the nodes represent workers, and the weights of the undirected edges correspond to the task completion effects between workers.    At this point, the worker recruitment problem is modeled as UCRG, i.e., recruiting the team with the best task completion effect from UCRG, based on the number of workers required for the task. This problem is proven to be NP-hard in Section 4. Lastly, the specific Tabu Search Recruitment (TSR) algorithm is proposed in this paper for UCRG, aiming to recruit the optimal execution team for each task. The algorithm also considers conflicts during the recruitment process and a privacy-based recruitment strategy, ensuring that the privacy loss constraints set by the task publishers are satisfied by the selected collaboration team from the execution team.
The primary contributions of this paper are summarized as follows:
\begin{itemize}
\item TREF is proposed based on GCNs in this paper. Expert knowledge satisfying trust properties is employed in TREF to reinforce trust propagation. Effective trust evaluation is achieved by capturing the asymmetric trust properties of workers through representing their active trust and passive trust.
\item To the best of our knowledge, this paper is the first to consider the workers' abilities, distances, and asymmetric trust values simultaneously in CMCS worker recruitment. The worker recruitment problem is modeled as UCRG, and TSR algorithm is proposed in this paper to search for approximate solutions. Additionally, by utilizing clustering algorithms to partition regions and deploying edge servers, distributed worker recruitment is achieved in this paper.
\item Extensive experiments are conducted on five real-world datasets in this paper to verify the performance of the proposed TSR algorithm and TREF. The results demonstrate that TSR algorithm outperforms other baselines. TREF outperforms the state-of-the-art in the literature in terms of F1-Score and MAE.
\end{itemize}
The rest of this paper is organized as follows: The related work is reviewed in Section 2. The system model and the worker recruitment problem are described in Section 3. The algorithm is introduced in Section 4. The performance is evaluated and discussed in Section 5. Finally, Section 6 summarizes the whole paper and future work.
\section{RELATED WORKS}
\subsection{Trust Evaluation}
Trust evaluation is a vital research field in the information society, and many researchers have proposed various trust evaluation models in recent years. Yao et al. \cite{r53} proposed a multi-faceted trust inference model named Matri, based on a collaborative filtering method. The trust evaluation problem was transformed into a recommendation problem by them. They considered multiple aspects of trust and directly modeled multiple latent factors for each trustor and trustee from locally generated trust relationships. However, overfitting could easily occur in their method when trust relationships are sparse, which limits its evaluation accuracy. Liu et al. \cite{r54} proposed a trust evaluation algorithm named OpinionWalk. Trust was modeled through a Dirichlet distribution by them, and the trust network was represented using matrices. The network is searched in a breadth-first manner by their method, and user trustworthiness is calculated iteratively. However, when dealing with large, highly interconnected trust networks, Dirichlet distribution modeling method is unable to fully capture complex trust properties. Liu et al. \cite{r55} introduced a trust evaluation method named NeuralWalk. The neural network of WalkNet is used by this method to simulate the process of single-hop trust propagation and fusion. The single-hop trust rules are applied to iteratively evaluate unknown multi-hop trust relationships between users. However, poor scalability is a drawback of their method due to its computational and memory complexity.
GNNs is an emerging machine learning model for handling graph-structured data \cite{r17, r42, r7, r74}. It's excellent information propagation modeling ability and non-linear modeling capability pave a new way for trust evaluation. Node feature representations are updated in GNNs as the aggregated result of considering its neighboring nodes' information, which allows effective information transmission and learning to be carried out by GNNs while considering the graph's topology. Some researchers have proposed efficient trust evaluation frameworks based on GNNs \cite{r5,r40}. Lin et al. \cite{r5} proposed a trust evaluation framework for online social networks based on GCNs named Guardian. Popularity trust propagation and participation trust propagation for nodes are separately conducted by them in the trust convolution layer. By stacking multiple trust convolution layers, the range of trust propagation is expanded, thereby leading to the achievement of efficient trust evaluation. Huo et al. \cite{r40} proposed a trust evaluation framework based on GNNs named TrustGNN. They integrated the propagative nature and composable nature of trust and introduced the concept of trust chains to simulate the propagation pattern of trust. They learned the importance coefficients of different types of trust chains through an attention mechanism and used these to aggregate information from multiple trust chains. However, both the Guardian and TrustGNN methods failed to adequately capture the trust properties of social networks, thereby impacting the accuracy of trust evaluation, particularly in the case of sparse social networks.
\subsection{Worker Recruitment in MCS}
The worker recruitment problem has remained a focal point of MCS research for a long time. In view of the diversity of workers and potential collaboration among them, the selection of appropriate workers from candidatesomplish tasks is seen as crucial. This is because the recruited workers directly affect the task completion effect. Wang et al. \cite{r1} proposed a graph theory-based algorithm and a multi-round User Recruitment strategy using the combinatorial Multi-armed Bandit model (URMB) to recruit workers with objective abilities and subjective collaboration possibilities. However, given the asymmetry of trust relationships among workers, their approach to defining the subjective collaboration possibilities between workers as symmetrical is insufficient. Estrada et al. \cite{r47} proposed a multi-level recruitment strategy, which includes a multi-objective task allocation algorithm based on the particle swarm optimization method, a queuing scheme for tasks, and a delegation mechanism through social networks. However, the possibility of collaboration among workers in executing tasks, which could affect the task completion effect, was not considered by their method. Hamrouni et al. \cite{r8} proposed CMCS recruitment strategies based on the platform and the leader, respectively. A low-complexity recruitment method using GNNs was developed by them. However, the social relationships between team members, which could affect team formation stability, were failed to be considered by their method. Wang et al. \cite{r51}  proposed an Acceptance-aware Worker Recruitment (AWR) approach for MCS. They utilized the Jaccard similarity coefficient to determine the probability of task propagation between workers, and propagated MCS tasks within the workers' social network. However, their method solely considers direct trust relationships within the workers' social network, limiting the scope of task propagation. Consequently, it hinders the participation of high-quality workers with potential trust relationships in collaborative task completion. Chen et al. \cite{r33} proposed a task recommendation algorithm for MCS, which uses the extroversion of social network users and the intimacy between users to maximize the number of tasks completed. However, they simply defined the intimacy between users as the proportion of common friends, which is too simplistic and unable to truly reflect the degree of intimacy between users. Wang et al. \cite{r49} used influence propagation strategies in social networks to address the worker recruitment problem in MCS. To achieve the goal of maximizing coverage, two algorithms were proposed, i.e., an iterative greedy worker selection algorithm based on predicted mobility and an accelerated worker selection algorithm based on the mutual geographical relationship between friends. However, as the abilities of the workers were not considered by them, their strategy was unable to achieve optimal results in terms of the task completion effect.
\section {SYSTEM MODEL AND PROBLEM FORMULATION}
\begin{table}[h!]
\caption{Explanation of formula symbols}
\label{table}
\setlength{\tabcolsep}{1.4mm}
\begin{tabular}{ll}         
\toprule 
Notation & Descriptions \\   	
\midrule
\itshape{$V$} & The node set\\
\itshape{$e_{u\rightarrow v}$} & The trust relationship from the trustor $u$ to the trustee $v$\\
\itshape{$\omega_{u\rightarrow v}$} & The trustworthiness of the trustor $u$ to the trustee $v$\\
\itshape{$\overline{\omega}_{i\rightarrow j}$} & The potential trustworthiness of the trustor $i$ to the trustee $j$\\
\itshape{$N_I(u)$} & The set of trustors who trust the trustee $u$\\
\itshape{$N_O(u)$} & The set of trustees trusted by the trustor $u$\\
\itshape{$R, W, T$} & The recruitment region set,worker set and task set\\
\itshape{$C_y$} & The candidate team of $t_y$\\
\itshape{$\theta_y$} & The execution team of $t_y$\\
\itshape{$\overline{\theta}_y$} & The collaboration team of $t_y$ \\
\itshape{$a_{iy}$} & The ability benefit of $w_i$ in executing $t_y$\\
\itshape{$t_{i\rightarrow{j}}$} & The trust value of $w_i$ to $w_j$ \\
\itshape{$s_{ij}$} & The trust benefit of the worker pair $(i,j)$ \\
\itshape{$p_{iy}$} & The distance benefit of $w_i$ in executing $t_y$\\
\itshape{$U_{ij}$} & The task completion effect of the worker pair $(i,j)$\\
\itshape{$\overline{U}_i$} & The total task completion effect of $w_i$ collaborating \\ & with other workers in the execution team\\
\itshape{$Q(\theta_y)$} & The Quality of Data (QoD) of the execution team of $t_y$\\
\itshape{$pl_y$} & The privacy loss of the collaboration team $\overline{\theta}_y$\\
\bottomrule
\end{tabular}
\label{tab1e}
\end{table}
\subsection{Trust Evaluation}
The missingness of trust relationships in online social networks has been proposed by scholars \cite{r5}, as shown in Fig. 2(a). Moreover, in practical social relationships, a potential trust relationship exists between any two users who lack a direct trust relationship. Therefore, trust relationships are classified into two categories in this paper, i.e., direct trust relationships and potential trust relationships.
In the context of CMCS, the problem of trust evaluation among workers is investigated in this paper. Specifically, the social network of workers is modeled using a directed graph structure $G=\left(V, E,\Omega\right)$, where any nodes $u,v \in V$ are represented as workers. The trust relationship from the trustor $u$ to the trustee $v$ is represented by $e_{u\rightarrow v}\in E$, and the trustworthiness of the trustor $u$ to the trustee $v$ is denoted by $\omega_{u\rightarrow v}\in \Omega$. The PGP and Advogato datasets are employed in this paper, with their trustworthiness categories being $w \in \{$Observer, Apprentice, Journeyer, Master$\}$, $|\omega|=4$. To facilitate modeling and to quantify trustworthiness into trust values, trustworthiness is represented using one-hot encoding and real-number in this paper, i.e., $\{$Observer, Apprentice, Journeyer, Master$\}=[0, 0, 0, 1]^T, [0, 0, 1, 0]^T, [0, 1, 0, 0]^T, [1, 0, 0, 0]^T = \{0.5, 1, 2, 3\}$. The potential trustworthiness of the trustor $i$ to the trustee $j$ to be evaluated is denoted by $\overline{\omega}_{i\rightarrow j} \in \overline{\Omega}$, where nodes $i, j\in V$ and $\overline{e}_{i\rightarrow j} \notin E$. This paper defines $N_I(u)$ to represent the set of trustors who trust the trustee $u$ and $N_O(u)$ to represent the set of trustees trusted by the trustor $u$. In this sense, the in-degree and out-degree of $u$ are represented by $|N_I(u)|$ and $|N_O(u)|$, respectively. 
Finally, the worker social trust evaluation problem is formally defined in this paper, i.e., to evaluate the potential trustworthiness $\overline{\omega}_{i\rightarrow j}$, where nodes $i,j\in V$ and $\overline{e}_{i\rightarrow j} \notin E$.

\subsection{ Worker Recruitment }
The recruitment algorithm proposed in this paper is applicable to general CMCS scenarios. Tasks are published by the platform in chronological order according to the requirements of the task publishers, and the platform recruits workers for each task. Workers are required to travel to a specific task execution area to accomplish a task. They can only execute one task at a time and are only able to accept a new task after the current one has been completed.\\
In the context of CMCS, the recruitment region set, worker set, and task set are represented as $R = \{r_1, r_2,\ldots, r_k\}$, $W = \{w_1, w_2,\ldots, w_m\}$, and $T=\{t_1, t_2,\ldots, t_n\}$, respectively.
Each worker $w_i$ $(1\leq i\leq m)$ is represented as a tuple of four elements: $<loc_i, len_i, num_i, trust_i>$, where $loc_i$ represents the location of $w_i$. $len_i$ and $num_i$ represent the total mileage and total number of tasks historically completed by $w_i$. $trust_i = \{t_{i \rightarrow j}, \forall j\in W, i\neq j\}$, where $t_{i \rightarrow j}$ represents the trust value of $w_i$ to $w_j$.
Similarly, each task $t_y$ $(1\leq i\leq y)$ is represented as a tuple of six elements: $<loc_y, \alpha_y, \beta_y, \zeta_y, |\theta_y|, z_y>$, where $loc_y$ represents the location of $t_y$. $\alpha_y$ and $\beta_y$ represent the ability weight parameters of $t_y$, satisfying the constraint $\alpha_y + \beta_y = 1$. $\zeta_y$ represents the privacy loss threshold of $t_y$. $|\theta_y|$ represents the number of execution team members recruited for $t_y$. $z_y$ represents the maximum recruitment range of $t_y$. $d_{iy}$ represents the distance from $w_i$ to $t_y$.\\ 
\textbf{Definition1: Candidate Team.} When $d_{iy}$ is less than $z_y$, worker $w_i$ is added to the candidate team $C_y$ for task $t_y$ by the platform.\\
\textbf{Definition2: Execution Team.} The execution team $\theta_y$ for task $t_y$ is recruited from $C_y$ by the platform.\\ 
\textbf{Definition3: Collaboration Team.} The collaboration team $\overline\theta_y$ for task $t_y$ is selected from $\theta_y$ by the platform. $|\overline{\theta}_y| \in [2, |\theta_y|]$ represents the number of collaboration team members for task $t_y$. The privacy loss of the collaboration team is required to be lower than the privacy loss threshold $\zeta_y$ of $t_y$. Information is shared among collaboration team members.\\
\textbf{Definition 4: Trust Benefit.} A stable social network is facilitated by good trust relationships. Through the positive network externality of a stable social network, information sharing is promoted, which in turn affects the task completion effect. This paper quantifies this impact of trust as the trust benefit. The trust benefit is greater when the trust value between workers is higher. Moreover, considering the asymmetry of trust relationships as proposed by scholars \cite{r5}, as shown in Fig. 2(b), the differences in asymmetric trust values among workers are considered by this paper when calculating trust benefits. For the worker pair $(i,j)$, the trust benefit $s_{ij}$ is shown as follows:
\begin{equation}
\begin{aligned}
{s}_{{ij}} = ({t}_{{i} \rightarrow {j}} + {t}_{{j} \rightarrow {i}}) \cdot {\exp}^{-|{t}_{{i} \rightarrow {j}} - {t}_{{j} \rightarrow {i}}|}
\label{eq. 1}
\end{aligned}
\end{equation}
When the difference in asymmetric trust values between workers $w_i$ and $w_j$ is large, their trust benefit during collaboration will decrease.
\begin{figure}
    \centering
    \includegraphics[height=3cm]{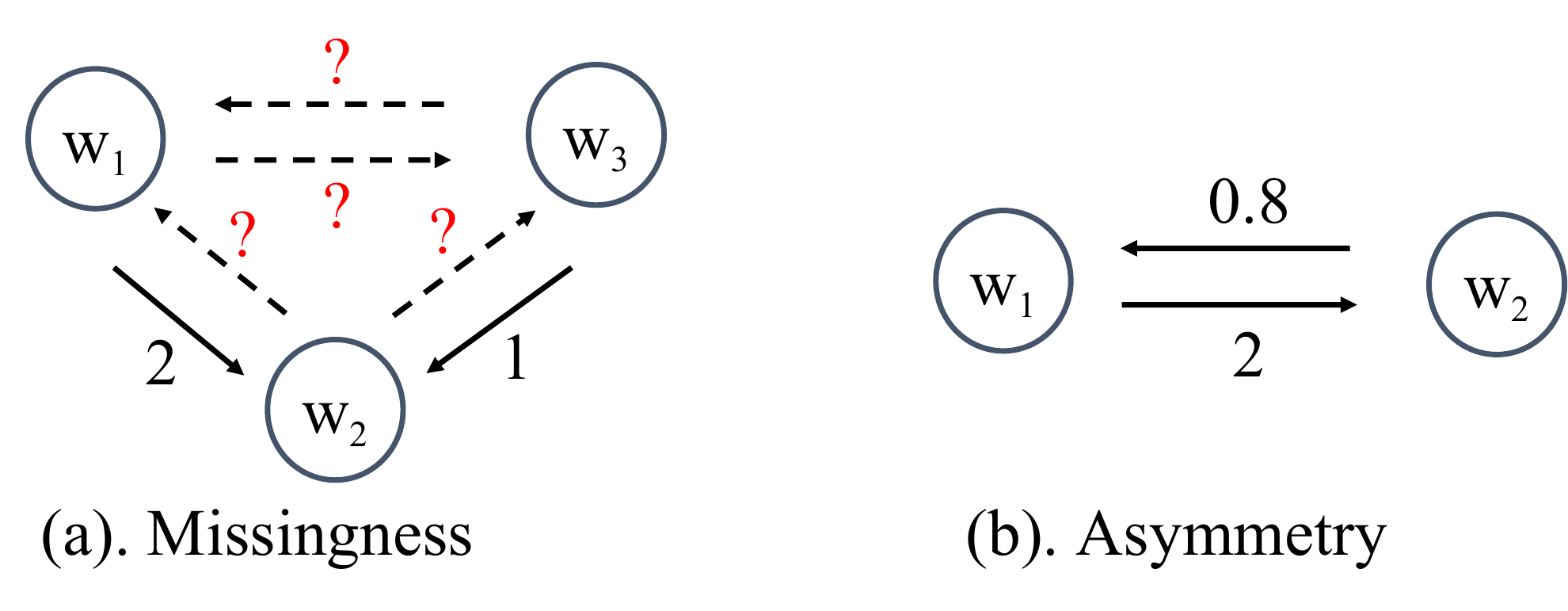} 
    \captionsetup{justification=centering}
    \caption{Property illustrations of trust relationships}
    \label{fig:figure2}
\end{figure}\\
\textbf{Definition 5: Ability Benefit.} The ability is influenced by different metrics.    For instance, in the case of Gaode Taxi, metrics such as total mileage and the total number of orders accepted serve as representations of a driver's ability, measuring the driver's business volume and activity level, thereby indirectly reflecting the driver's ability.    Similarly, these metrics represent the general CMCS worker's ability, and a direct impact on the task completion effect is made by the worker's ability.    In this paper, the impact of this ability is quantified as the ability benefit.  The total number of tasks completed and total mileage are pre-processed using Min-Max normalization in this paper, utilizing the worker's historical data. Then, according to the task publisher's ability weight parameters for different metrics, the worker's ability benefit is calculated.    The ability benefit $a_{iy}$ for worker $w_i \in W$ in executing task $t_y\in T$ is shown as follows:
\begin{equation}
\begin{split}
    a_{iy} = \epsilon\{\frac{\alpha_y \cdot [\mathrm{len}_i - \mathrm{len}_{\min}]}{\mathrm{len}_{\max} - \mathrm{len}_{\min}} + \frac{\beta_y \cdot [\mathrm{num}_i - \mathrm{num}_{\min}]}{\mathrm{num}_{\max} - \mathrm{num}_{\min}}\}
\label{eq. 2}
\end{split}
\end{equation}
where $\epsilon$ represents the ability benefit coefficient.\\
\textbf{Definition 6: Distance Benefit.} In CMCS scenarios, the task completion effect is directly influenced by the distance. This paper quantifies this impact as the distance benefit $p_{iy} \in [0, 1]$ for each worker $w_i\in W$ in executing task $t_y \in T$. When the distance $d_{iy}$ is smaller, worker $w_i$ could arrive at the task location earlier, enabling them to provide more shared information and task data, resulting in $p_{iy}$ being closer to $1$. In contrast, as the distance $d_{iy}$ increases, $p_{iy}$ tends to approach $0$. The calculation of the distance benefit $p_{iy}$ is shown as follows:
\begin{equation}
\begin{aligned}
p_{iy} &= \frac{1}{{\exp}^{\frac{d_{iy}}{\kappa}}}, \quad {d}_{iy} < z_y
\label{eq. 3}
\end{aligned}
\end{equation}
where $\kappa$ represents the distance attenuation ratio.\\
In conclusion, the task completion effect of worker pair is calculated in this paper, taking into account three objectives simultaneously, i.e., the worker's ability benefit, distance benefit, and trust benefit. For the worker pair $(i, j)$, the task completion effect $U_{ij}$ in executing task $t_y \in T$ is calculated as follows:
\begin{equation}
\begin{aligned}
U_{ij} & = \left(a_{iy}\cdot p_{iy} + a_{jy}\cdot p_{jy}\right) \cdot s_{ij}
\label{eq. 4}
\end{aligned}
\end{equation}
This paper assumes that each worker pair has a similar task completion effect. Therefore, the Quality of Data (QoD)\cite{r62, r1} is introduced to measure the average task completion effect of the execution team. The QoD of the execution team $\theta_y$ is represented as $Q(\theta_y)$, and its calculation is as shown below:
\begin{equation}
Q\left(\theta_y\right)=\sum_{i \in \theta_y} \overline{U}_i
\label{eq. 5}
\end{equation}
where $\overline{U}_i = \sum_{j \in \theta_y \setminus \{i\}} \frac{U_{ij}}{\left|\theta_y\right| \cdot \left(\left|\theta_y\right|-1\right)}$ represents the total task completion effect of worker $w_i$ collaborating with other workers in the execution team $\theta_y$.
To mitigate the risk of worker privacy leakage, the privacy loss threshold is set by the task publisher, requiring the privacy loss of the collaboration team to stay below this threshold. For the collaboration team $\overline{\theta}_y$, the calculation of its privacy loss $pl_y$ is as follows:
\begin{equation}
pl_y=\exp \left(-\frac{|\overline{\theta}_y|}{\sum_{i,j \in \overline{\theta}_y}\left(1-\frac{s_{ij}}{6}\right)\left(1-\prod_{i,j \in \overline{\theta}_y} AUC_{i\rightarrow j}\right)}\right)
\label{eq. 6}
\end{equation}
where $AUC_{i\rightarrow j}$ represents the trustworthiness accuracy from the worker $w_i$ to the worker $w_j$. If there is a direct trust relationship between workers, the trustworthiness accuracy is 1, and there is no need for trust evaluation. If there is a lack of direct trust relationship between workers, it is necessary to obtain potential trustworthiness through TREF, and the trustworthiness accuracy is equivalent to the accuracy of TREF trust evaluation. Table 1 summarizes the main mathematical symbols used in this paper.
\subsection{Problem Formulation}
When a new task is published, the task completion effect of each worker pair within the candidate team is calculated by the platform by considering the ability, trust, and distance benefits of the workers, which is then used as the weight to construct UCRG. Ultimately, the challenge faced in this paper is to be able to recruit a specified number of workers from the UCRG of each task to form the optimal execution team for each task, maximizing the task completion effect for all tasks. This constitutes the CMCS worker recruitment problem at a given moment, which could be expressed as:
\begin{equation}
\begin{array}{ll}
\text { Maxmize } & \sum_{y=1}^n Q\left(\theta_y\right)
\label{eq. 7}
\end{array}
\end{equation}
\begin{equation}
\begin{array}{l}
\text { subject to } 
\theta_c \cap \theta_f = \emptyset \quad \forall c, f \in [1, n], c \neq f
\label{eq. 8}
\end{array}
\end{equation}
Under the privacy loss constraint of task, the platform determines the collaboration team for each execution team. \\
This paper presents a detailed example of the CMCS worker recruitment problem. Firstly, a region is chosen by the task publisher to publish the task $t_y$. According to the worker recruitment requirements for task $t_y$, the number of execution team members $|\theta_y|$ is set at 10, the ability weight coefficients $\alpha_y$ and $\beta_y$ set at 0.8 and 0.2, the distance attenuation ratio$\kappa$  is set at 10, the privacy loss threshold  $\zeta$ is set at 0.7, and the maximum recruitment range $z_y$ is set at 50 kilometers. It should be noted that the size of the maximum recruitment range is dependent on the task publisher's sensitivity to task completion time.
Next, workers are recruited by the platform through the edge servers located in the region where the task is located. Specifically, workers located within 50 kilometers of $t_y$ are added to the candidate team by the platform. Then, the distance benefits of candidate team members are calculated by the platform. Additionally, the ability and trust benefits of candidate team members are calculated through the pre-processing of abilities and the trust evaluation by TREF. Furthermore, the task completion effects among candidate team members are calculated, and the worker recruitment problem is modeled as UCRG by the platform. Through the utilization of the specific TSR algorithm proposed in this paper, the execution team with the highest QoD for the task is successfully recruited by the platform, ensuring that the privacy loss of the collaboration team is less than 0.7. At this point, the worker recruitment problem has been resolved.
\section{ALGORITHM INTRODUCTION}
\subsection{Trust Evaluation}
TREF is proposed in this paper, as shown in Fig. 3. The strengths of GCNs in processing graph-structured data, information propagation, and node embedding learning are utilized by TREF, enabling the accurate extraction of each worker's trust characteristics. Moreover, expert knowledge is utilized by TREF to enhance trust evaluation. The trustworthiness and trust benefits among workers are ultimately obtained in this paper. TREF is composed of five parts:
\begin{figure}
    \centering
    \includegraphics[width=9cm, height=9.36cm]{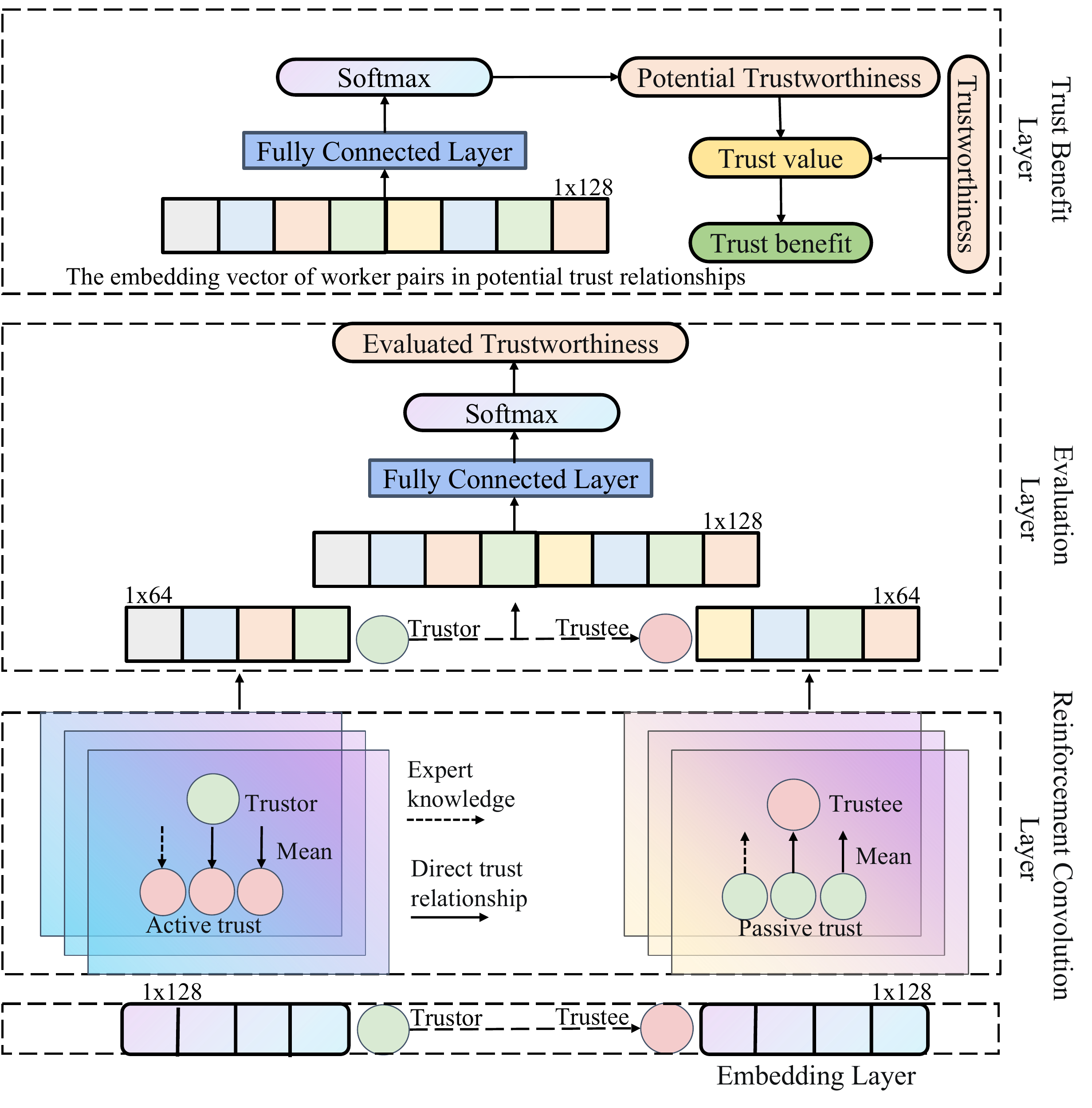} 
    \captionsetup{justification=centering}
    \caption{The architecture of TREF}
    \label{fig:figure3}
\end{figure}\\
\textbf{Embedding Layer}\\ The embedding layer aims to map workers to a low-dimensional space so that similar workers have similar representations. This aids in identifying potential similarities among workers and provides a basis for evaluating trustworthiness.  In this paper, the initial state of each node in the embedding layer, represented as $x[u]\in R^{D_e \times 1}$, is obtained using the Node2Vec pre-training model\cite{r9}.\\
\textbf{Reinforcement Convolution Layer}\\ In this paper, the reinforcement convolution network, inspired by the expert knowledge in reinforcement learning, is proposed to enhance trust propagation. Combining expert knowledge with trust propagation allows for better utilization of the nature of trust, ultimately improving trust evaluation. By introducing the propagative nature and composable nature of trust in this paper\cite{r40, r57}, an attempt is made to explore and apply expert knowledge to enhance the trust evaluation of worker social networks. Specifically, $w_1\rightarrow w_2\rightarrow w_3$, as shown in Fig. 4a, represents a trust propagation path. It is determined, according to the propagative nature of trust, that the trust propagation value from node $w_1$ to $w_3$ is $min(2,1)=1$. This nature has been confirmed in real social datasets \cite{r5}. There exist multiple trust propagation paths between $w_1$ and $w_3$, as shown in Fig. 4b. A Bernoulli random variable ($X$) is defined in this paper to decide which trust propagation value to select, based on the composable nature of trust. The maximum value event ($X = 1$) represents the choice of the maximum value among all propagation trust values, while the minimum value event ($X = 0$) represents the selection of the minimum value among all propagation trust values. The maximum probability of Bernoulli random variables, represented as Pr, is obtained through verification conducted on the Advogato and PGP datasets in this paper. It should be noted that the specific value of Pr could differ based on the dataset being used.
\begin{figure}
    \centering
    \includegraphics[height=2.7cm]{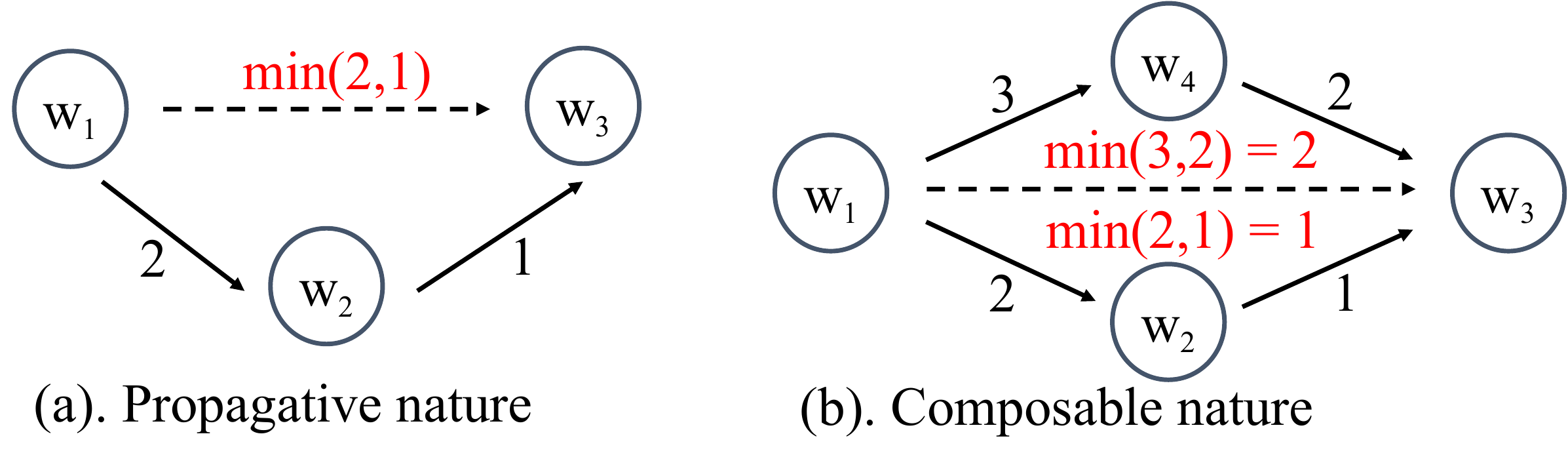} 
    \captionsetup{justification=centering}
    \caption{Illustrations of the nature of trust}
    \label{fig:figure4}
\end{figure}
The probability mass function of the Bernoulli random variable is presented as follows:
\begin{equation}
P(X = x) = 
\begin{cases} Pr, & \text{if } x = 1 \\ 
1 - Pr, & \text{if } x = 0 
\end{cases}
\label{eq. 9}
\end{equation}
After obtaining the probability mass function, based on the propagative nature and composable nature of trust, this paper constructs an incoming and outgoing edge with trustworthiness for each node and uses it as expert knowledge to reinforce trust evaluation. Taking the expert knowledge of node $u\in V$ as an example, if there exists a node $v\in V$ that has no direct trust relationship with node $u$, and the nature of trust is satisfied between node $u$ and node $v$, the expert knowledge of node $u$ is shown as follows:
\begin{equation}
\omega_{u \leftarrow v} = \begin{cases}
\min \left(t_{u \leftarrow {N}_I(u)}, t_{{N}_I(u) \leftarrow v}\right) & \text{if } |{N}_I(u)| = 1 \\
\max\left(\max_{{N}_I(u)},\min_{{N}_I(u)}\right) & \text{if } |{N}_I(u)| > 1,\\& \hfill Rd < Pr \\
\min\left(\max_{{N}_I(u)},\min_{{N}_I(u)}\right) & \text{if } |{N}_I(u)| > 1,\\& \hfill Rd > Pr \\
\end{cases}
\label{eq. 10}
\end{equation}
\begin{equation}
\omega_{u \rightarrow v} = \begin{cases}
\min\left(t_{u \rightarrow {N}_O(u)}, t_{{N}_O(u) \rightarrow v}\right) & \text{if } |{N}_O(u)| = 1 \\
\max\left(\max_{{N}_O(u)},\min_{{N}_O(u)}\right) & \text{if } |{N}_O(u)| > 1,\\& \hfill Rd < Pr \\
\min\left(\max_{{N}_O(u)},\min_{{N}_O(u)}\right) & \text{if } |{N}_O(u)| > 1,\\& \hfill Rd > Pr \\
\end{cases}
\label{eq. 11}
\end{equation}
where $Rd \in [0,1]$ is defined as the random variable that participates in the discrimination of the Bernoulli random variable.  $\max_{{N}_I(u)}$ and $\min_{{N}_I(u)}$ are the maximum and minimum of all propagation trust values from nodes $v\in V$ to node $u$, respectively. $\max_{{N}_O(u)}$ and $\min_{\hat{N}_O(u)}$ are the maximum and minimum of all propagation trust values from node $u$ to nodes $v\in V$, respectively.
The types of trust propagation between the trustor and the trustee are further studied in this paper. Trust propagation is categorized into two types: active trust $h_O [u]$ represents the average degree to which node $u$ actively trusts other nodes, and passive trust $h_I [u]$ represents the average degree to which node $u$ is trusted by other nodes. These two types of trust propagation remain independent during the message passing. As shown in Fig. 5(a), the passive trust value of node $B$ is determined to be $2$ by calculating the average of the trust values of all incoming edges of $B$. In Fig. 5(b), the active trust value of node $B$ is calculated to be $1.8$ by averaging the trust values of all outgoing edges of node $B$.
\begin{figure}
    \centering
    \includegraphics[height=1.8cm]{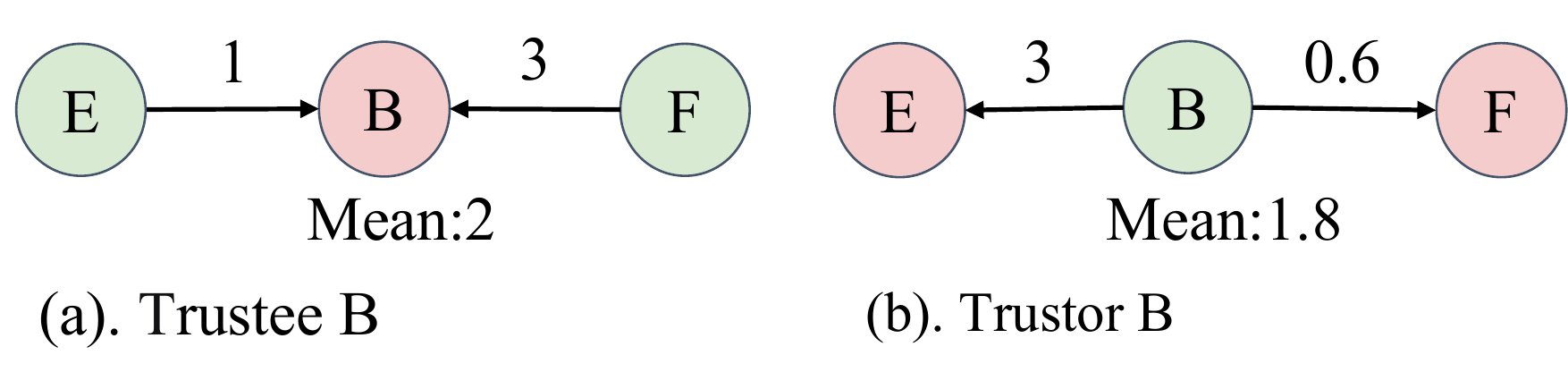} 
    \captionsetup{justification=centering}
    \caption{Trust propagation of trustee and trustor}
    \label{fig:figure5}
\end{figure}
Moreover, to capture multi-layer trust relationships, multiple reinforcement convolution layers are stacked together in this paper. This allows the model to receive trust relationships from $l$ layer neighbors. Higher-order trust propagation aids in the learning of complex trust relationships between nodes, which in turn enhances the precision of trustworthiness evaluation. The equation for higher-order trust propagation is as follows:
\begin{equation}
h_O^0[u] = x[u], \quad h_I^0[u] = x[u]
\label{eq. 12}
\end{equation}

\begin{equation}
\operatorname{At}_{u \rightarrow v}^l=h_O^{l-1}[v] \otimes\left\{W_{u \rightarrow v}^l \cdot \omega_{u \rightarrow v}\right\}
\label{eq. 13}
\end{equation}

\begin{equation}
\operatorname{Pt}_{u \leftarrow v}^l=h_I^{l-1}[v] \otimes\left\{W_{u \leftarrow v}^l \cdot w_{u \leftarrow v}\right\} 
\label{eq. 14}
\end{equation}

\begin{equation}
h_O^l[u]=\sigma\left(W_O^l\left(D_O^{-1} A_O {At}_{u \rightarrow v}^l\right)+b_O^l\right)
\label{eq. 15}
\end{equation}

\begin{equation}
h_I^l[u]=\sigma\left(W_I^l\left( A_I D_I^{-1}\operatorname{Pt}_{u \leftarrow v}^l\right)+b_I^l\right)
\label{eq. 16}
\end{equation}
where $l$ denotes the layers of trust propagation. $At_{u\rightarrow v}^l$ and $Pt_{u\leftarrow v}^l$ represent the feature representations of active trust and passive trust between node $u$ and node $v$ at layer $l$, respectively. $W_{u\leftarrow v}^l$, $W_{u\rightarrow v}^l$, $W_I^l$, and $W_O^l$ are the model's trainable weight parameters. $b_I^l$ and $b_O^l$ are the model's trainable bias term parameters. $D_O^{-1}$ and $D_I^{-1}$ represent the out-degree and in-degree matrix of node $u$, respectively. $A_I$ and $A_O$ represent the adjacency matrix of node $u$ in terms of in-degree and out-degree, respectively. $\otimes$ denotes the concatenation operator of two vectors.\\
\textbf{Evaluation Layer}\\The embedding vectors of node active trust and passive trust at layer $l$ are connected, then input into a fully connected layer and use the softmax function $\sigma$ to evaluate trustworthiness.
The evaluation layer could be formulated as:
\begin{equation}
\tilde{h}_{u \rightarrow v}=\sigma\left(W_{f c} \cdot(h_O^l[u] \otimes h_I^l[v])\right)
\label{eq. 17}
\end{equation}

\begin{equation}
\tilde{\omega}_{u \rightarrow v}={\operatorname{argmax}}\left(\tilde{h}_{u \rightarrow v}\right)
\label{eq. 18}
\end{equation}

\begin{equation}
\tilde{\omega}_{u \rightarrow v} \neq \tilde{\omega}_{v \rightarrow u}
\label{eq. 19}
\end{equation}
where $\tilde{\omega}_{u \rightarrow v}$ represents the evaluated trustworthiness of trustor $u$ to trustee $v$. $W_{fc}$ represents the trainable weight matrix for the four trustworthiness categories in the fully connected layer.\\
\textbf{Optimization Objective}\\ The model parameters are trained using the cross-entropy loss as the optimization objective, which measures the difference between the evaluated trustworthiness and the true trustworthiness. Throughout the training process, the Adam optimizer is employed to adjust the model parameters, minimizing the cross-entropy loss and thus achieving the objective of learning the trustworthiness between nodes. The loss function is as follows:
\begin{equation}
\tilde{L}=-\frac{1}{|\Omega|} \sum_{\left(\langle u, v\rangle, \omega_{u \rightarrow v}\in \Omega \right)} \log _{\tilde{\omega}_{u \rightarrow v}, {\omega_{u \rightarrow v}}}+\lambda \cdot\|\Theta\|_2^2
\label{eq. 20}
\end{equation}
where $\Theta=\left\{\left\{W_{u \leftarrow v}^l, \omega_{u \rightarrow v}^l, W_I^l, b_I^l, W_O^l, b_O^l\right\}_{l=1}^L, W_{f c}\right\}$ represents the model parameters. $|\Omega|$ represents the number of trustworthiness in $\Omega$. $\lambda$ controls $L_2$ regularization.\\
\textbf{Trust Benefit Layer}\\ After completing TREF training, the embedding vectors of node active trust and passive trust are obtained. These vectors are then connected and fed into the fully connected layer to evaluate potential trustworthiness between workers. Subsequently, both potential trustworthiness and trustworthiness are converted into trust values. Finally, the trust benefits among the workers are calculated and obtained in this paper.\\
The entire process is shown in Algorithm 1. Initially, the directed graph structure $G$ of the worker social network is inputted, and the initial state of the node embedding vector is obtained through Node2Vec pre-training (line 1). The expert knowledge is generated for each node in $V$ that satisfies the nature of trust (line 2). Subsequent to this, through forward propagation, the embedding vectors of active trust and passive trust are obtained for each node (lines 3-11). The trustworthiness in $\Omega$ is then evaluated. The cross-entropy loss between the evaluated trustworthiness and the true trustworthiness is calculated, and backward propagation is performed using the Adam optimizer to update the model parameters (lines 12-16). Upon completion of model training, using the embedding vectors of active trust and passive trust of nodes, along with weight matrices for the four trustworthiness categories, the potential trustworthiness among nodes is evaluated.(lines 17-20). Finally, trustworthiness and potential trustworthiness are mapped to trust values using the real-number mapping $[0, 0, 0, 1]^T, [0, 0, 1, 0]^T, [0, 1, 0, 0]^T, [1, 0, 0, 0]^T = {0.5, 1, 2, 3}$. The trust benefits among all workers are then calculated and outputted (lines 21-26).
\begin{algorithm}
\caption{Trust Evaluation Algorithm}
\label{alg: Trust_Evaluation_ Algorithm}
\begin{algorithmic}[1]
\Require G(V, E, $\Omega$)
\Ensure $\{s_{{ij}} | \forall w_i,w_j \in W, i \neq j\}$
\State Generate initial states of the embedding vectors of nodes for G through the pre-training with Node2Vec
\State Generate the expert knowledge for each node in $V$ through Eqs. \ref{eq. 9}--\ref{eq. 11}
\For{all $u \in V$}
  \For{$l = 1...L$}
    \State $h_O^0[u] = x[u],\quad h_I^0[u] = x[u]$
    \State $\operatorname{At}_{u \rightarrow v}^l=h_O^{l-1}[v] \otimes\left\{W_{u \rightarrow v}^l \cdot \omega_{u \rightarrow v}\right\}$
    \State $\operatorname{Pt}_{u \leftarrow v}^l=h_I^{l-1}[v] \otimes\left\{W_{u \leftarrow v}^l \cdot w_{u \leftarrow v}\right\}$
    \State $h_O^l[u]=\sigma\left(W_O^l\left(D_O^{-1} A_O {At}_{u \rightarrow v}^l\right)+b_O^l\right)$
    \State $h_I^l[u]=\sigma\left(W_I^l\left( A_I D_I^{-1}\operatorname{Pt}_{u \leftarrow v}^l\right)+b_I^l\right)$
  \EndFor
\EndFor
\For{all $\omega_{u\rightarrow v} \in \Omega$}
    \State $\tilde{h}_{u \rightarrow v}=\sigma\left(W_{f c} \cdot(h_O^l[u] \otimes h_I^l[v])\right)$
    \State $\tilde{\omega}_{u \rightarrow v}={\operatorname{argmax}}\left(\tilde{h}_{u \rightarrow v}\right)$
    \State $\tilde{L}=-\frac{1}{|\Omega|} \sum_{\left(\langle u, v\rangle, \omega_{u \rightarrow v}\in \Omega \right)} \log _{\tilde{\omega}_{u \rightarrow v}, {\omega_{u \rightarrow v}}}+\lambda \cdot\|\Theta\|_2^2$
\EndFor
\For{all $\overline{\omega}_{i\rightarrow j} \in \overline{\Omega}$}
    \State $\tilde{h}_{i \rightarrow j}=\sigma\left(W_{f c} \cdot(h_O^l[i] \otimes h_I^l[j])\right)$
    \State $\overline{\omega}_{i \rightarrow j}={\operatorname{argmax}}\left(\tilde{h}_{i \rightarrow j}\right)$
\EndFor
\For{all $\forall i,j\in V$}
    \State $\omega_{i\rightarrow j}$ or $\overline{\omega}_{i\rightarrow j}$ is mapped to $t_{i\rightarrow j}$
    \State $\omega_{j\rightarrow i}$ or $\overline{\omega}_{j\rightarrow i}$ is mapped to $t_{j\rightarrow i}$
    \State ${s}_{{ij}} = ({t}_{{i} \rightarrow {j}} + {t}_{{j} \rightarrow {i}}) \cdot {\exp}^{-|{t}_{{i} \rightarrow {j}} - {t}_{{j} \rightarrow {i}}|}$
\EndFor
\State \Return $\{s_{{ij}} | \forall w_i,w_j \in W, i \neq j\}$
\end{algorithmic}
\end{algorithm}
\subsection{Update Algorithm}    
In the current CMCS environment where mobile devices are pervasive and there's rapid expansion in the number of workers and tasks, a large amount of data needs to be handled by platform. Traditional CMCS recruitment strategies struggle with this computational load. To alleviate this burden on platform servers, a task region partitioning method based on clustering algorithms is proposed in this paper, and edge servers are deployed within these regions. Traditional clustering algorithms, however, when confronted with large-scale datasets, have problems such as high computational complexity and slow convergence speed. The Mini-Batch K-Means algorithm is adopted in this paper to resolve these problems. By reducing the amount of computation in each iteration through random sampling and online updating, this algorithm effectively accomplishes the clustering and regional partition of large-scale datasets. Following this, in this paper, the multiple ability indicators of workers are subjected to min-max normalization based on historical data. Subsequently, the platform updates the worker's ability benefit based on task requirements and ability indicators, and the distance benefit based on the distance between the worker and the task.
The entire process is shown in Algorithm 2. Firstly, the number of clusters $k$, the task set $T$, and the candidate team $C_y (\forall t_y \in T)$ are input. Then, the regions of task publishing are subsequently partitioned using the Mini-Batch K-Means clustering algorithm. This approach speeds up convergence by randomly selecting a subset of historical task samples, referred to as a mini-batch $M$, in each iteration. Each task point in $M$ is assigned to its nearest centroid $c_i \in \mathcal{C}$, forming the centroid-task set $M_i$. Subsequently, the location of $c_i \in \mathcal{C}$ is updated based on the mean location of task points in $M_i$ (lines 2-10). Once the location of the centroids stop changing or the iteration is complete, the spatial extent of each cluster is calculated. This is done by determining the maximum and minimum longitude and latitude of the task points within each centroid-task set, forming a region $r_i$ for each cluster $i$ (lines 11-13).
This ultimately results in a partition of the regions of task publishing into $k$ distinct regions, and the region set of task publishing is represented as $R = \{r_1,r_2,...,r_k\}$. The edge servers are then deployed in each region. (line 14). Finally, the workers' ability benefits and distance benefits for all tasks are then calculated and output (lines 15-21).
\begin{algorithm}
\caption{Update Algorithm}
\label{alg:worker_weights}
\begin{algorithmic}[1]
\Require $k$, $T$, $C_y ( \forall t_y \in T )$
\Ensure $\{(a_{iy}, p_{iy}) | \forall w_i \in C_y, t_y \in T\}$
\State Initialize centroids $\mathcal{C} = \{c_1, c_2, ..., c_k\}$
\While{$\Delta \mathcal{C} \neq 0 \vee Iteration < Iteration_{\text{max}}$}
    \State Select a mini-batch $M$ of size $b$ randomly
    \For{all $x \in M$}
        \State assign $x$ to nearest $c_i \in \mathcal{C}$
    \EndFor
    \For{all $c_i \in \mathcal{C}$}
        \State $c_i = \frac{1}{|M_i|}\sum_{x \in M_i} x$
    \EndFor
\EndWhile
\For{$i = 1 \cdot\cdot\cdot k$}
    \State $r_i = \{\max\limits_{x \in M_i} \text{x}_{lon}, \max\limits_{x \in M_i} \text{x}_{lat}, \min\limits_{x \in M_i} \text{x}_{lon}, \min\limits_{x \in M_i} \text{x}_{lat}\}$
\EndFor
\State $R = \{r_1,r_2,...,r_k\}$
\For{all $t_y\in T$}
    \For{all $w_i \in C_y$}
      \State $a_{iy} = \epsilon\{\frac{\alpha_y \cdot [\mathrm{len}_i - \mathrm{len}_{\min}]}{\mathrm{len}_{\max} - \mathrm{len}_{\min}} + \frac{\beta_y \cdot [\mathrm{num}_i - \mathrm{num}_{\min}]}{\mathrm{num}_{\max} - \mathrm{num}_{\min}}\}$
      \State $p_{iy} = \frac{1}{{\exp}^{\frac{d_{iy}}{\kappa}}}$
    \EndFor
\EndFor
\State \Return $\{(a_{iy}, p_{iy}) | \forall w_i \in C_y, t_y \in T\}$
\end{algorithmic}
\end{algorithm}
\subsection{TSR Algorithm}
Upon the acquisition of trust benefits, ability benefits, and distance benefits for the task's candidate team members, the platform calculates the values of the task completion effect among all candidate team members. These values are then mapped into the weights of undirected edges, leading to the construction of UCRG $G'$ with candidate team members as nodes. It is assumed that the task publisher is required to recruit $k$ workers, modeling the worker recruitment problem into finding the subgraph with the highest QoD amongst all subgraphs in $G'$ with an equal number of nodes to k. This paper uses a specific example to detail UCRG. As shown in Fig. 6, the platform needs to recruit the optimal execution team consisting of two workers from the three workers in $G'$. Upon comparison, it is discovered that the task completion effect of the worker pair $(w_1, w_3)$ outperforms all other teams composed of two workers, leading to their selection as the execution team for the task. As the scale of the worker recruitment problem expands, identifying the best execution team to maximize QoD becomes challenging, and the existing recruitment strategies fail to effectively function \cite{r8, r49}. Prior to the resolution of the worker recruitment problem, the problem was first proven to be NP-hard in this paper.
\begin{figure}[!t]
\centerline{\includegraphics[width=0.5 \columnwidth]{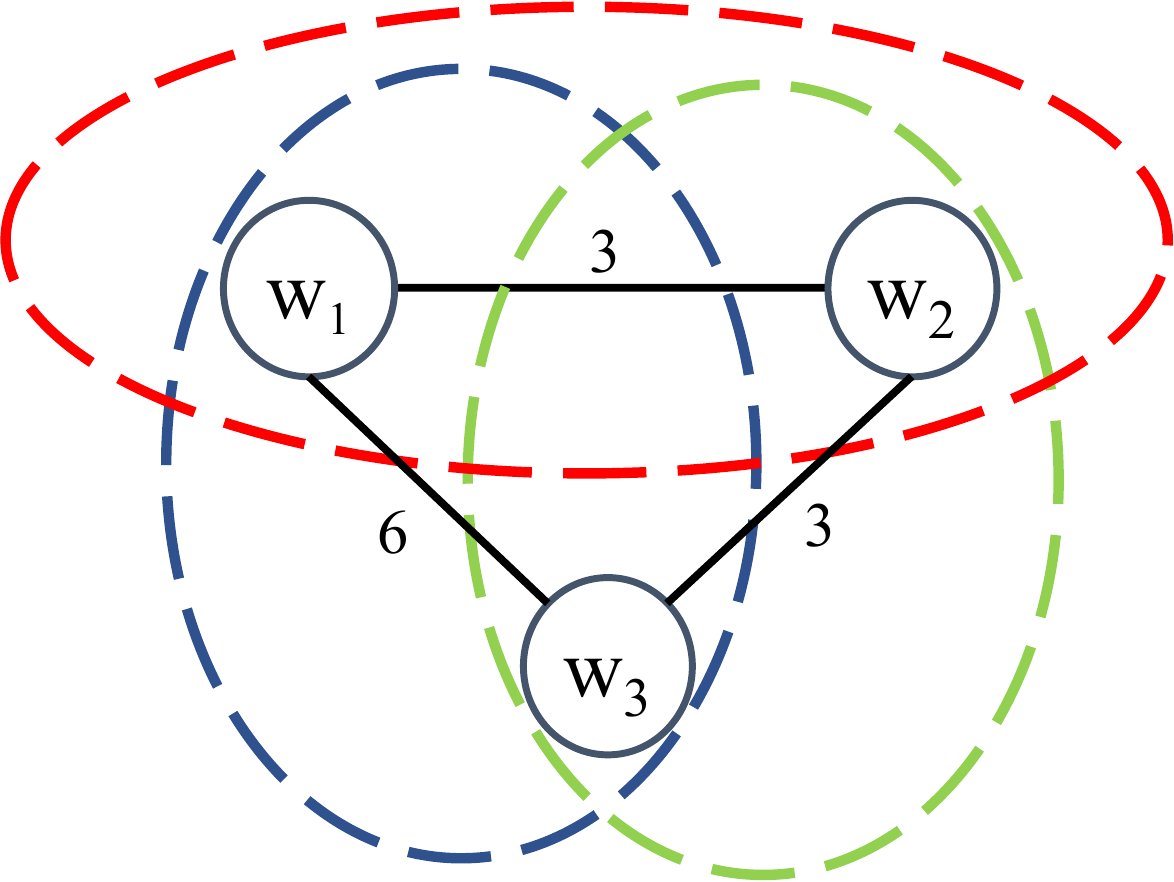}}
\caption{An example: UCRG $G'$}
\label{fig:figure7}
\vspace{-1.2em}
\end{figure}\\
Proof. The worker recruitment problem is viewed as the classic Knapsack problem to ensure the optimal QoD of the execution team, which is recognized as NP-hard. In this problem, a series of items exist (corresponding to workers), and each knapsack (corresponding to tasks) possesses a capacity (corresponding to the number of execution team members for the task) and a value (corresponding to the QoD of the execution team). The goal is to select items with the maximum value and place them into the knapsack, given the capacity limit. As the Knapsack problem is known to be NP-hard, the worker recruitment problem is transformed into the Knapsack problem through polynomial time reduction, thereby establishing that the worker recruitment problem is also NP-hard.\\
An approximate optimal solution to the worker recruitment problem is sought in this paper by proposing TSR algorithm based on the idea of taboo search\cite{r58}, as shown in Algorithm 3. The initial input includes the ability benefits, distance benefits, and trust benefits of the task's candidate team members, as well as the privacy loss thresholds of the tasks. Following this, the taboo list $T_l$ and other related parameters are initialized (line 1). The task completion effects of the worker pairs are calculated (lines 2-5) and serve as the weight of the undirected edges, thus modeling the worker recruitment problem as UCRG $G'$. A randomly generated execution team, $\theta_y$, in $G$ becomes the current solution, and its QoD is calculated (lines 6-7). During each iteration, neighbors of the current solution, which are similar solutions obtained by making minor changes to the current solution, are generated, and their respective QoDs are calculated (lines 8-9). The neighbor, represented as $\theta^{\prime}_y$, with the highest QoD is selected (line 10). If the neighbor's QoD exceeds that of the current solution's QoD and the neighbor is not in the taboo list $T_l$, which logs previously explored solutions to avoid redundant exploration, the current solution and its QoD are updated. Additionally, $\theta^{\prime}_y$ is added to the taboo list $T_l$. Upon completion of the iterations, the current solution $\theta_y$ is recruited as the execution team for each task $t_y \in T$ (lines 11-17). If a worker is recruited by multiple execution teams, a conflict is said to occur. In each conflicting execution team, the sum of the task completion effects of this worker with other workers in the team is calculated. The execution team with the highest task completion effect will recruit this worker, and the other teams will need to re-recruit workers until no conflicts occur (line 18). To select the collaboration team members for each execution team, the collaboration team initially consists of the execution team. If the privacy loss of the collaboration team surpasses the privacy loss threshold of the task, the platform removes the worker with the lowest average trust benefits with the other workers in the collaboration team. This removal continues until the privacy loss of the collaboration team falls below the privacy loss threshold of the task (lines 19-26). Finally, the execution team and collaboration team for each task are outputted (line 27).
\begin{algorithm}
\caption{TSR Algorithm}
\label{alg:tsr_algorithm}
\begin{algorithmic}[1]
\Require $(s_{ij}, a_{iy}, p_{iy}, \zeta_y) | \forall w_i,w_j \in C_y, t_y \in T, i\neq j\}$
\Ensure $\{\theta_y, \overline{\theta}_y) | \forall t_y \in T\}$
\State Initialize parameters and an empty tabu list $T_l$
\For{all $t_y\in T$}
  \For{all $w_i,w_j \in C_y, i\neq j$}
    \State $U_{ij} = \left(a_{iy}\cdot p_{iy} + a_{jy}\cdot p_{jy}\right) \cdot s_{ij}$
  \EndFor
  \State  Construct UCRG $G'$ and generate $\theta_y$ in $G'$
  \State $Q\left(\theta_y\right)=\sum_{i \in \theta_y} \sum_{j \in \theta_y \setminus \{i\}} \frac{\left(a_{iy}\cdot p_{iy} + a_{jy}\cdot p_{jy}\right) \cdot s_{ij}}{\left|\theta_y\right| \cdot \left(\left|\theta_y\right|-1\right)}$
  \For{Iteration 1, 2, $\dots$}
    \State Generate the neighbors of $\theta_y$ in $G'$
    \State Select the neighbor $\theta^{\prime}_y$ with the highest QoD
    \If{$Q\left(\theta^{\prime}_y\right) > Q\left(\theta_y\right)  \land \theta^{\prime}_y \notin T_l$}
      \State $\theta_y = \theta^{\prime}_y$
      \State $Q\left(\theta_y\right) = Q\left(\theta^{\prime}_y\right)$
      \State $T_l =T_l \cup \{\theta^{\prime}_y\}$
    \EndIf
  \EndFor
\EndFor
\State Resolve conflicts among all the execution teams
\State $\overline{\theta}_y = \theta_y$
\State $pl_y=\exp \left(-\frac{|\overline{\theta}_y|}{\sum_{i,j \in \overline{\theta}_y}\left(1-\frac{s_{ij}}{6}\right)\left(1-\prod_{i,j \in \overline{\theta}_y}AUC_{i\rightarrow j}\right)}\right)$
\If{$pl_y > \zeta_y$}
  \Repeat
    \State $w_i = \arg\min_i \left\{\sum_{j \in \overline{\theta}_y \setminus \{i\}} \frac{s_{ij}}{\left|\overline{\theta}_y\right| \cdot \left(\left|\overline{\theta}_y\right|-1\right)} , \forall i \in \overline{\theta}_y \right\}$
    \State $\overline{\theta}_y = \overline{\theta}_y \setminus \{w_i\}$

  \Until{the if condition does not hold}
\EndIf
\State \Return $\{\theta_y, \overline{\theta}_y) | \forall t_y \in T\}$
\end{algorithmic}
\end{algorithm}

\section{PERFORMANCE EVALUATION}
In this section, the settings of the experiment are primarily introduced. These include the datasets, baselines, performance metrics, and experimental parameters. A comprehensive analysis is then carried out to compare the performance of the algorithm proposed in this paper with the baselines. The experiments were conducted on a computer running the Windows 10 operating system. The hardware configuration of this computer included an Intel Core i5-12490F CPU, an NVIDIA GeForce RTX 4060Ti GPU with 8GB of video memory, and 32GB of RAM.
\subsection{Datasets}
Experiments are conducted based on five real datasets in this paper, aiming to evaluate the efficiency of TSR algorithm and TREF. Two real social network datasets, namely Advogato and PGP, are used as trust evaluation datasets. The Advogato dataset, specifically, originates from an online social network dedicated to open-source developers and encompasses four distinct trustworthiness categories. The PGP dataset, derived from an encryption program, also includes four different trustworthiness categories. The trustworthiness categories of these two datasets stand at \{$Observer, Apprentice, Journeyer, Master$\}, with their detailed data information shown in Table 2.
\begin{table}
  \centering
  \caption{Statistics of the trust evaluation datasets}
  \label{table2}
  \begin{tabular}{lcccc}
    \toprule
    Dataset & Nodes & Edges & Density & Average Degree \\
    \midrule
    ADVOGATO & 5.2K & 47.1K & 0.0035 & 18 \\
    PGP & 10.7K & 24.3K & 0.0004 & 4 \\
  \bottomrule
\end{tabular}
\end{table}\\
The worker recruitment datasets include an Advogato social dataset as well as three global check-in datasets: Brightkite, Gowalla, and Foursquare. Brightkite and Gowalla are early applications for geographic location sharing, while Foursquare is famous for its location check-in and recommendation features. Through these datasets, the location information of workers and tasks in the real world, including longitude, latitude, and timestamps, is acquired. The attributes of these three check-in datasets are shown in Table 3.
\begin{table}
  \centering
  \caption{Statistics of the check-in datasets}
  \label{table3}
  \begin{tabular}{lcccc}
    \toprule
    Dataset & Nodes & Locations & Check-ins & Collection Period \\
    \midrule
    Brightkite & 58,228 & 772,966& 4,491,143 & 2008 - 2010 \\
    Gowalla & 196,591 & 1,280,969& 6,442,890 & 2009 - 2010 \\
    Foursquare & 266,909 & 3,680,126 & 33,278,683 & 2012 - 2013 \\
\bottomrule    
\end{tabular}
\end{table}
It's worth noting that the trustworthiness among workers in the check-in datasets, which only includes "trust" and "distrust", could not represent the complex trust relationships between workers in reality. For this reason, the Advogato dataset, which includes four trustworthiness categories, is selected as the source of the trust relationships among workers in this paper. Trust evaluation is conducted on the Advogato dataset via TREF, obtaining the trust values among all workers.
\subsection{Baselines}
\textbf{Trust Evaluation Baselines}\\
TEF: This is a trust evaluation method based on GCNs that achieves effective trust assessment by capturing the asymmetric trust properties of workers. The only distinction between this method and TREF is that it does not use expert knowledge to enhance trust propagation.
TrustGNN\cite{r40}: This is a trust evaluation method based on GNNs. It defines trust chains to simulate trust propagation patterns and uses attention mechanisms to learn the importance coefficients of different types of chains, distinguishing the contributions of different propagation processes.
Guardian\cite{r5}: This represents the pioneering method within trust evaluation by using GNNs. It simulates trust interactions in social networks through popularity and participation, two types of trust relationships.
Matri\cite{r53}: This method leverages matrix factorization to achieve trust evaluation. It improves the accuracy of trust evaluation by introducing prior knowledge and trust propagation.
Opinionwalk\cite{r54}: In this approach, trust evaluation is conducted based on the modeling of trust via Dirichlet distribution.
Neuralwalk\cite{r55}: This is a trust evaluation method predicated on neural networks.\\
\textbf{Worker Recruitment Baselines}\\
Differential Evolution\cite{r6} (DE): This algorithm has been adapted in this paper to accommodate the worker recruitment requirements in CMCS scenarios.  Execution teams are recruited using differential mutation and crossover strategies in this optimization algorithm.
Particle Swarm Optimization\cite{r47} (PSO): This classic optimization algorithm has been modified in this paper to tackle worker recruitment challenges. By simulating bird foraging behavior, the swarm of particles is empowered to explore the search space and find the nearly optimal execution team for tasks.
Variable Neighborhood Search (VNS)\cite{r64}: It was a metaheuristic optimization algorithm that combined local and global search strategies. In the context of worker recruitment, this algorithm is adjusted in this paper to systematically explore multiple neighborhood structures to search for the approximately optimal execution team.
Simulated Annealing\cite{r65} (SA): This optimization algorithm simulates the cooling process of a solid. In this paper, this algorithm is adapted to accommodate worker recruitment cases. The identification of the approximately optimal execution team is facilitated by accepting suboptimal solutions to avoid settling into local optima.
Greedy for Minimum Privacy Loss (GMPL): The platform recruits a group of workers with the highest trust level to form the execution team, with the goal of minimizing the team's privacy loss.
Greedy for Maximum Distance Benefit (GMDB): The execution team closest to the location of task is recruited by the platform, aiming to maximize the distance benefit of the team.
Greedy for Maximum Ability Benefit (GMAB): The execution team with the best ability is recruited by the platform, aiming to maximize the ability benefit of the team.
Random: A group of workers is randomly recruited by the platform from candidate teams for the task, forming the execution team.
\subsection{Performance Metrics}
\textbf{Trust Evaluation Performance Metrics}\\
F1-Score: This metric, used for evaluating the performance of classification models, considers both the precision and recall of the model. The higher the F1-Score, the more precise the model's evaluations are deemed to be.\\
MAE: It is a commonly employed metric for evaluating regression models. The average absolute error between the model's evaluations and the actual values is measured by it. A smaller MAE value indicates less evaluation error, which means the evaluation result is closer to the real value.\\
\textbf{Worker Recruitment Performance Metrics}\\
QoD: This metric, calculated according to Eq. 5, is used to represent the average task completion effect of the execution team.\\
Privacy Loss: This metric, calculated according to Eq. 6, is used to represent the privacy loss of the collaboration team.\\
Running Time: This metric is used to indicate the running time of the recruitment algorithm.
\subsection{Parameter Settings}
\textbf{Trust Evaluation Parameter Settings}\\
In this paper, TREF is implemented via PyTorch, and a $128$-dimensional initial embedding vector is generated for each node using the Node2Vec method. The hyperparameters are set as follows: the learning rate is $0.01$, the dropout rate is $0.0$, the normalization coefficient is $10^{-5}$, and the output dimensions of the three layers of reinforcement convolution layers are $[64, 64, 64]$. The parameter settings for the baselines TrustGNN, Guardian, Matri, Neuralwalk, and Opinionwalk could be referenced in \cite{r40}, \cite{r5}, \cite{r53}, \cite{r55}, and \cite{r54}, respectively.\\
\textbf{Recruitment Algorithm Parameter Settings}\\
The number of clusters $k$ is set to $100$ and the mini-batch size $M$ is set to $3100$. The locations of both workers and tasks are randomly generated within the recruitment regions. For each task $t_y \in T $, the ability weight coefficients $\alpha_y$ and $\beta_y$ are set to $0.2$ and $0.8$, respectively. The maximum recruitment range $z_y$ is set to $200$, with measurements in kilometers. The distance attenuation ratio $\kappa$ is set to $10$.\\
\begin{table}[ht]
\centering
\caption{Performance evaluation of $80\%$ training set sizes on PGP and ADVOGATO: identifying the best results}
\begin{tabular}{lcccc}
\toprule
Methods & \multicolumn{2}{c}{Advogato} & \multicolumn{2}{c}{PGP} \\
\cmidrule(lr){2-3} \cmidrule(lr){4-5}
& F1-Score & MAE & F1-Score & MAE \\
\midrule
    TREF & \pmb{76.6}\% & \pmb{0.075} & \pmb{87.4}\% & \pmb{0.081} \\
    TrustGNN & 74.4\%  & 0.081 & 87.2\%  & 0.083 \\
    TEF & 74.6\% & 0.081 & 86.7\% & 0.087 \\    
    Guardian & 73.1\% & 0.087 & 86.7\% & 0.086 \\
    Matri & 65.3\% & 0.141 & 67.9\% & 0.136 \\
    NeuralWalk & 74.0\% &0.082& – &–\\
    OpinionWalk & 63.3\% &0.232& 66.8\% &0.251\\
\bottomrule
\end{tabular}
\end{table}

\begin{table}
  \centering
  \caption{Performance evaluation of different training set sizes on ADVOGATO}
  \label{tab:dataset_metrics}
  \begin{tabular}{lcccc}
    \toprule
    Method & Trainning Set(\%) & F1-Score & MAE \\
    \midrule
    TREF & 80\% & $\mathbf{76.0\%}\pm \mathbf{0.5\%}$ & $\mathbf{0.076}\pm \mathbf{0.002}$ \\
    TREF & 60\% & $\mathbf{74.9\%}\pm \mathbf{0.2\%}$ & $\mathbf{0.079}\pm \mathbf{0.001}$ \\
    TREF & 40\% & $\mathbf{73.1\%}\pm \mathbf{0.2\%}$ & $\mathbf{0.087}\pm \mathbf{0.001}$ \\
    TrustGNN & 80\%  & $74.4\%\pm 0.1\%$  & $0.081\pm 0.001$ \\
    TrustGNN & 60\%  & $72.6\%\pm 0.1\%$  & $0.088\pm 0.001$ \\
    TrustGNN & 40\%  & $70.1\%\pm 0.1\%$  & $0.096\pm 0.001$\\
    Guardian & 80\%  & $73.0\%\pm 0.1\%$  & $0.087\pm 0.001$ \\
    Guardian & 60\%  & $71.7\%\pm 0.2\%$  & $0.091\pm 0.001$ \\
    Guardian & 40\%  & $69.7\%\pm 0.0\%$  & $0.100\pm 0.000$ \\
    Matri & 80\%  & $65.0\%\pm 0.4\%$ & $0.141\pm 0.001$ \\
    Matri & 60\%  & $63.9\%\pm 0.3\%$ & $0.145\pm 0.001$ \\
    Matri & 40\%  & $61.7\%\pm 0.3\%$ & $0.153\pm 0.001$ \\
  \bottomrule
\end{tabular}
\end{table}
\begin{table}
  \centering
  \caption{Performance evaluation of different training set sizes on PGP}
  \label{tab:dataset_metrics}
  \begin{tabular}{lcccc}
    \toprule
    Method & Training Set (\%) & F1-Score & MAE \\
    \midrule
    TREF & 80\% & $\mathbf{87.4\%}\pm \mathbf{0.0\%}$ & $\mathbf{0.081}\pm \mathbf{0.000}$ \\
    TREF & 60\% & $\mathbf{86.7\%}\pm \mathbf{0.1\%}$ & $\mathbf{0.087}\pm \mathbf{0.001}$ \\
    TREF & 40\% & $\mathbf{85.8\%}\pm \mathbf{0.1\%}$ & $\mathbf{0.094}\pm \mathbf{0.001}$ \\
    TrustGNN & 80\%  & $87.2\%\pm 0.1\%$  & $0.083\pm 0.001$ \\
    TrustGNN & 60\%  & $86.3\%\pm 0.1\%$  & $0.090\pm 0.001$ \\
    TrustGNN & 40\%  & $85.4\%\pm 0.1\%$  & $0.097\pm 0.001$\\
    Guardian & 80\%  & $86.7\%\pm 0.1\%$  & $0.086\pm 0.001$ \\
    Guardian & 60\%  & $85.9\%\pm 0.1\%$  & $0.091\pm 0.001$ \\
    Guardian & 40\%  & $84.6\%\pm 0.1\%$  & $0.100\pm 0.001$ \\
    Matri & 80\%  & $67.3\%\pm 0.7\%$ & $0.136\pm 0.003$ \\
    Matri & 60\%  & $64.7\%\pm 0.1\%$ & $0.143\pm 0.004$ \\
    Matri & 40\%  & $60.5\%\pm 0.1\%$ & $0.164\pm 0.001$ \\
    \bottomrule
  \end{tabular}
\end{table}
\subsection{Evaluation Results and Analysis}
\subsubsection{Comparison of Trust Evaluation}
In this paper, $80\%$ and $20\%$ of the dataset are divided into the training set and test set, respectively, for the purpose of validating the trust evaluation performance of TREF. The experimental results are shown in Table 4, with the best outcomes highlighted in bold. Firstly, TREF was compared with all baselines in terms of F1-Score and MAE on the Advogato dataset .  Compared to the best baseline TrustGNN, TREF improved the F1-Score by $2.2\%$.  In terms of MAE, TREF also surpasses the best baseline, TrustGNN. Secondly, on the PGP dataset, TREF is compared with all baselines. TREF outperforms all baselines in both F1-Score and MAE. The Neuralwalk method cannot be evaluated on the PGP dataset due to memory limitations. Influenced by the potent learning abilities of neural networks, methods leveraging neural networks (TREF, TEF, TrustGNN, Guardian, Neuralwalk) surpass other methods in trust evaluation. Among these neural network-based methods, TREF, proposed in this paper, boasts a more refined design. The ablation experimental results regarding expert knowledge provide further validation that TREF outperforms TRE in terms of F1-Score and MAE. These results demonstrate that when guided by expert knowledge, TREF is able to capture workers' trust properties more effectively, resulting in superior evaluation performance.
In addition, a more detailed split is performed on the dataset in this paper to validate the performance of TREF. Specifically, the training set is divided into $80\%$, $60\%$, and $40\%$, with corresponding divisions of $20\%$, $40\%$, and $60\%$ for the test set. TrustGNN, Guardian, and Matri are selected as baselines. The experimental results are shown in Table 5 and Table 6.    It is evident that TREF has obtained the best results in all divisions of both the Advogato and PGP datasets, thereby further validating its superior trust evaluation performance.
\subsubsection{Comparison of QoD and Privacy Loss for Different Algorithms under Different Tasks}
When $\kappa = 10$, in order to compare the QoD and privacy loss performance of different algorithms under different tasks, six tasks were published in different regions by the task publishers, each region containing 200 workers. The recruitment of 10 workers was required for each task to form an execution team. Furthermore, to enable a more direct comparison of the privacy loss performance among different algorithms, this section and the subsequent experimental sections directly compare the privacy loss of the execution teams recruited by different algorithms. A lower privacy loss of the execution team allows the platform to select more collaboration team members from the execution team based on the task's privacy loss threshold.\\
Based on the experimental results in Fig. 7, it can be observed that the TSR algorithm is consistently outperformed by other algorithms in terms of QoD performance, which is reasonable. Due to the randomness of worker occurrence, there is no definite linear relationship among trust benefits, ability benefits, and distance benefits between workers. Therefore, the execution team that excels in one benefit may not necessarily excel in other benefits. The execution team is greedily recruited by the baselines through optimizing specific benefits, resulting in poor QoD performance. In contrast, the TSR algorithm optimizes QoD performance by comprehensively considering the recruitment of ability benefits, trust benefits, and distance benefits. Furthermore, according to the experimental results in Fig. 8, it is evident that the TSR algorithm outperforms all baselines in terms of privacy loss, except for the GMPL algorithm. This is because the TSR algorithm and the GMPL algorithm employ different strategies regarding trust benefits. The GMPL algorithm greedily recruits workers with the highest trust level, while the TSR algorithm merely considers trust benefits as one of the factors during worker recruitment. Hence, the overall trust benefit performance of the TSR algorithm is inferior to that of the GMPL algorithm. Additionally, there exists an inverse relationship between trust benefits among workers and privacy loss. Consequently, the TSR algorithm exhibits lower privacy loss performance compared to the GMPL algorithm while surpassing other baselines. Furthermore, the GMPL algorithm also has certain limitations, as it defines the worker's trust level as the average trust value among all other workers. This implies that the execution team, consisting of workers with the highest trust level, may not necessarily acquire the highest trust benefits. Even if the trust level of the execution team members is the highest, it cannot guarantee that they will have the highest trust values among themselves. Furthermore, having the highest trust level among the execution team members does not guarantee a smaller difference in asymmetric trust values between them. Therefore, this leads to higher privacy loss values in the GMPL algorithm compared to the TSR algorithm, as shown in Fig. 8(b) for task $t_1$.
\begin{figure*}[!t]
\centering
\begin{minipage}{\textwidth}
  \includegraphics[width=0.33\linewidth]{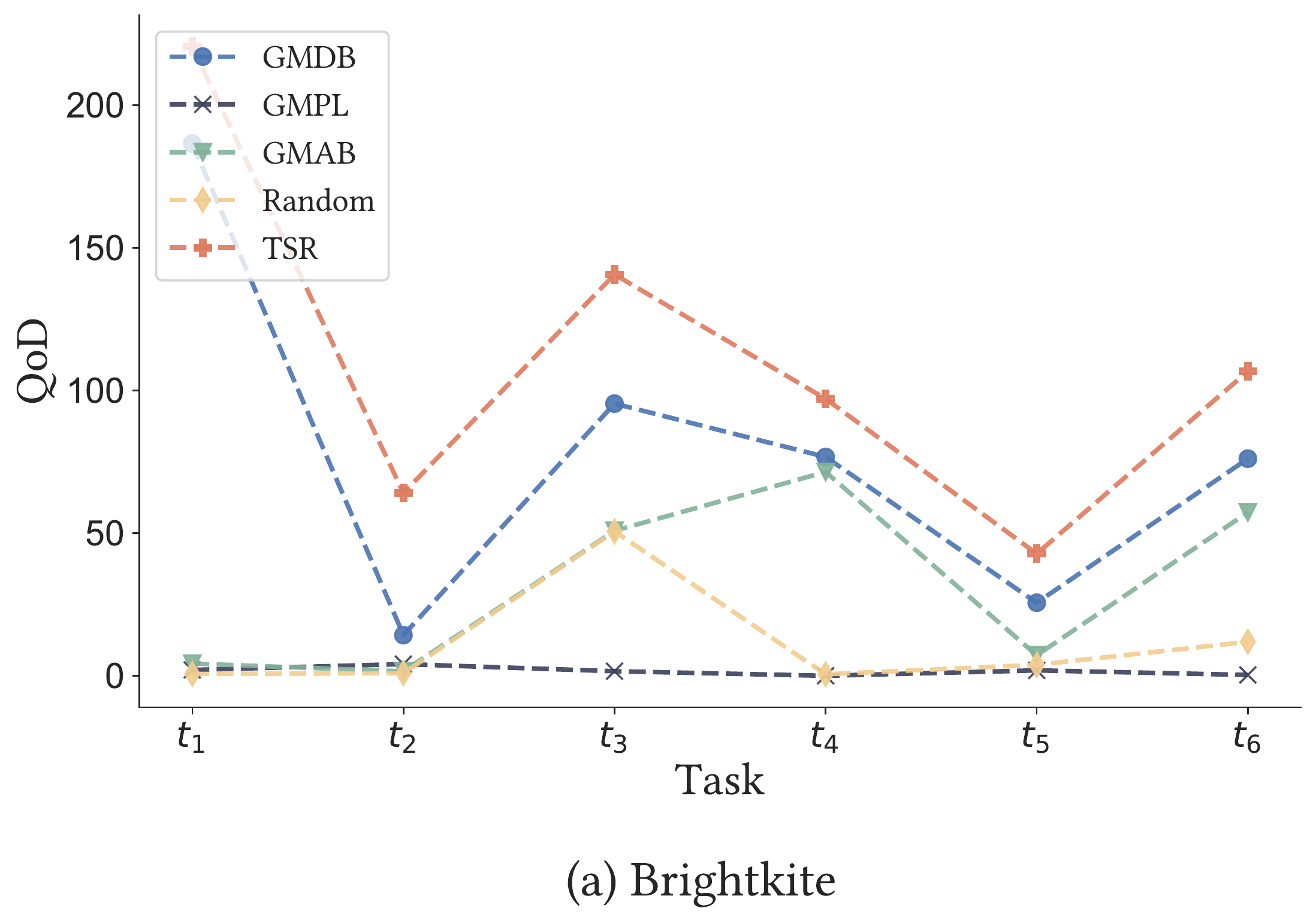}
  \hfill
  \includegraphics[width=0.33\linewidth]{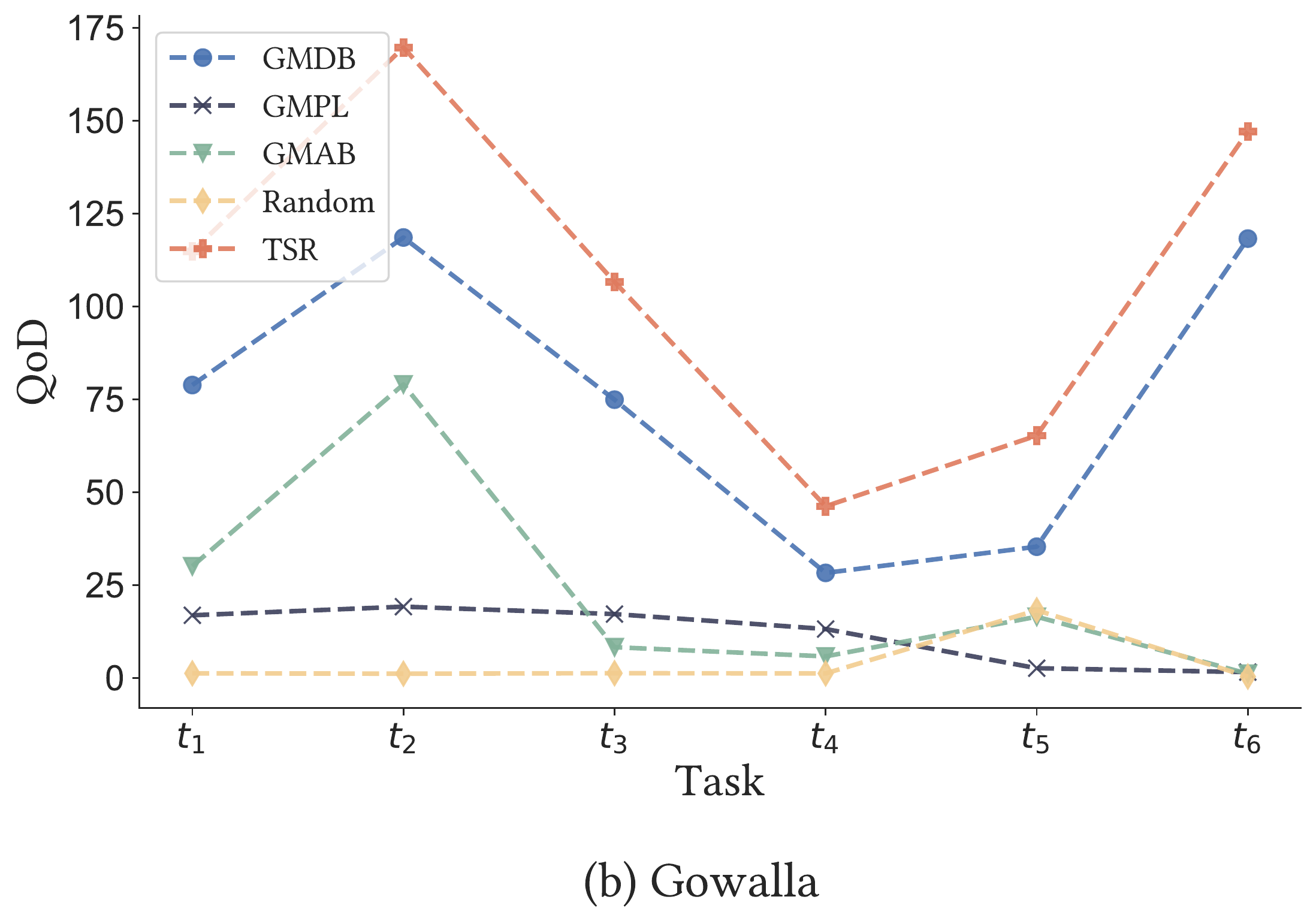}
  \hfill
  \includegraphics[width=0.33\linewidth]{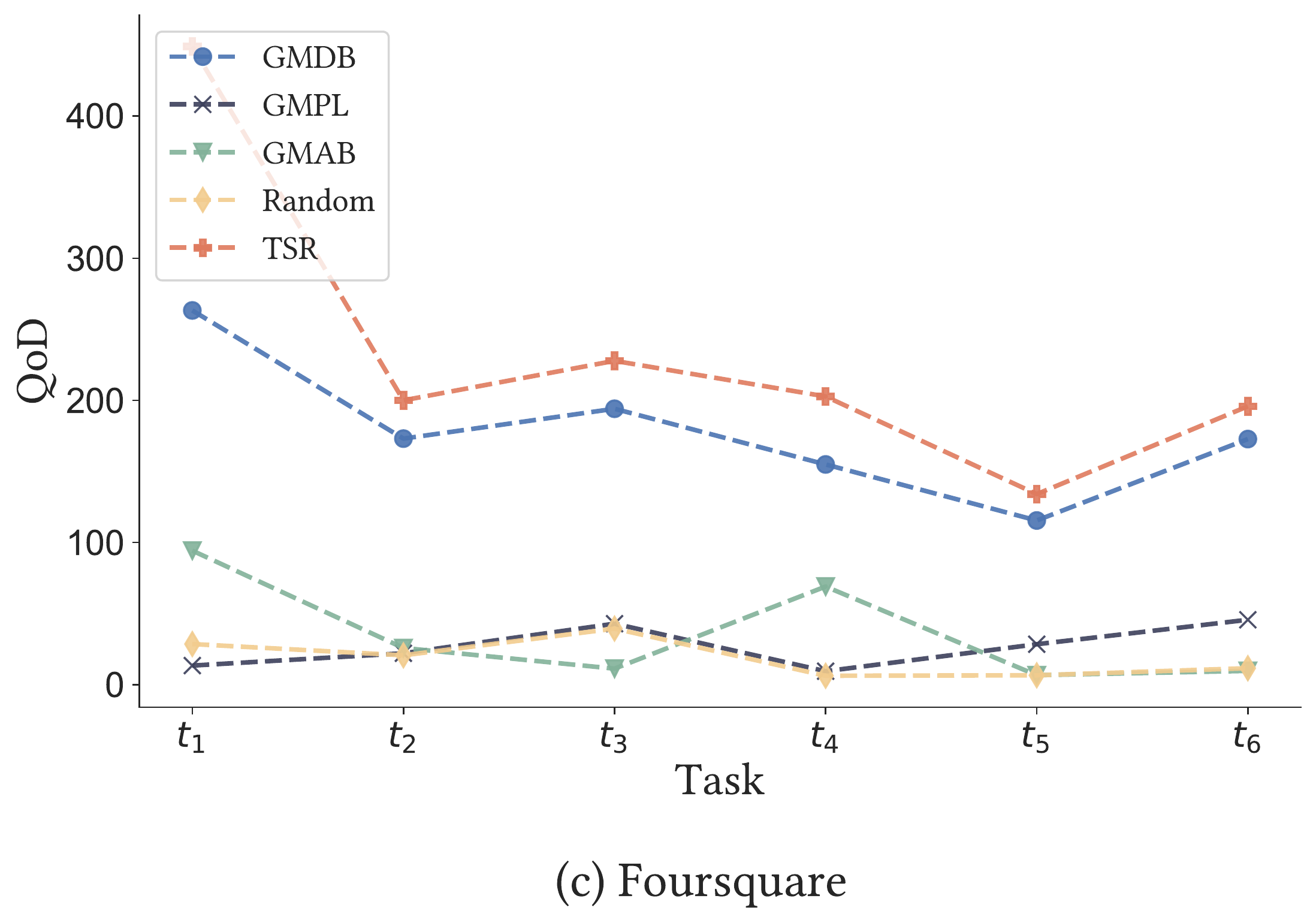}
  \captionsetup{justification=centering, singlelinecheck=false}
  \caption{Comparison of QoD for different algorithms under different tasks}
  \label{fig:both_figures}
\end{minipage}
\vspace{-1.2em}
\end{figure*}
\begin{figure*}[!t]
\centering
\begin{minipage}{\textwidth}
  \includegraphics[width=0.33\linewidth]{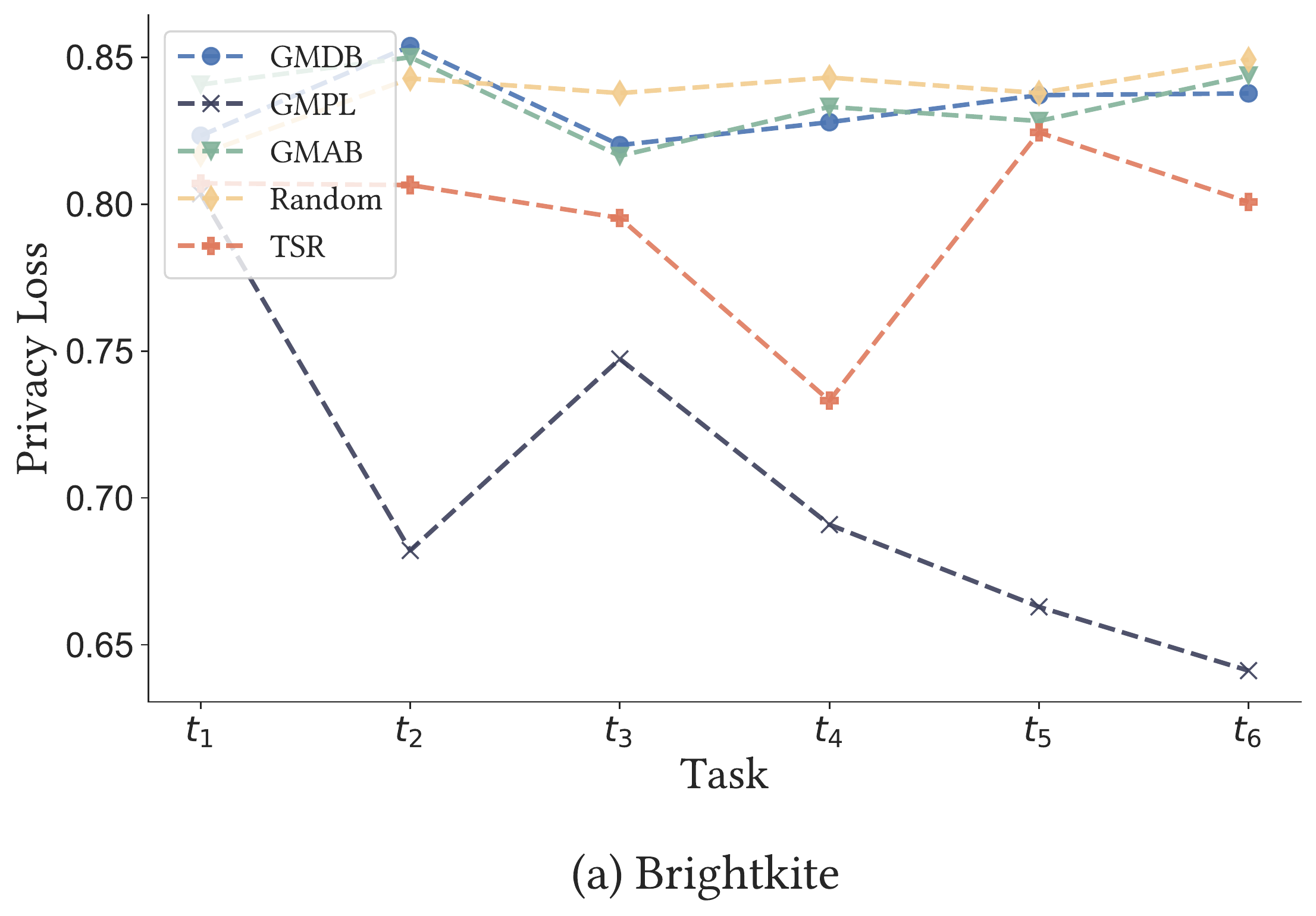}
  \hfill
  \includegraphics[width=0.33\linewidth]{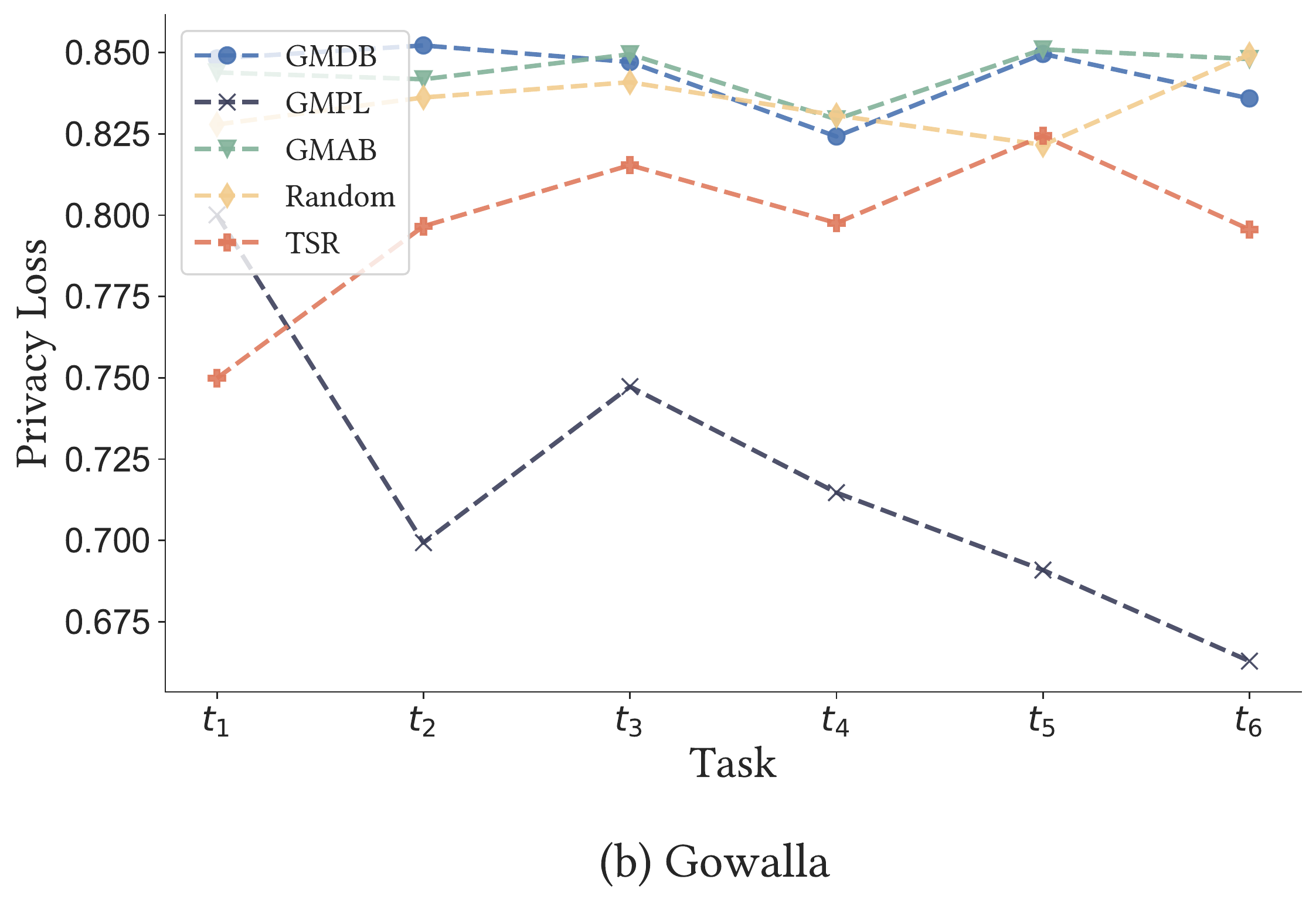}
  \hfill
  \includegraphics[width=0.33\linewidth]{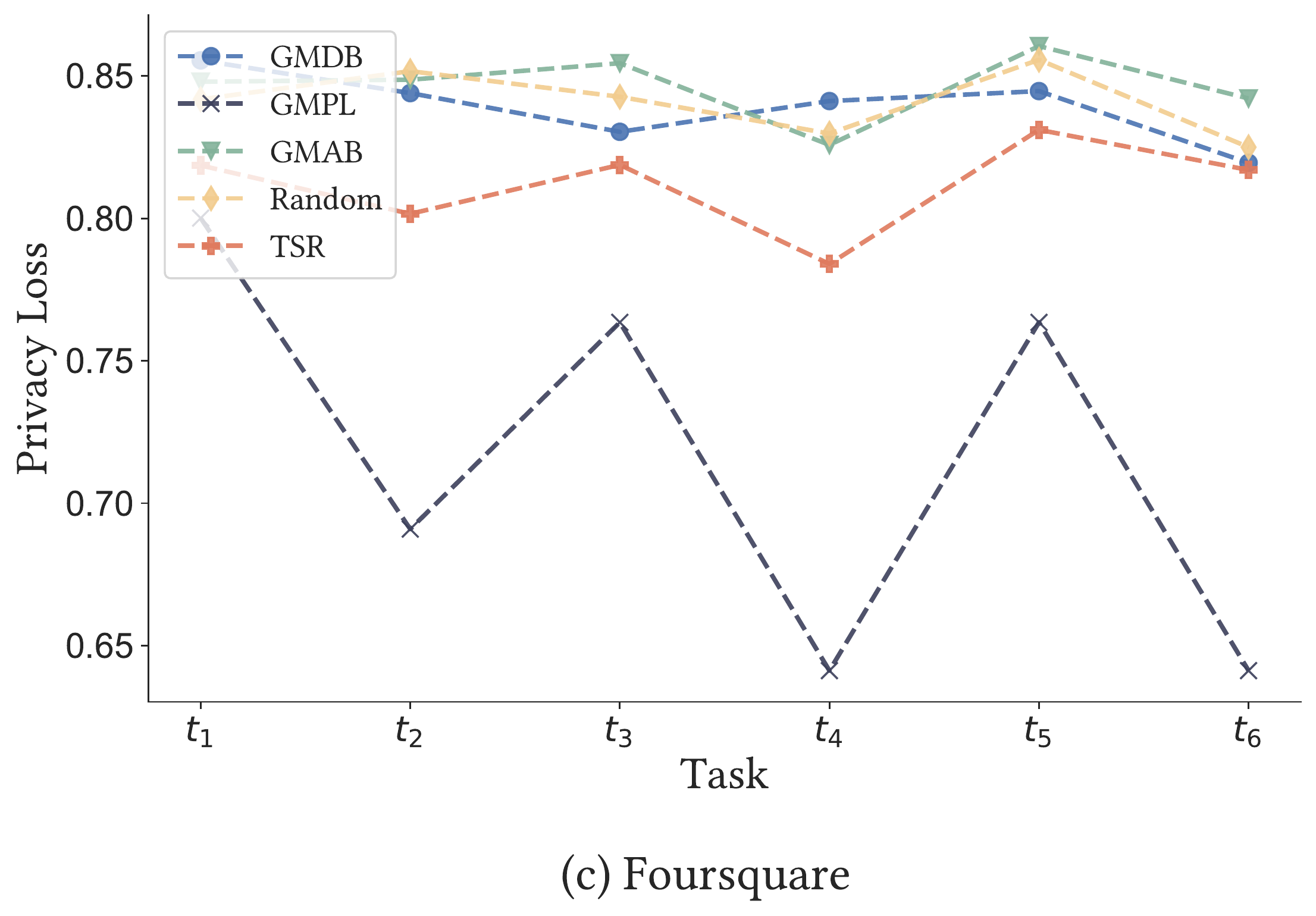}
  \captionsetup{justification=centering, singlelinecheck=false}
  \caption{Comparison of privacy loss for different algorithms under different tasks}
  \label{fig:both_figures}
\end{minipage}
\vspace{-1.2em}
\end{figure*}
\subsubsection{Comparison of QoD and Privacy Loss for Different Algorithms under Different Worker Numbers}
Under the condition of $\kappa = 10$, a task requiring the recruitment of 10 workers was published by the task publisher. The platform simulated an increase in worker numbers from 200 to 1200 to compare the QoD and privacy loss performance of different algorithms. It should be noted that the locations of previously increased workers remained unchanged.
From the experimental results of the three datasets in Fig. 9, it can be observed that as the worker numbers increased, the QoD performance of the TSR algorithm surpassed other baselines and exhibited an initial increase followed by a slowdown. This is because the TSR algorithm could recruit an execution team with optimal comprehensive ability benefits, distance benefits, and trust benefits. As the number of workers increased, the TSR algorithm had a larger solution space and could recruit more exceptional candidate workers. However, after reaching a certain threshold, the quality of newly increased workers is difficult to surpass the quality of previously increased workers, resulting in a gradual flattening of the QoD growth for the TSR algorithm.
Furthermore, from the experimental results of the three datasets in Fig. 10, it can be observed that as the worker numbers increased, the TSR algorithm exhibited fluctuations in privacy loss and outperformed all baselines except the GMPL algorithm. This further validated the rationale of considering trust benefits in the TSR algorithm to maximize QoD. The TSR algorithm did not blindly recruit newly appeared workers with higher trust values but adopted a strategy that integrated different benefits to maximize QoD. Additionally, as the number of workers increased, the trust level of the execution team recruited by the GMPL algorithm was not lower than that of the previously recruited execution team. However, the experimental results showed fluctuations in privacy loss for the GMPL algorithm as the worker numbers increased. This also indicated that recruiting the execution team with the highest trust level greedily did not necessarily result in optimal privacy loss performance.
\begin{figure*}[!t]
\begin{minipage}{\textwidth}
  \includegraphics[width=0.33\linewidth]{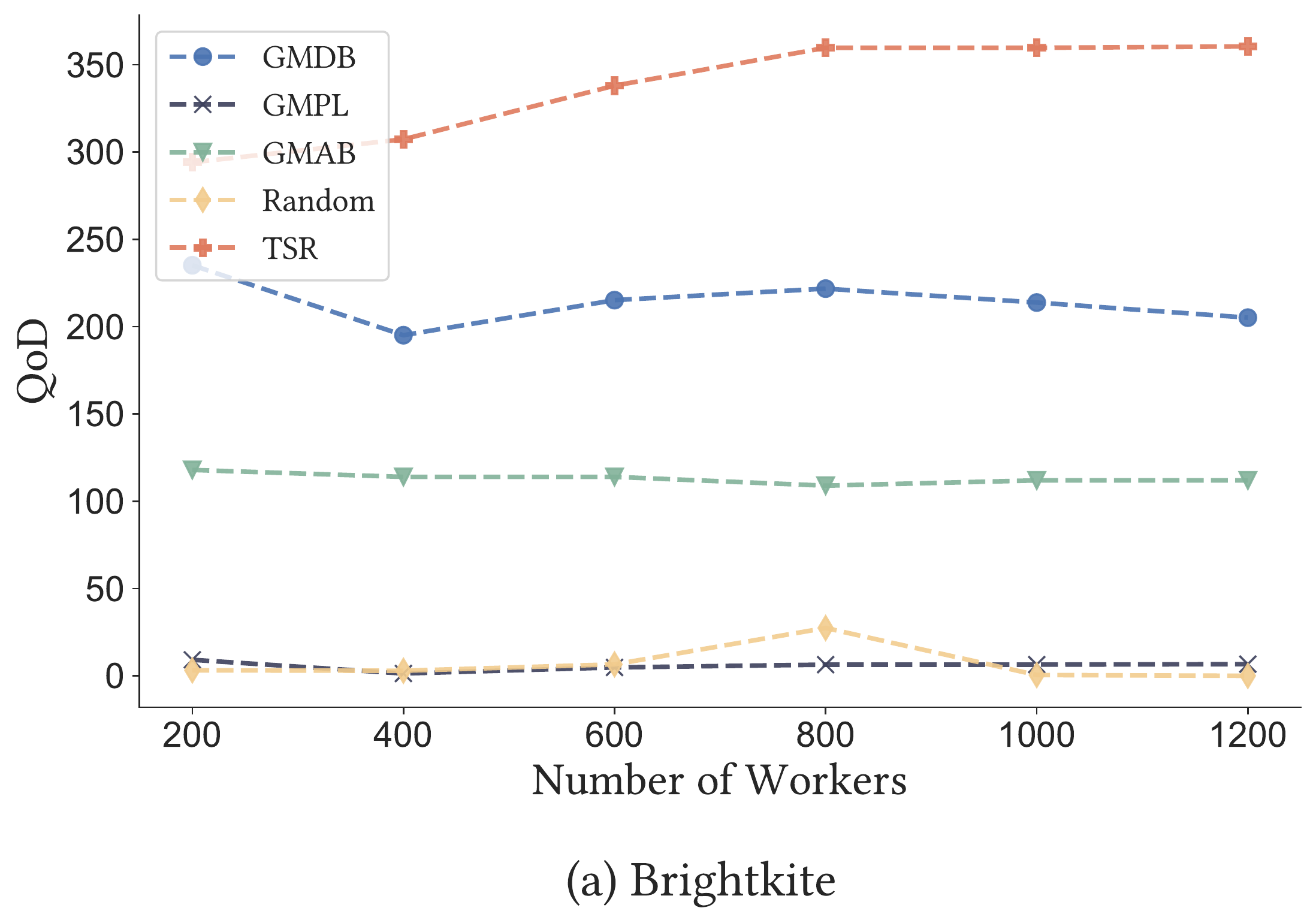}
  \hfill
  \includegraphics[width=0.33\linewidth]{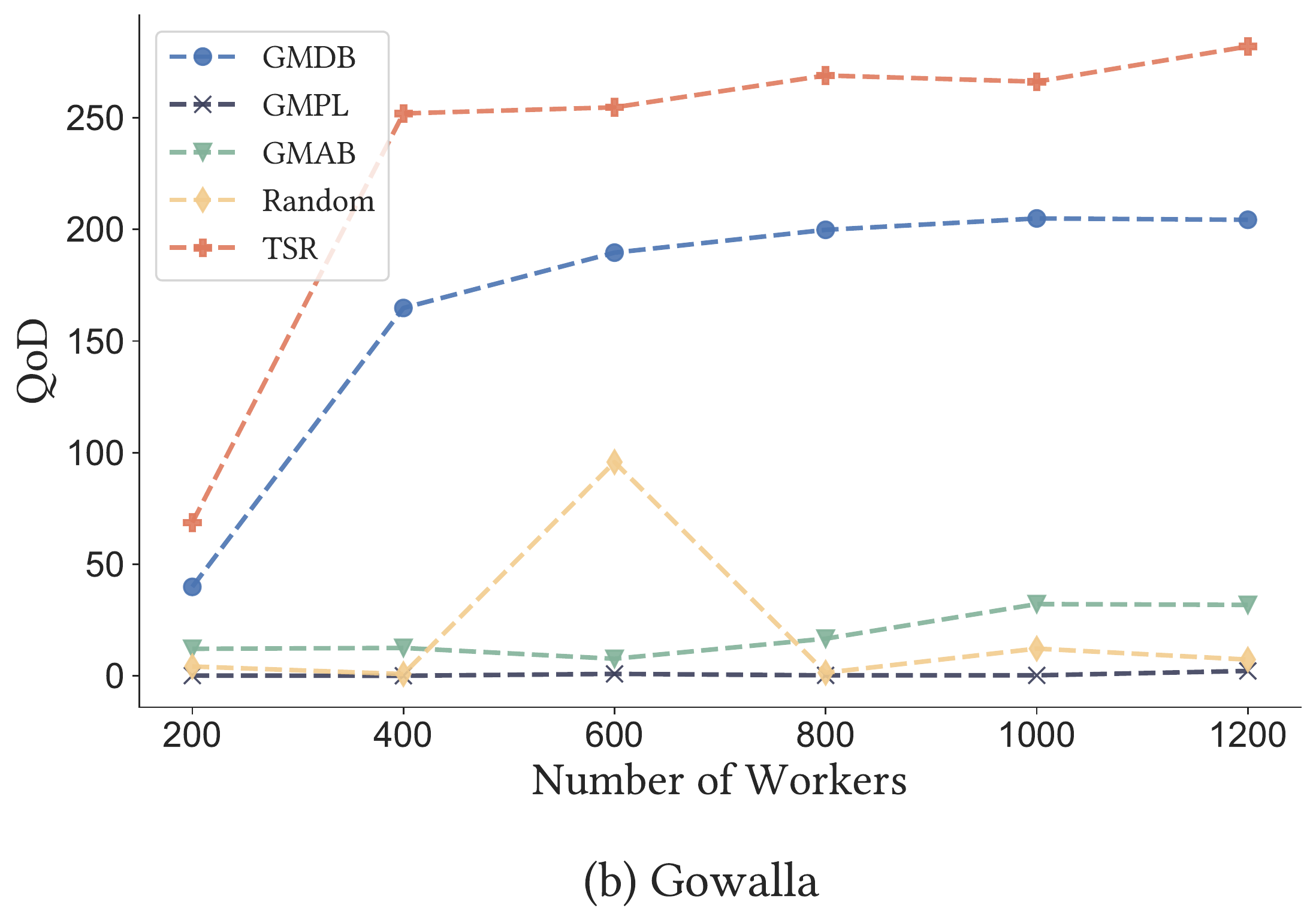}
  \hfill
  \includegraphics[width=0.33\linewidth]{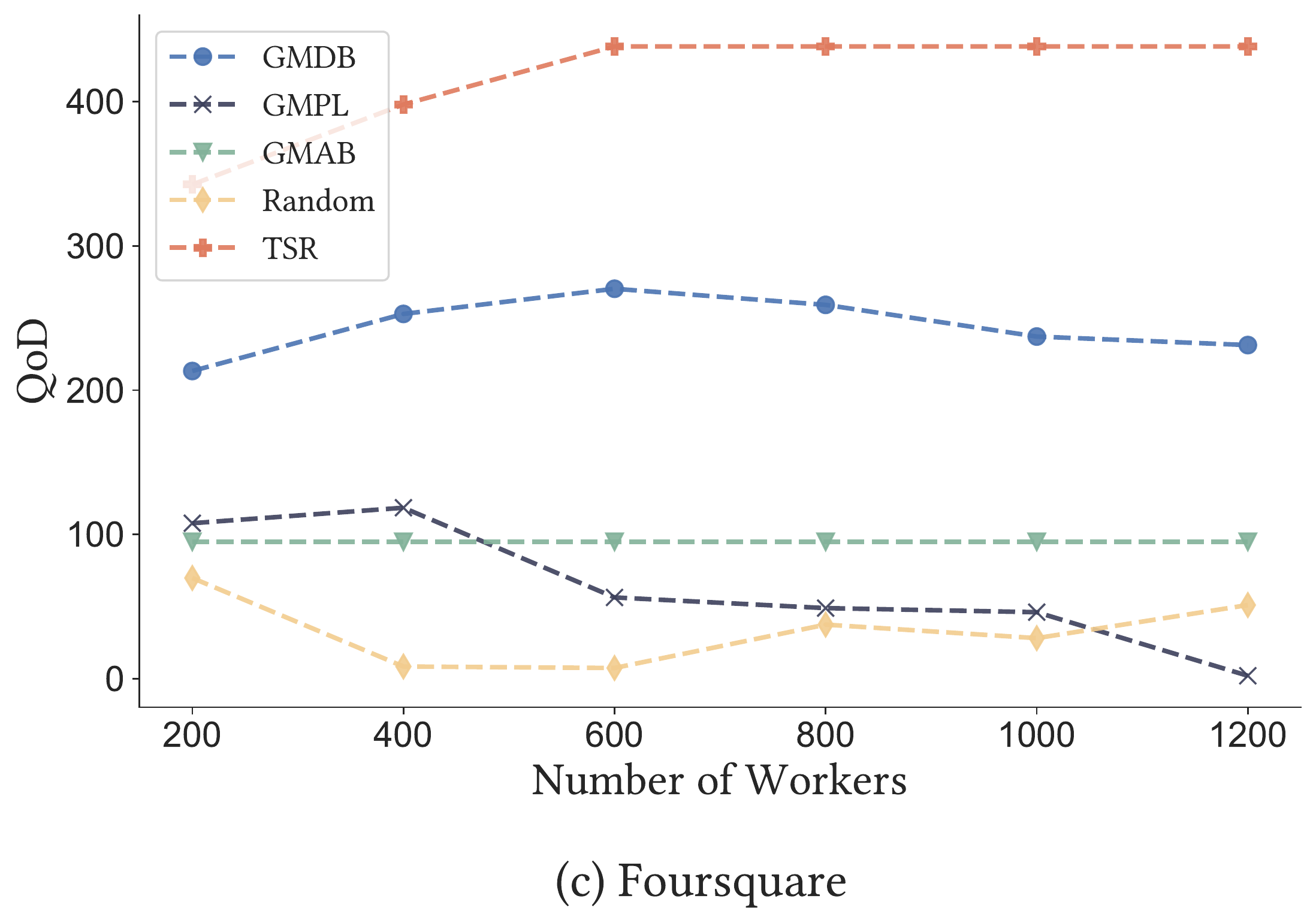}
  \captionsetup{justification=centering, singlelinecheck=false}
  \caption{Comparison of QoD for different algorithms under different worker numbers}
  \label{fig:both_figures}
\end{minipage}
\vspace{-1.2em}
\end{figure*}
\begin{figure*}[!t]
\begin{minipage}{\textwidth}
  \includegraphics[width=0.33\linewidth]{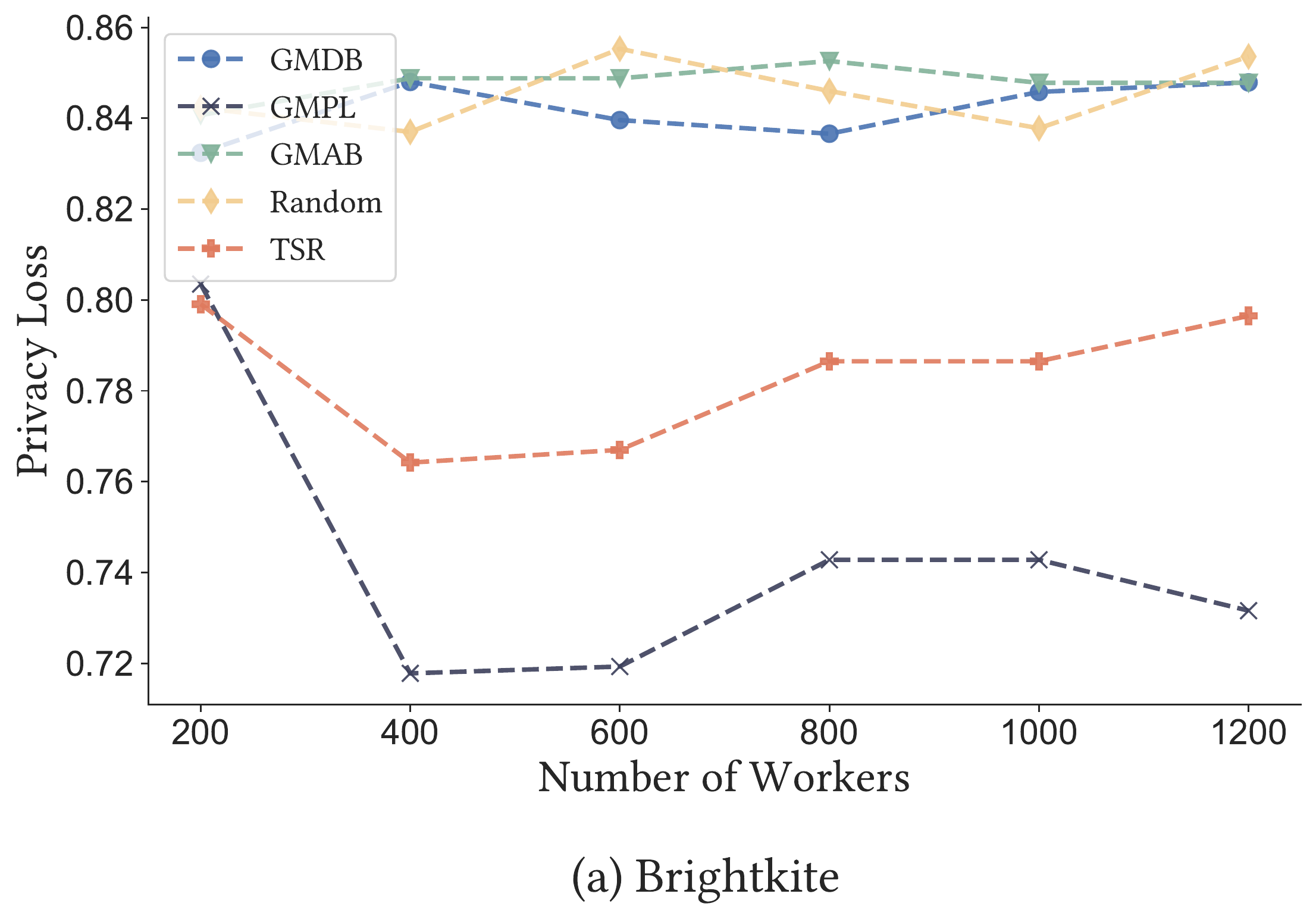}
  \hfill
  \includegraphics[width=0.33\linewidth]{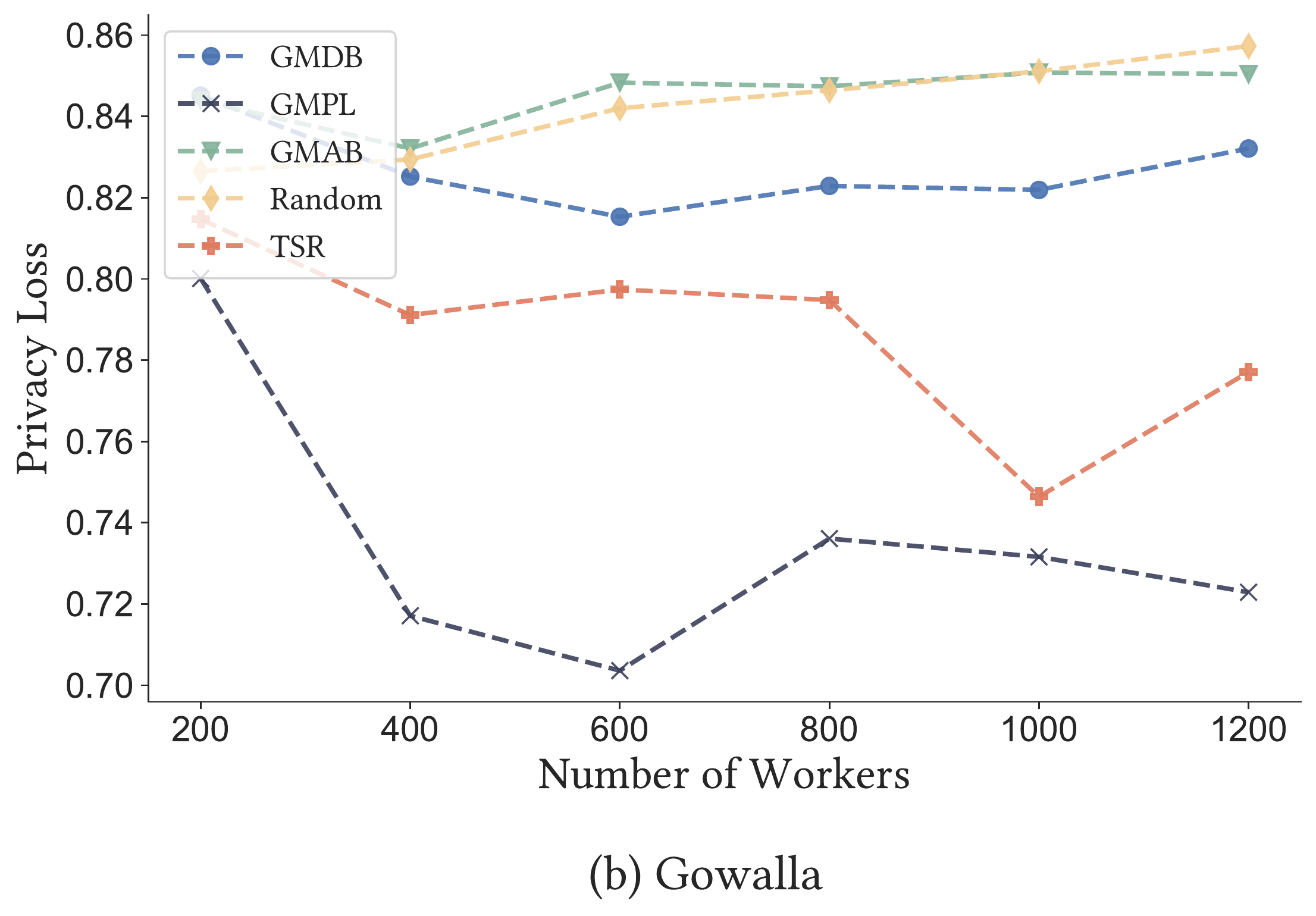}
  \hfill
  \includegraphics[width=0.33\linewidth]{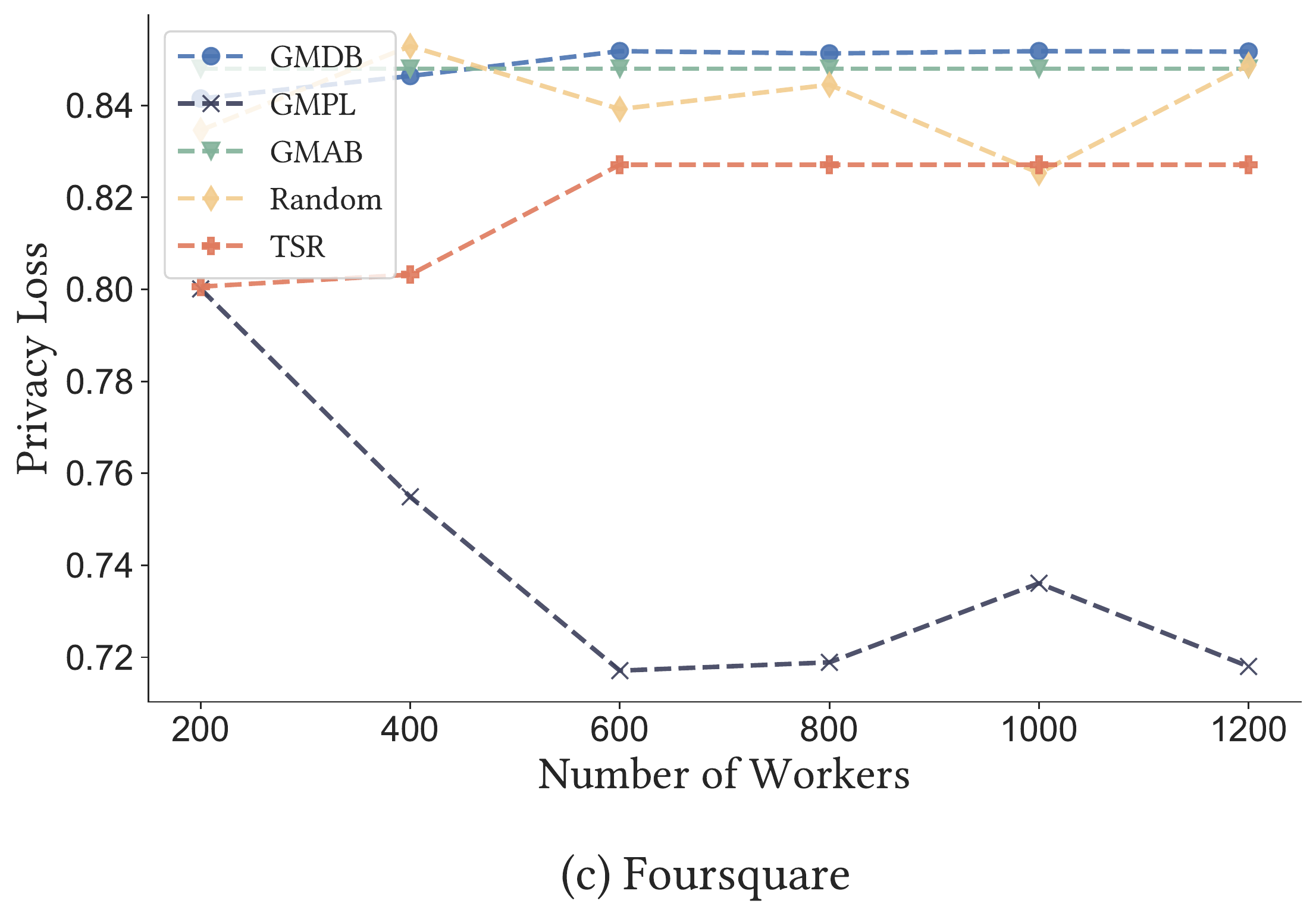}
  \captionsetup{justification=centering, singlelinecheck=false}
  \caption{Comparison of privacy loss for different algorithms under different worker numbers}
  \label{fig:both_figures}
\end{minipage}
\vspace{-1.2em}
\end{figure*}
\begin{figure*}[!t]
\begin{minipage}{\textwidth}
  \includegraphics[width=0.33\linewidth]{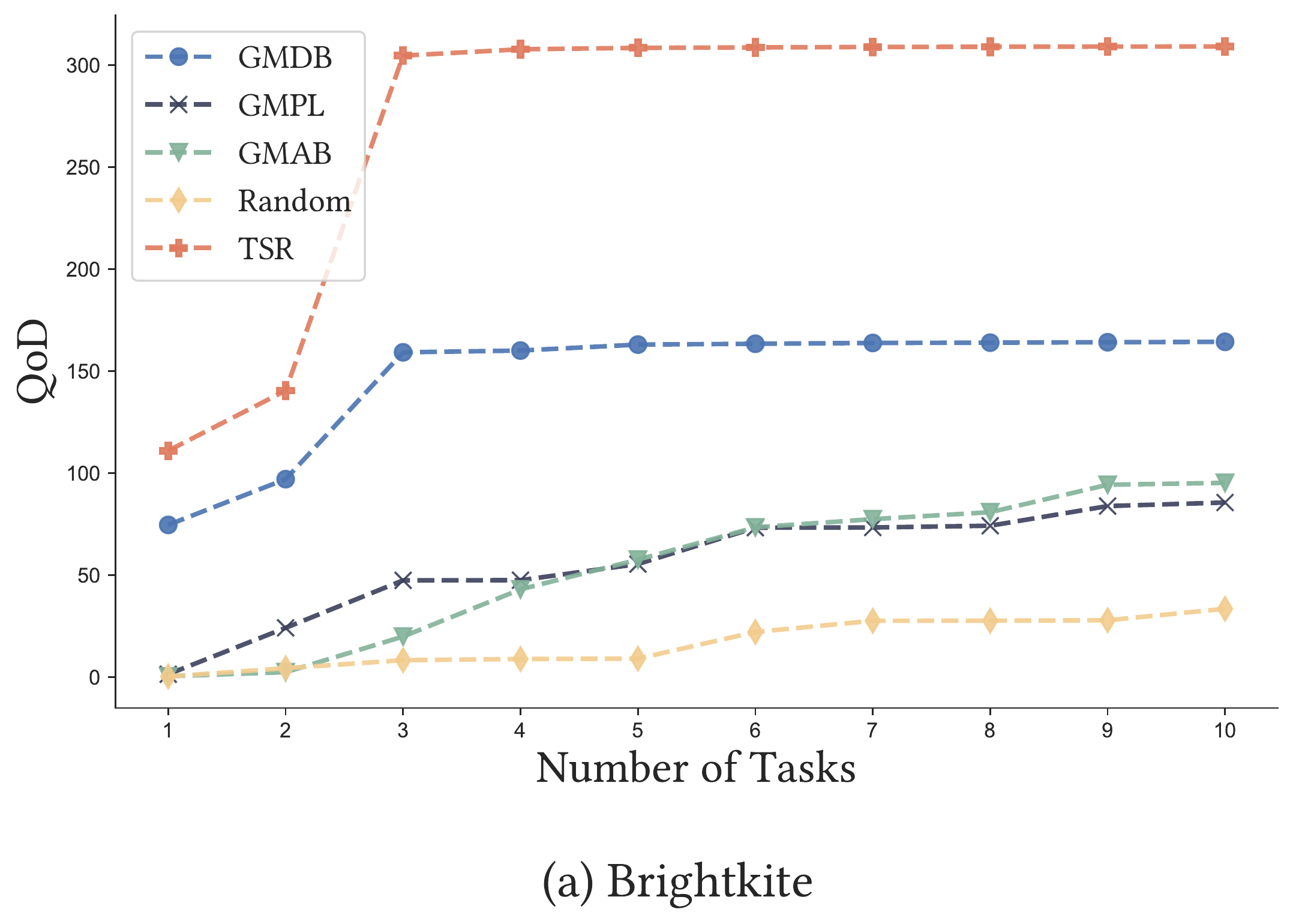}
  \hfill
  \includegraphics[width=0.33\linewidth]{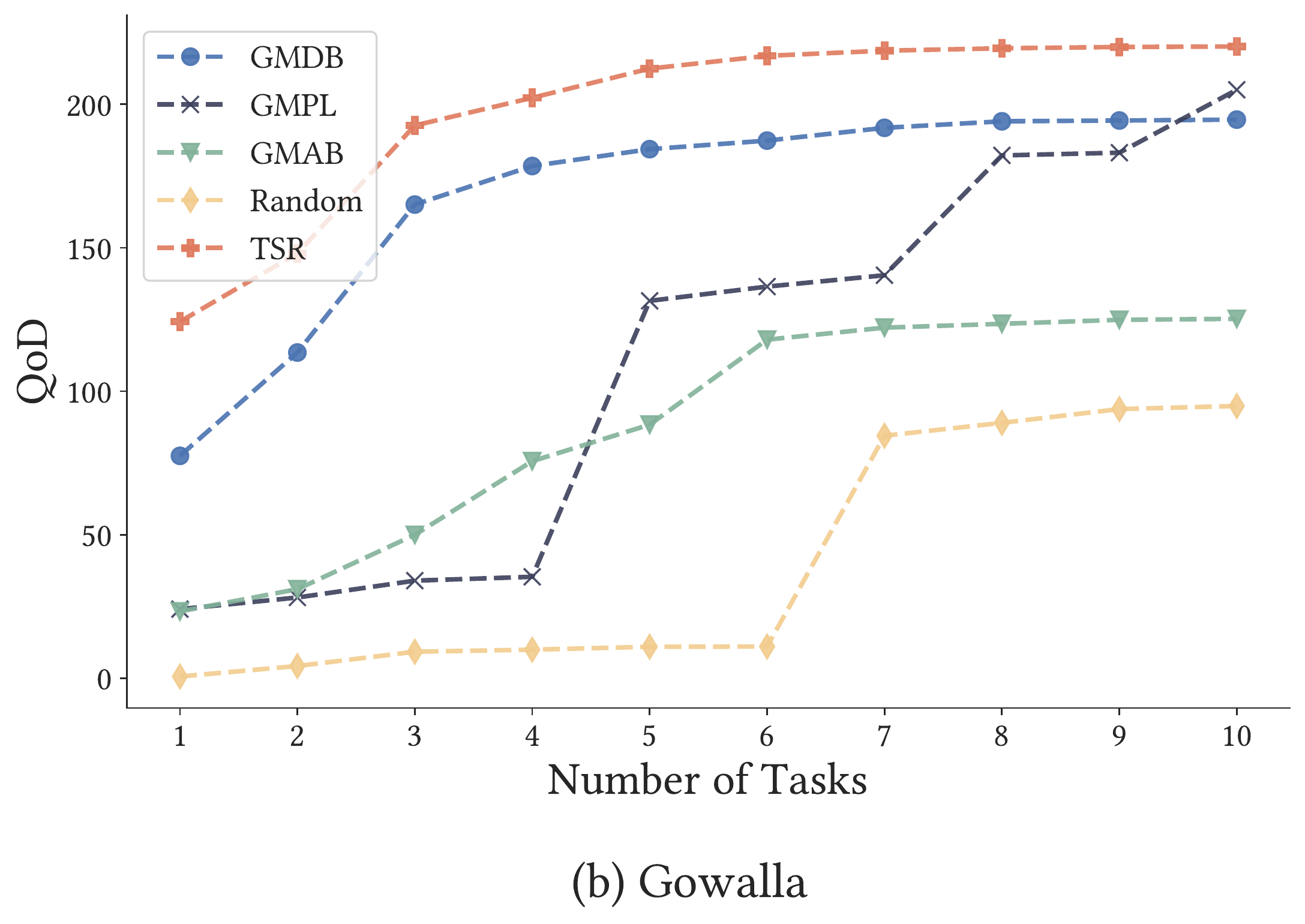}
  \hfill
  \includegraphics[width=0.33\linewidth]{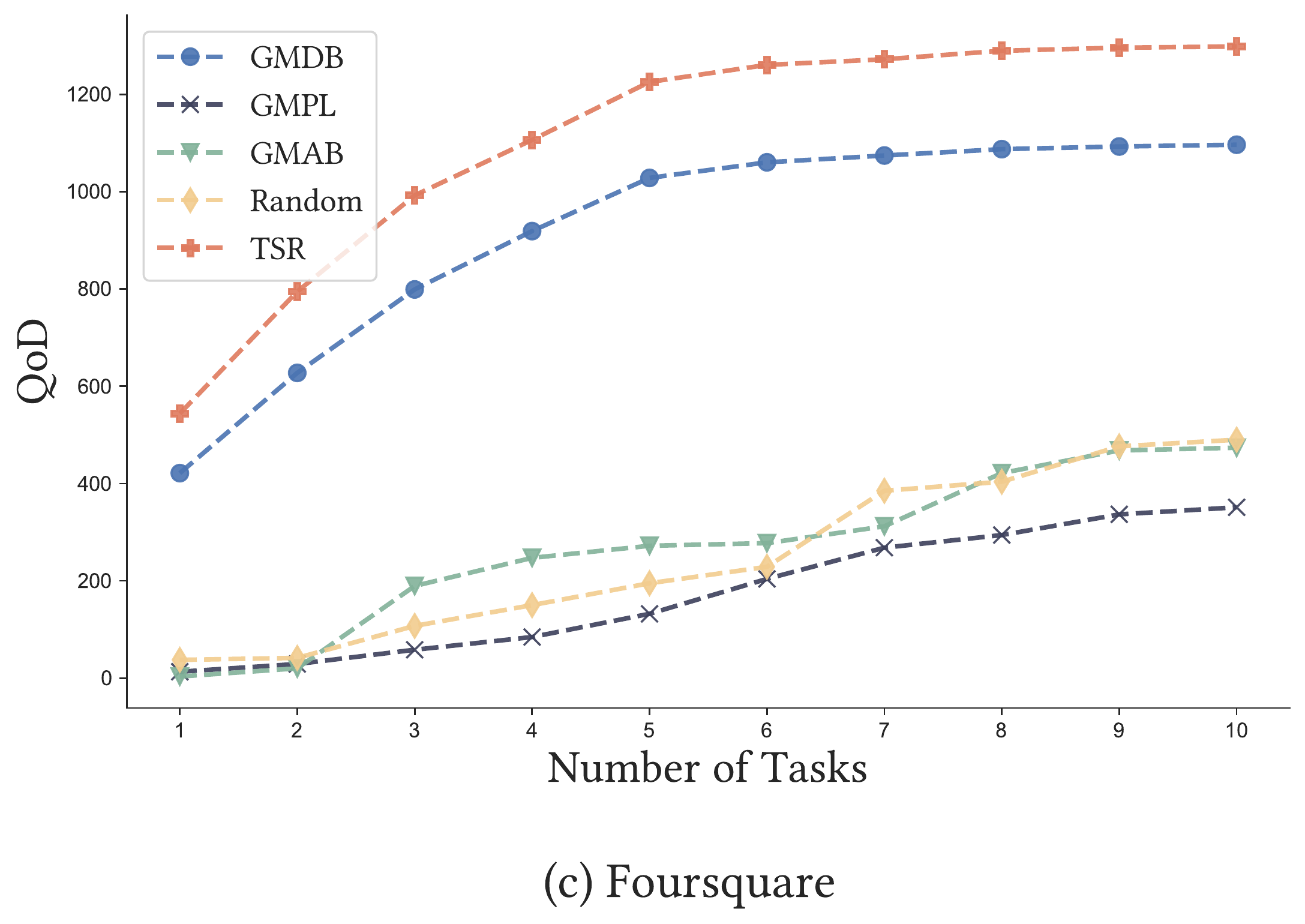}
  \captionsetup{justification=centering, singlelinecheck=false}
  \caption{Comparison of QoD for different algorithms under different task numbers}
  \label{fig:both_figures}
\end{minipage}
\vspace{-1.2em}
\end{figure*}
\begin{figure*}[!t]
\begin{minipage}{\textwidth}
  \includegraphics[width=0.33\linewidth]{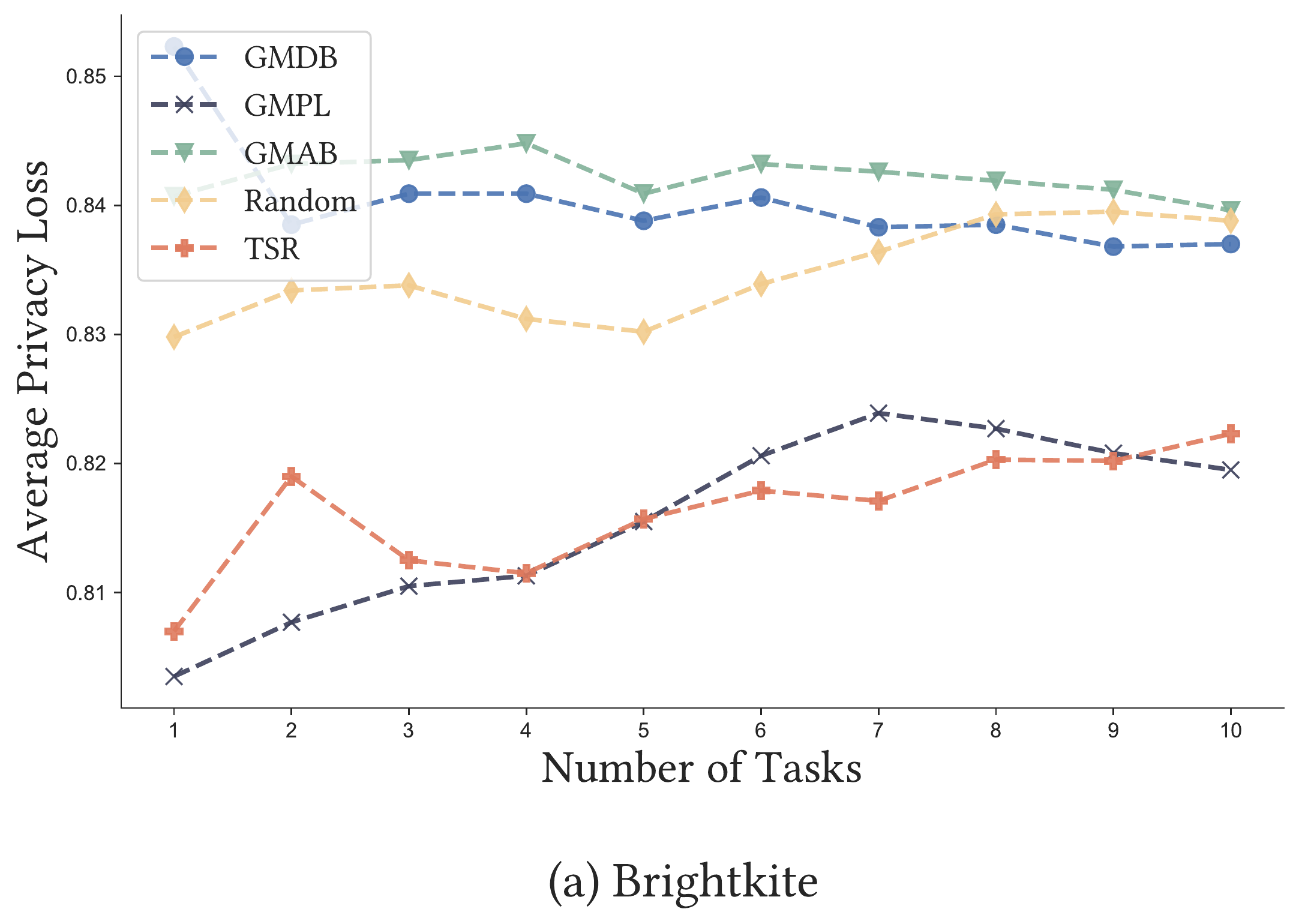}
  \hfill
  \includegraphics[width=0.33\linewidth]{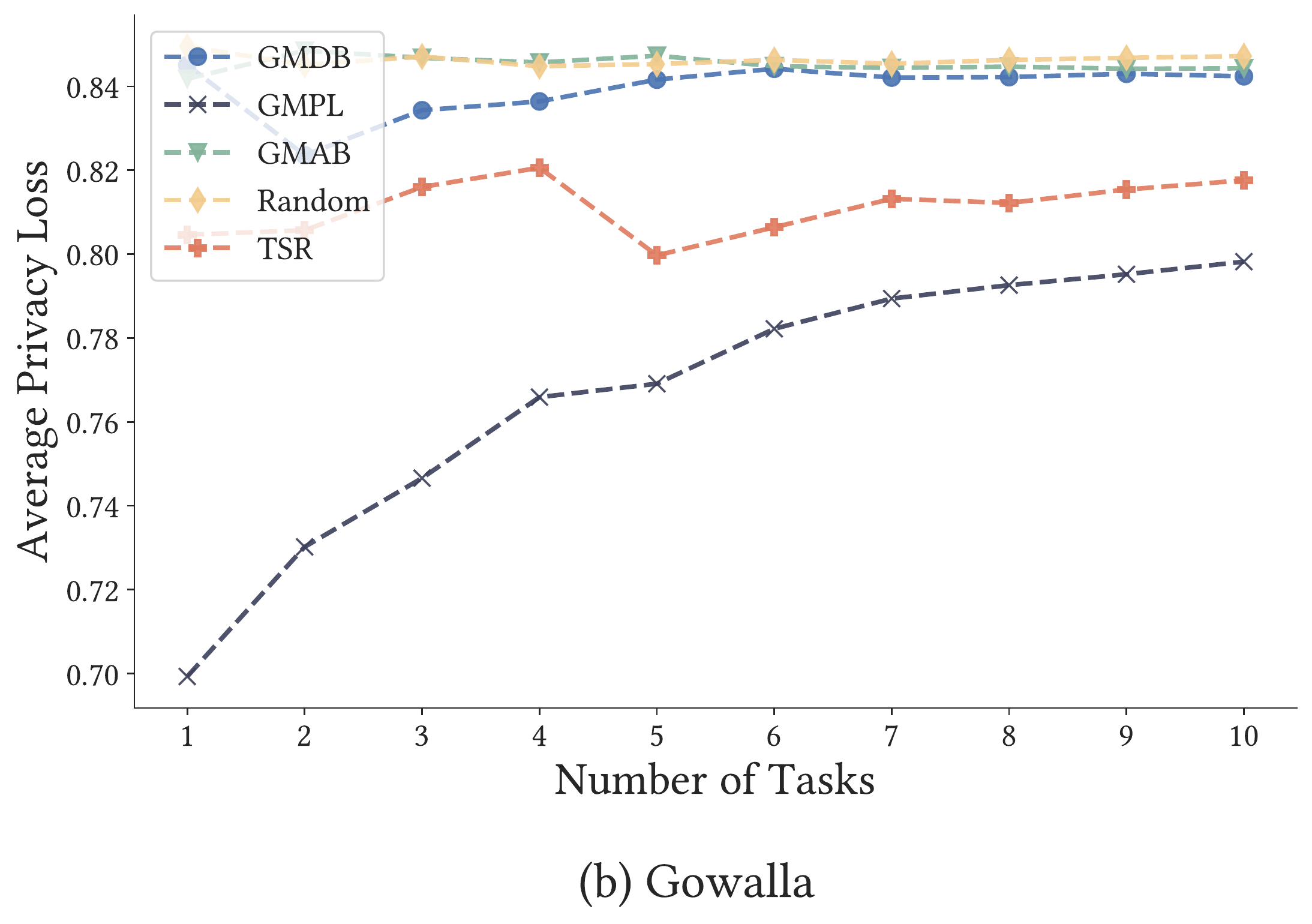}
  \hfill
  \includegraphics[width=0.33\linewidth]{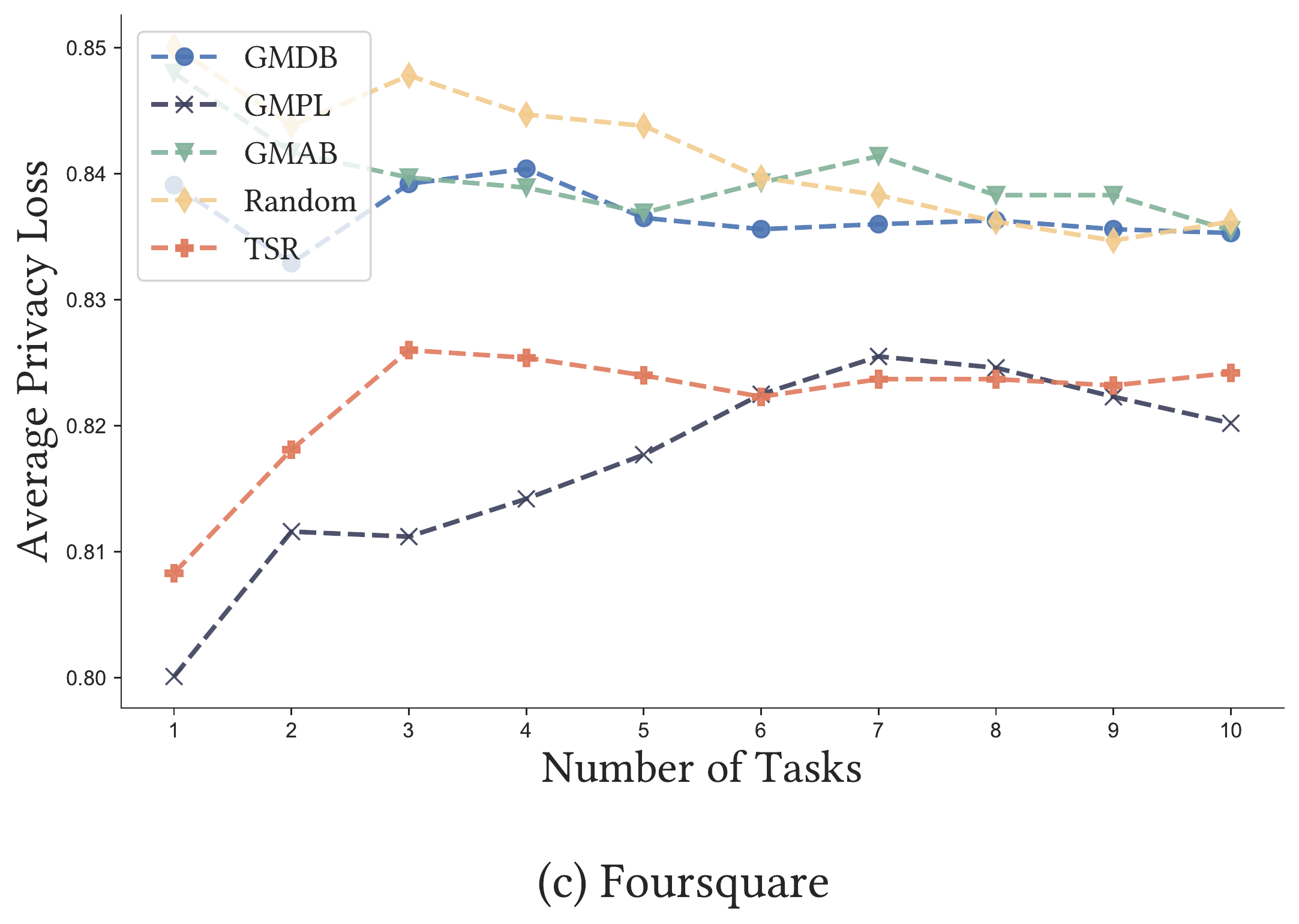}
  \captionsetup{justification=centering, singlelinecheck=false}
  \caption{Comparison of average privacy loss for different algorithms under different task numbers}
  \label{fig:both_figures}
\end{minipage}
\vspace{-1.2em}
\end{figure*}
\begin{figure*}[!t]
\centering
\begin{minipage}{\textwidth}
  \includegraphics[width=0.33\linewidth]{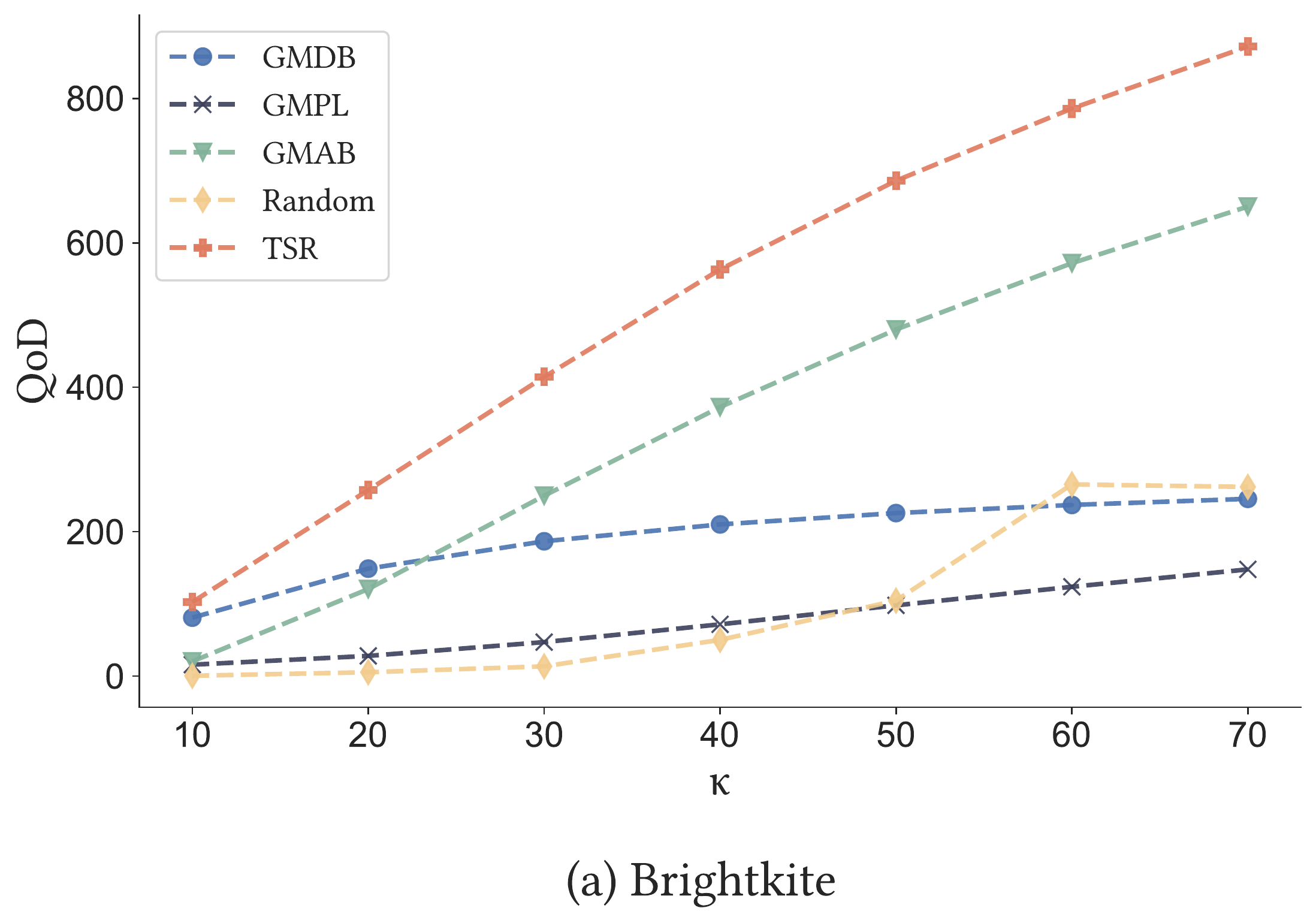}
  \hfill
  \includegraphics[width=0.33\linewidth]{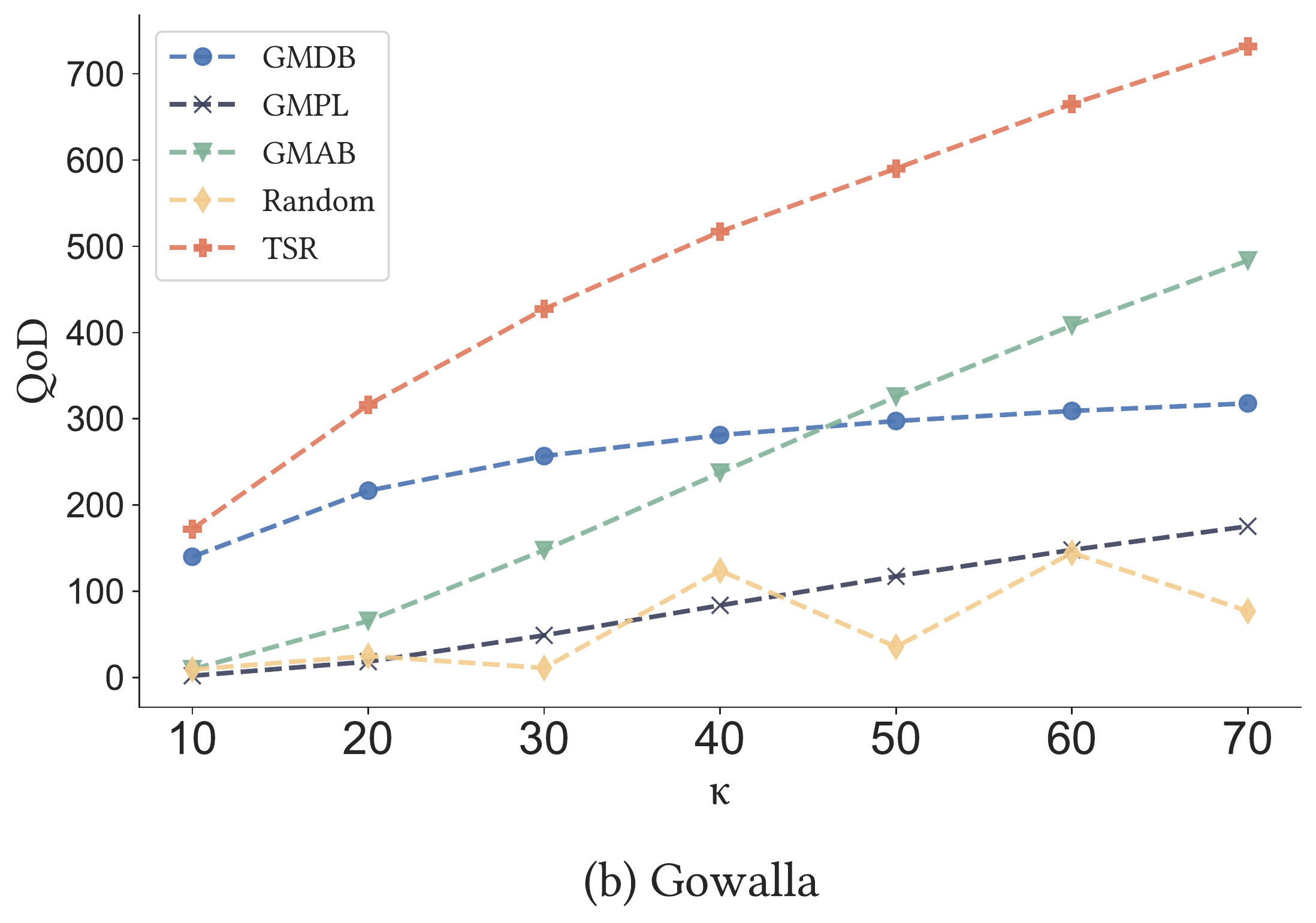}
  \hfill
  \includegraphics[width=0.33\linewidth]{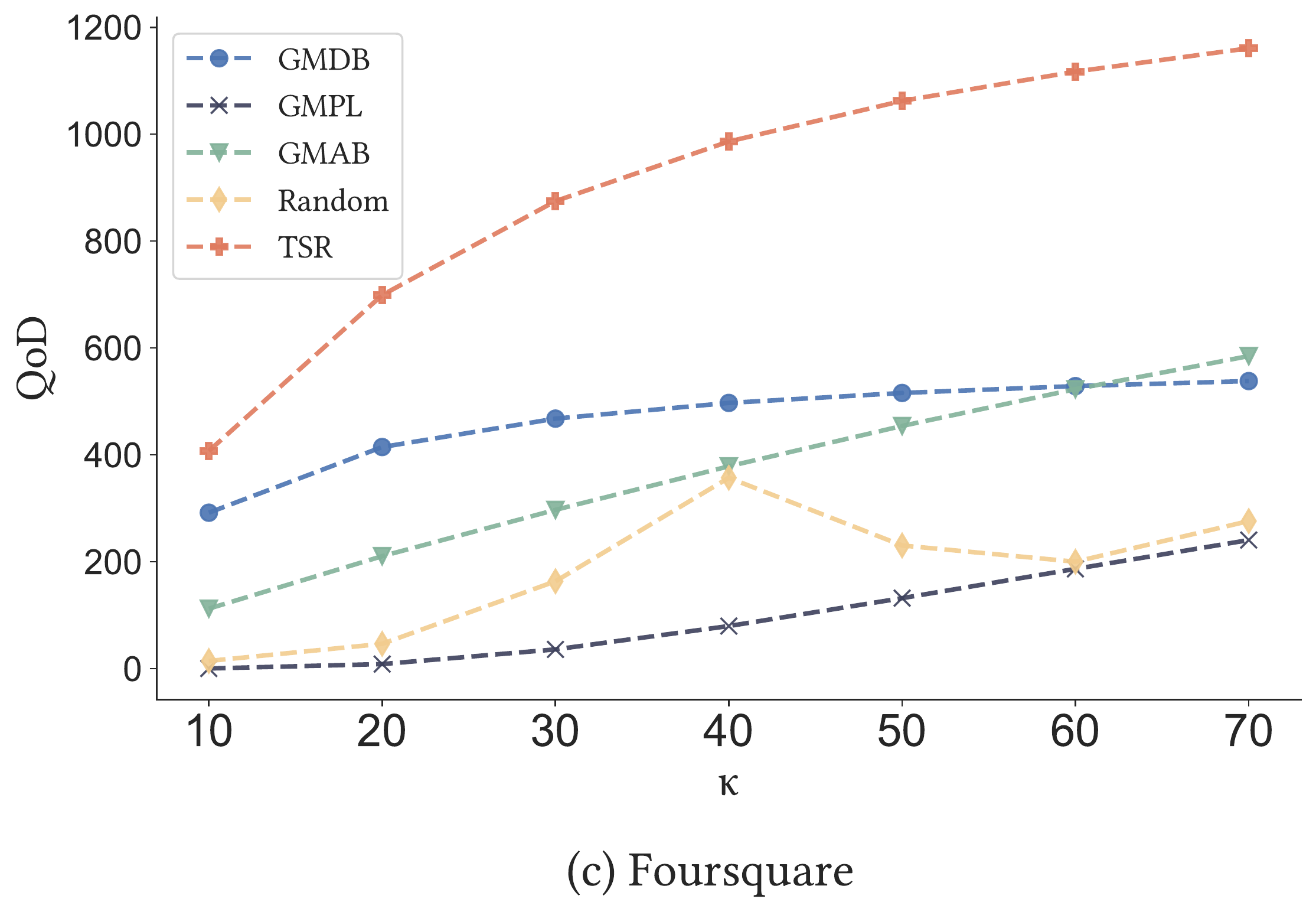}
  \captionsetup{justification=centering, singlelinecheck=false}
  \caption{Comparison of QoD for different algorithms under different $\kappa$ values}
  \label{fig:both_figures}
\end{minipage}
\vspace{-1.2em}
\end{figure*}
\begin{figure*}[!t]
\centering
\begin{minipage}{\textwidth}
  \includegraphics[width=0.33\linewidth]{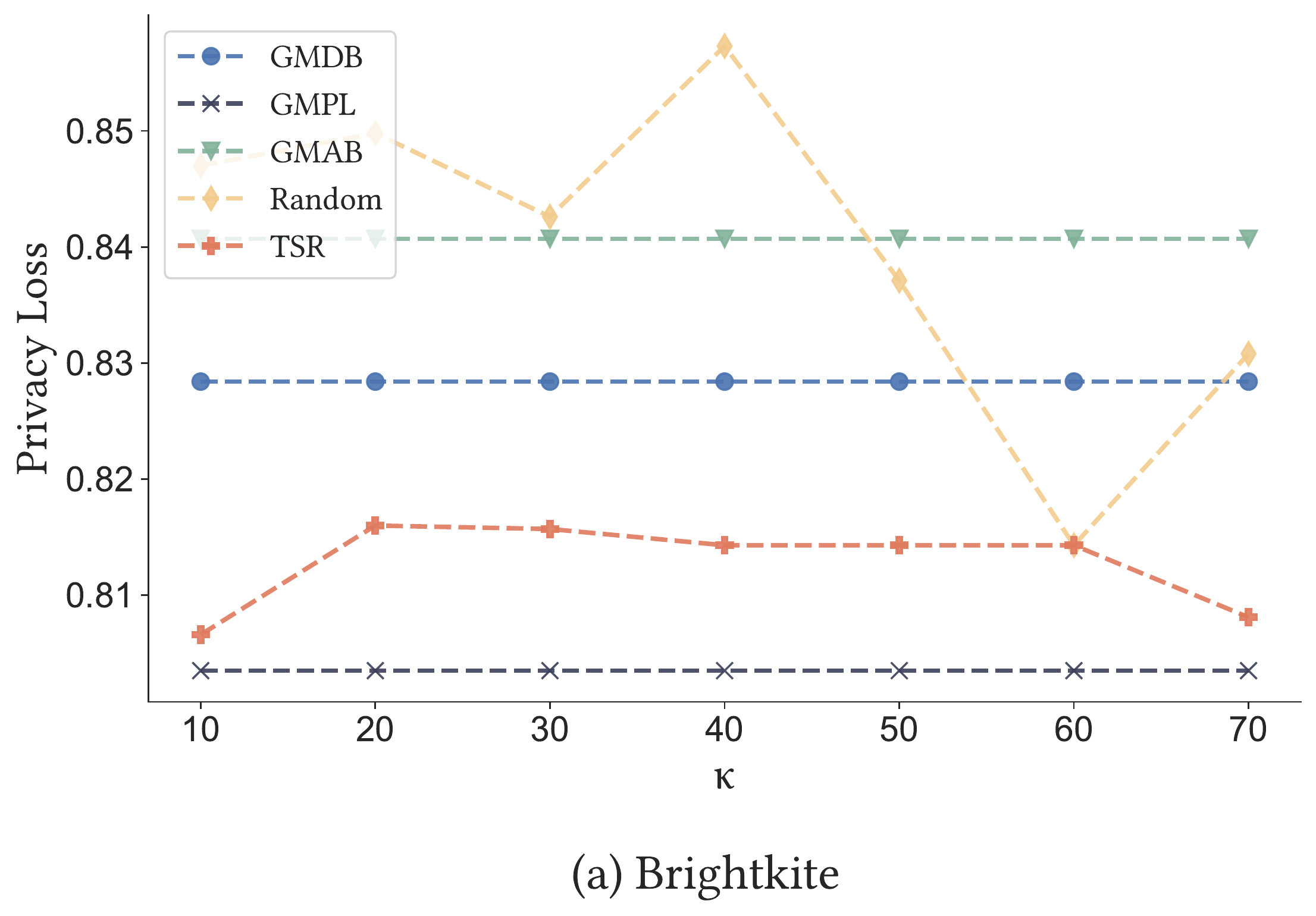}
  \hfill
  \includegraphics[width=0.33\linewidth]{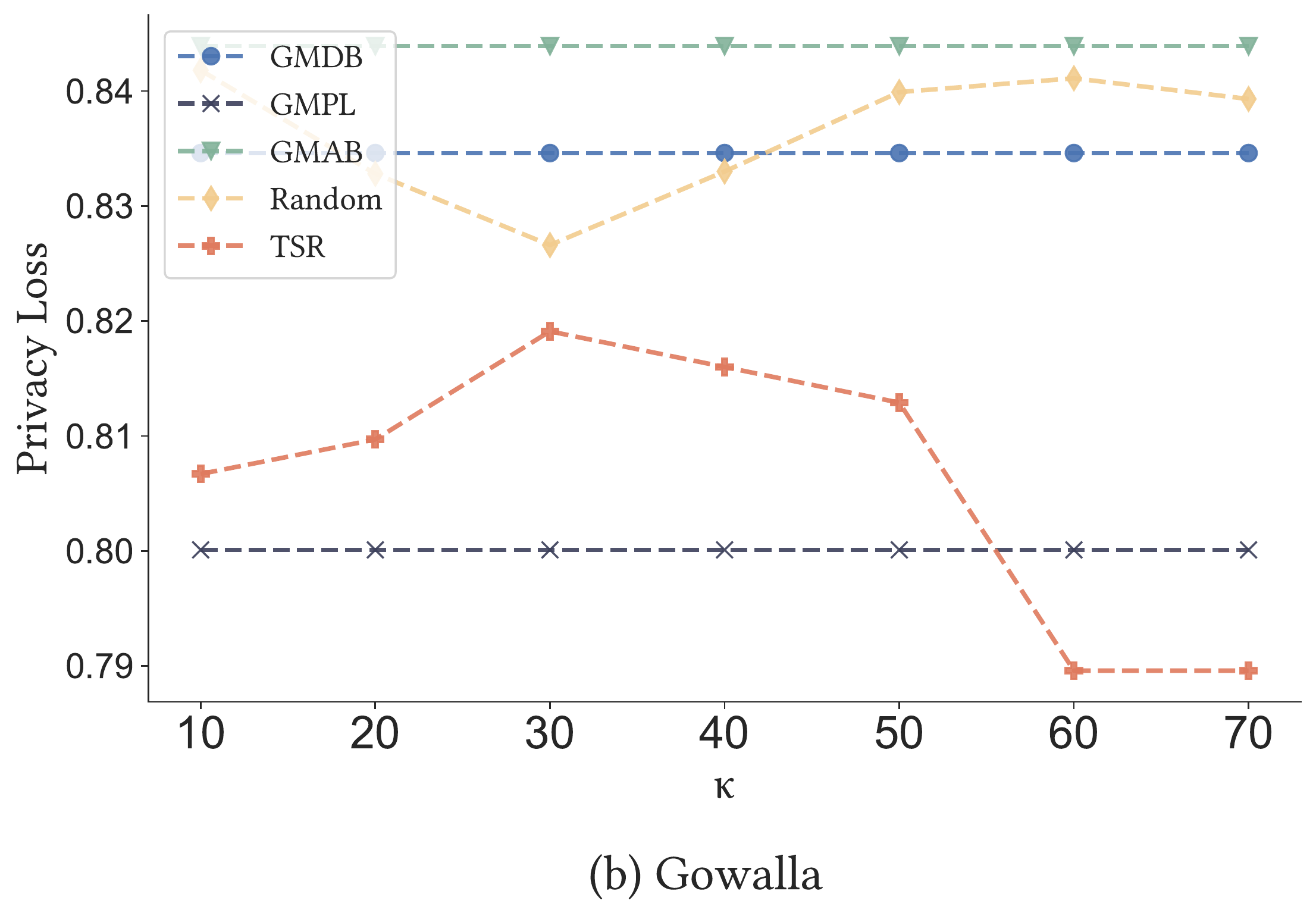}
  \hfill
  \includegraphics[width=0.33\linewidth]{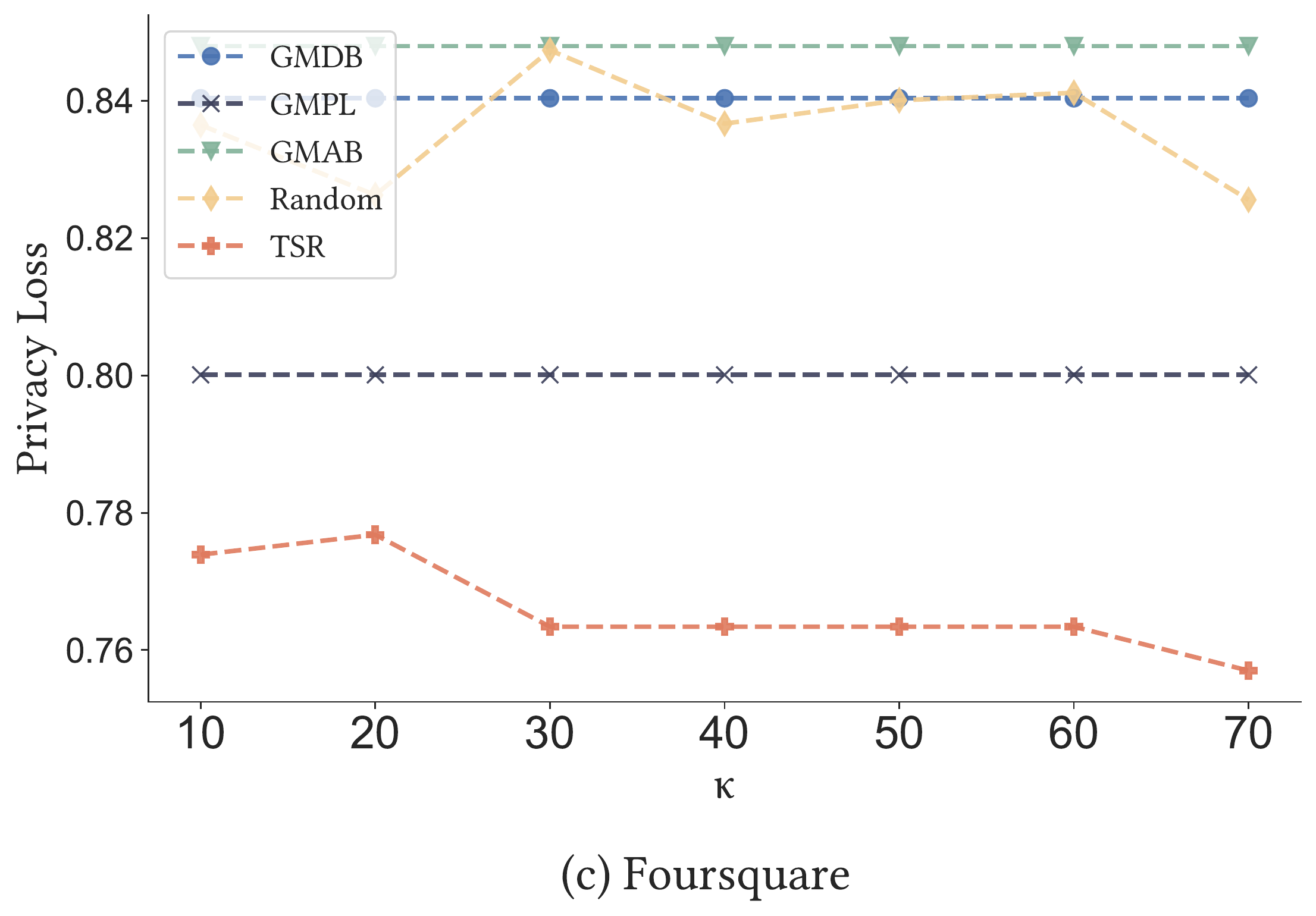}
  \captionsetup{justification=centering, singlelinecheck=false}
  \caption{Comparison of privacy loss for different algorithms under different $\kappa$ values}
  \label{fig:both_figures}
\end{minipage}
\vspace{-1.2em}
\end{figure*}

\begin{figure*}[!t]
    \centering
    \begin{minipage}{0.4\textwidth}
        \centering
        \includegraphics[width=\textwidth]{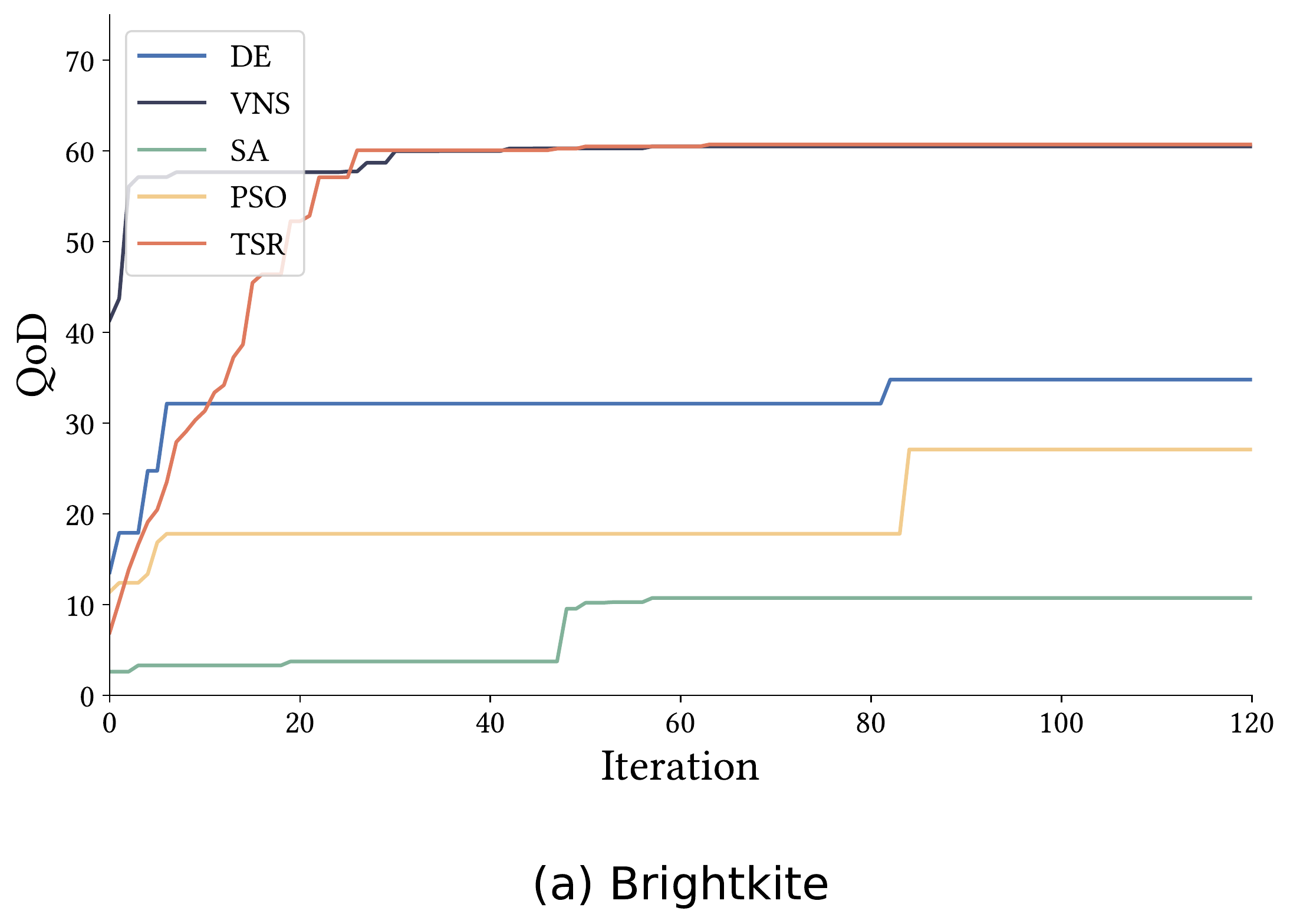}
    \end{minipage}\hspace{1.3cm}%
    \begin{minipage}{0.4\textwidth}
        \centering
        \includegraphics[width=\textwidth]{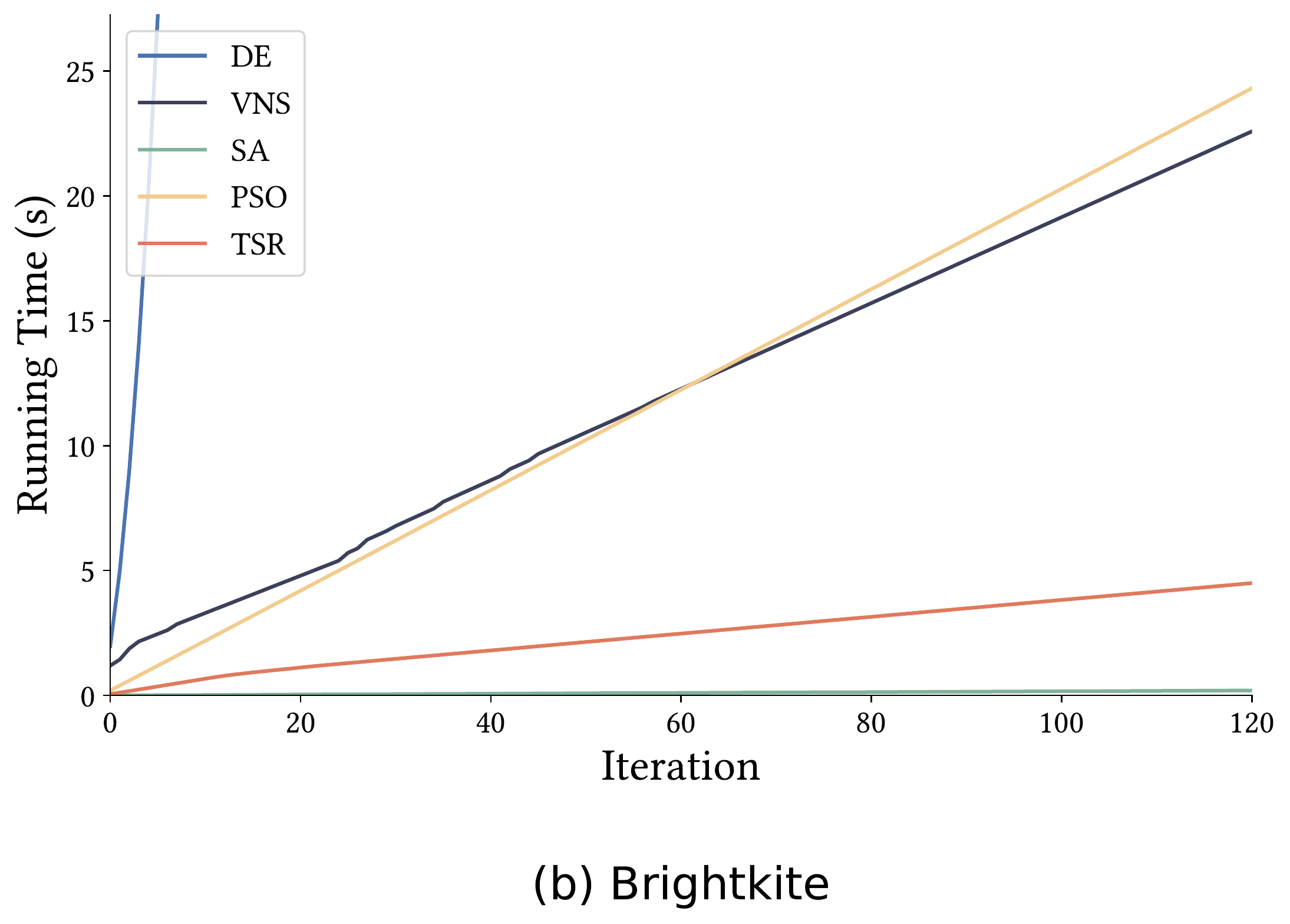}
    \end{minipage}
    \caption{Comparison of the performance for different algorithms in the same recruitment strategy}
\end{figure*}
\subsubsection{Comparison of QoD and Average Privacy Loss for Different Algorithms under Different Task Numbers}
When $\kappa = 10$, 200 workers randomly appeared in a region. Tasks were sequentially published in this region by task publishers to explore the QoD and average privacy loss performance of different algorithms under varying task numbers. Each task required the recruitment of 10 workers to form an execution team.
By observing the experimental results of the three datasets in Fig. 11, it is evident that with an increasing number of tasks, the TSR algorithm demonstrates significant superiority in QoD performance compared to other baselines, showing an increasing-then-stabilizing trend in QoD. This is because, during the first task publish, the TSR algorithm recruited high-quality workers to compose the execution team, resulting in better QoD than other baselines. However, as high-quality workers are recruited for the first task, the solution space performance of subsequent tasks deteriorates, leading to a slowdown in the growth rate of the TSR algorithm's QoD as tasks increase.
On the other hand, based on the results shown in Fig. 12, as the task numbers increase, both the TSR algorithm and the GMPL algorithm exhibit an upward trend in average privacy loss. The average privacy loss performance of the TSR algorithm is comparable to that of the GMPL algorithm and superior to other baselines. This is because, with the increase in task numbers, the quality of the trust benefit of the remaining workers decreases, increasing the average privacy loss for the execution teams recruited by the TSR and GMPL algorithms. Moreover, due to the consideration of workers' trust values during the recruitment process, the TSR and GMPL algorithm performs better than other baselines in terms of average privacy loss performance.
\subsubsection{Comparison of QoD and Privacy Loss for Different Algorithms under Different $\kappa$ Values}
To delve into the QoD and privacy loss of different algorithms under various $\kappa$ values, a task was published by the task publisher in a region with 200 workers. the task required the recruitment of 10 workers to form an execution team. The platform incrementally raised the $\kappa$ values from 10 to 70, increasing by 10 each time to explore the QoD and privacy loss performance of different algorithms under varying $\kappa$ values.\\
According to the experimental results shown in Fig. 13, as the $\kappa$ values increased, the quality of the execution teams recruited by the TSR algorithm significantly surpassed other baselines. This is because the TSR algorithm comprehensively considered different benefits, resulting in optimal QoD performance. Additionally, except for the Random algorithm, all algorithms demonstrated an upward trend in terms of QoD. The increased $\kappa$ values led to heightened distance benefits among workers, thus continuously improving the QoD of the algorithms. In addition, as the $\kappa$ values increase, the distance benefits of the workers approach $1$, and the influence of distance benefits on the QoD of execution teams gradually decreases. Consequently, ability benefits and trust benefits become key factors influencing the QoD performance. As a result, the GMAB algorithm outperforms the GMDB algorithm in terms of QoD as the $\kappa$ value increases.
Moreover, the experimental results in Fig. 14 demonstrated that as the $\kappa$ values increased, the privacy loss of the TSR and Random algorithms exhibited fluctuations, while the privacy loss of the other baselines remained constant. This further confirmed the adaptability of the TSR algorithm, highlighting its flexibility in recruiting the optimal execution team based on the changing $\kappa$ values. The workers recruited by the GMDB, GMAB algorithms remained unchanged as the $\kappa$ values increased. However, the QoD performance of the GMAB algorithm gradually surpassed that of the GMDB algorithm, providing further support for the notion that the influence of distance benefits on QoD weakens as the $\kappa$ values increase.
\subsubsection{Comparison of the Performance of Different Algorithms in the Same Recruitment Strategy}
The aim of this section is to evaluate the performance of different algorithms by comparing their convergence and runtime when adopting the same recruitment strategy, which considers ability benefits, distance benefits, and trust benefits. A task was published by the task requester, requiring the formation of an execution team of 10 workers from a pool of 200 candidate workers. 
As could be seen in Fig. 15(a), an approximate optimal solution could be quickly converged by both TSR and VNS algorithms, with an increase in the number of iterations. However, as shown in Fig. 15(b), when dealing with problems of the same scale, less running time is required by the TSR algorithm compared to the VNS algorithm. In addition, slower convergence speeds are exhibited by PSO, DE, and SA algorithms, and they are more prone to falling into local optimal solutions. Moreover, the DE algorithm has a long execution time. After considering all factors, the best performance when solving the worker recruitment problem is demonstrated by the TSR algorithm proposed in this paper.
\section{CONCLUSION}
TREF based on GCNs has been designed in this paper to evaluate the potential trustworthiness among all workers. When dealing with the worker recruitment problem in CMCS, the impacts of trust benefits, ability benefits, and distance benefits on task completion effects have been considered. The worker recruitment problem has been modeled as UCRG, and the TSR algorithm solution has been proposed for UCRG. For each task, a multi-objective comprehensive optimal execution team is recruited by the TSR algorithm, and under the constraint of privacy loss of the task, the collaboration team members are selected from the execution team. Moreover, the Mini-Batch K-Means clustering algorithm has been adopted to partition regions while deploying edge servers, thereby achieving distributed worker recruitment. 
In our planned future work, Graph Attention Network and Reinforcement Learning methods \cite{r30,r35} will be introduced to recommend task execution paths for workers, aiming to achieve Pareto optimal solutions for the spatiotemporal worker recruitment problem.
Finally, extensive experiments have been conducted on five real-world datasets in this paper. Experimental results indicate that the TSR algorithm and TREF proposed in this paper outperform other baselines.
\bibliographystyle{unsrt}
\bibliography{zzw}

\begin{IEEEbiography}[{\includegraphics[width=1in,height=1.25in,clip,keepaspectratio]{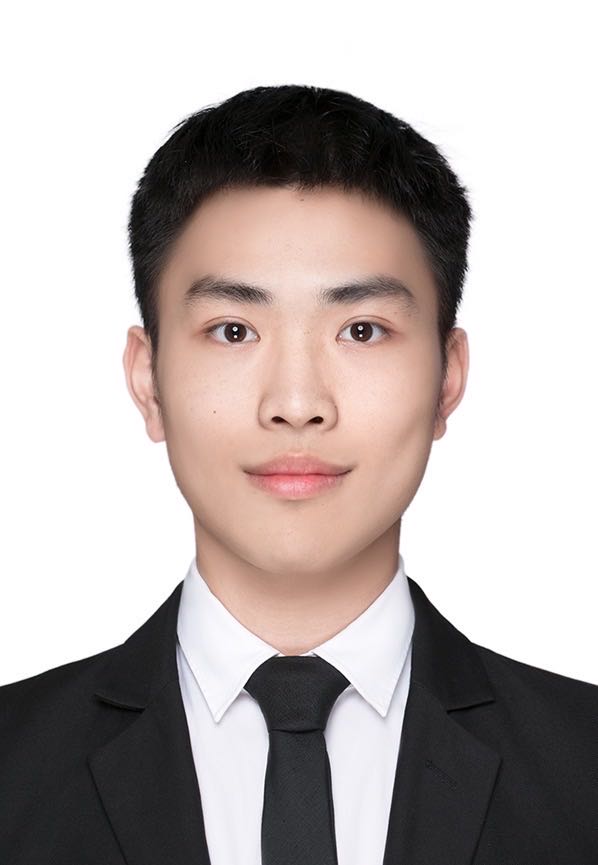}}]{Zhongwei Zhan}
received the Bachelor degree in Computer Science and Technology from Yantai Nanshan University. He is currently pursuing the Master degree in the School of Computer and Control Engineering, Yantai University. His research interests include mobile crowdsourcing and social networks.
\end{IEEEbiography}

\begin{IEEEbiography}[{\includegraphics[width=0.9in,height=1.125in,clip,keepaspectratio]{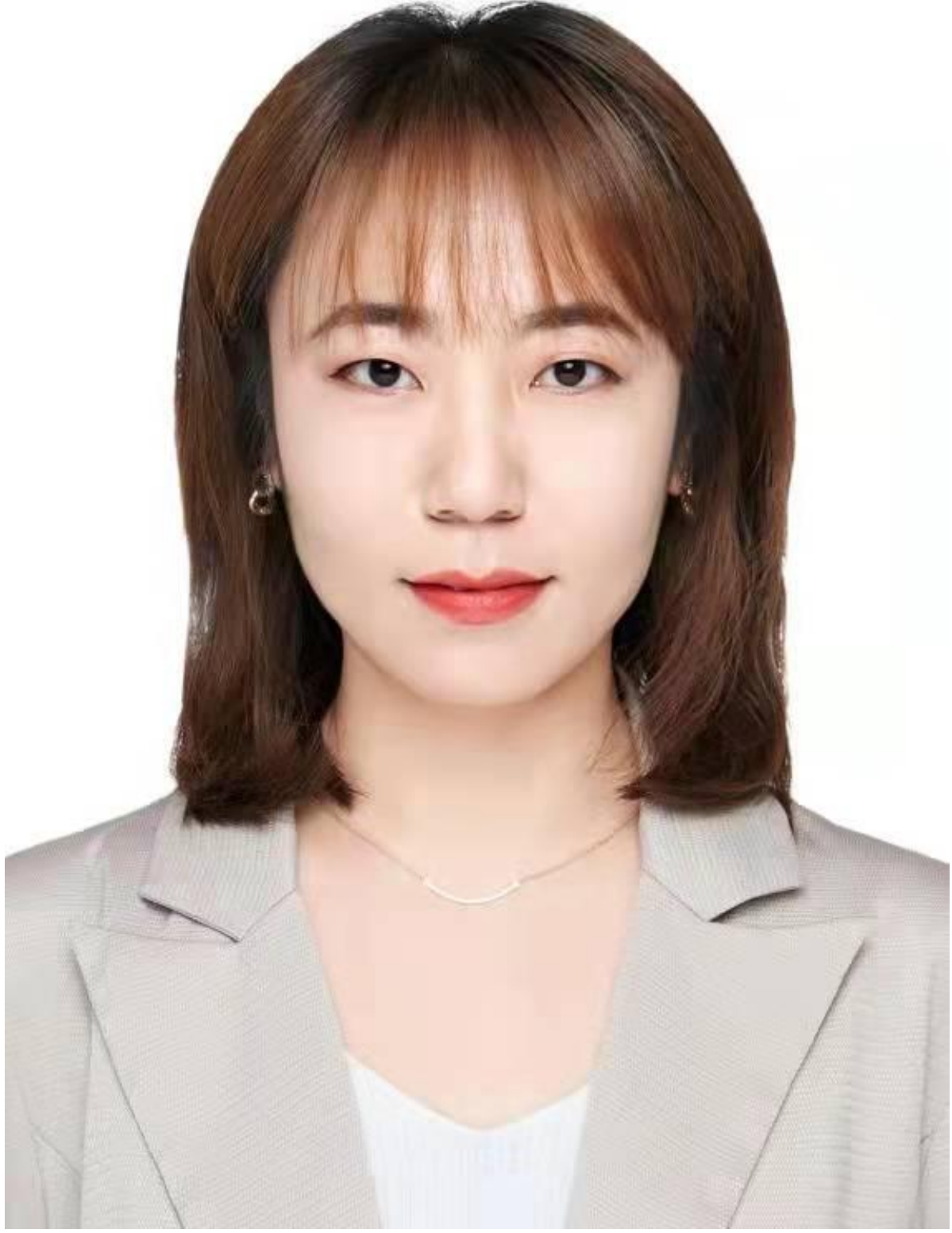}}]{Yingjie Wang}
(Member, IEEE) received the Ph.D. degree from the College of Computer Science and Technology, Harbin Engineering University, Harbin, China, in 2015. She visited Georgia State University, Atlanta, GA, USA, from September 2013 to September 2014, as a Visiting Scholar. She is currently an Associate Professor with the School of Computer and Control Engineering, Yantai University, Yantai, China. She holds a post-doctoral position with South China University of Technology, Guangzhou, China. She has published more than 50 papers in well-known journals and conferences in her research field, which includes two ESI high cited papers. In addition, she has presided one National Natural Science Foundation of China project and two China Postdoctoral Science Foundation projects. Her research interests include service computing, mobile crowdsourcing, and trust computing. Dr. Wang received the Shandong Province Artificial Intelligence Outstanding Youth Award.
\end{IEEEbiography}

\begin{IEEEbiography}[{\includegraphics[width=1in,height=1.25in,clip,keepaspectratio]{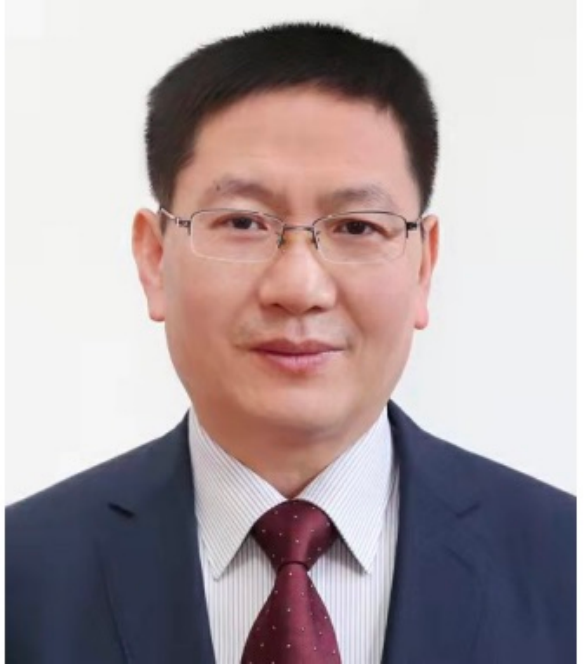}}]{Peiyong Duan}
Peiyong Duan received the M.S. degree from the Shandong University of Technology (merged into Shandong University, in 2000), Shandong, China, in 1996, and the Ph.D. degree from Shanghai Jiaotong University, Shanghai, China, in 1999. From 1999 to 2014, he was with the School of Information and Electrical Engineering, Shandong Jianzhu University, where he was appointed as an Associate Professor and a Full Professor, in 1999 and 2002, respectively. From 2017 to 2019, he was with the School of Information Science and Engineering, Shandong Normal University, Shandong. He is currently a Full Professor in the Faculty of Electronics, Electrical and Control at Qilu University of Technology (Shandong Academy of Sciences). He has authored over 140 refereed articles. His research interests include discrete optimization and scheduling, privacy protection and service computing.
\end{IEEEbiography}

\begin{IEEEbiography}[{\includegraphics[width=1in,height=1.5in,clip,keepaspectratio]{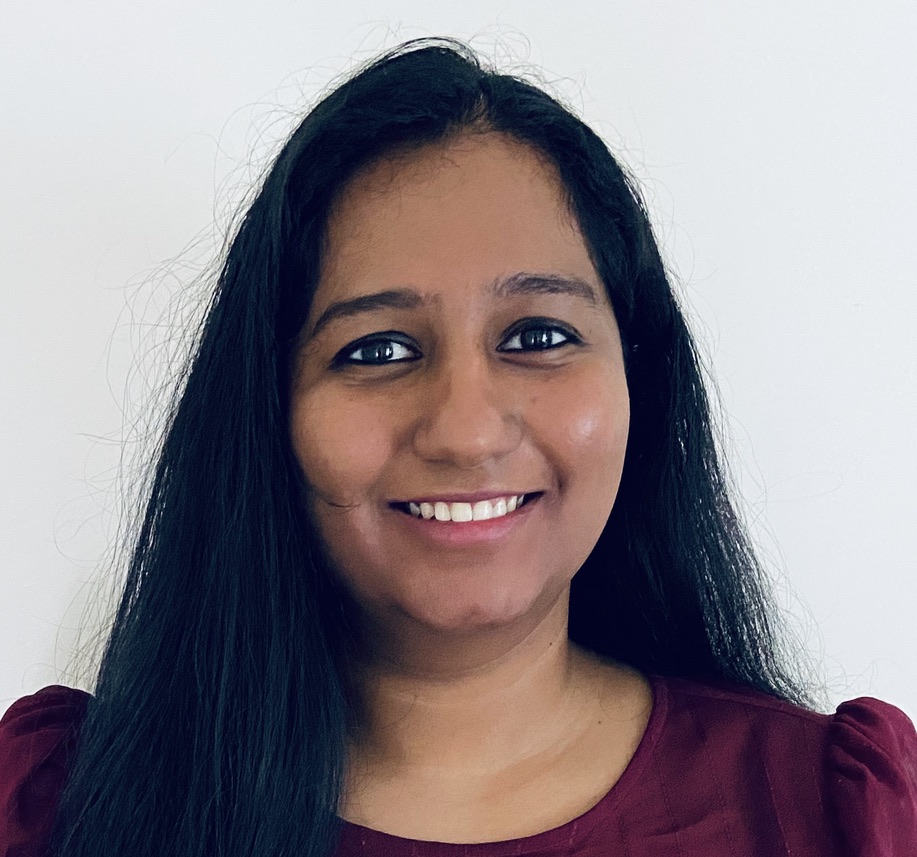}}]{AKSHITA MARADAPU VERA VENKATA SAI}
received the B.S. degree in computer science and engineering from GITAM University, India, in 2016, and the M.S. degree in computer science from Georgia State University, USA, in 2017, where she is currently pursuing the Ph.D. degree with the Department of Computer Science. Her research interests include mobile social networks, security and privacy, digital twin networks, edge computing and fedearated learning.
\end{IEEEbiography}

\begin{IEEEbiography}[{\includegraphics[width=1in,height=1.25in,clip,keepaspectratio]{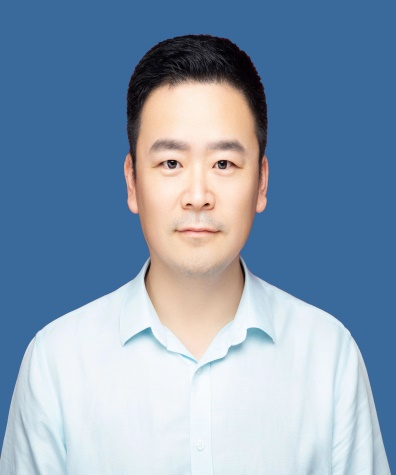}}]{Zhaowei Liu}
(Member, IEEE) was born in Yantai, Shandong, China, in 1979. He received the B.S., M.S., and Ph.D. degrees from the College of Control Science and Engineering, Shandong University, Jinan, China, in 2002, 2005, and 2018, respectively. He is currently an Associate Professor with the School of Computer Science and Control Engineering, Yantai University, Yantai, China. His research interests include graph learning, and its applications in physics simulation and blockchain.
\end{IEEEbiography}

\begin{IEEEbiography}[{\includegraphics[width=1in,height=1.25in,clip,keepaspectratio]{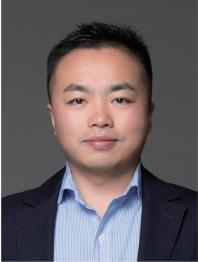}}]{Chaocan Xiang}
(Member, IEEE)received the B.S. and Ph.D. degrees in computer science and engineering from the Nanjing Institute of Communication Engineering, Nanjing, China, in 2009 and 2014, respectively. He studied at the Real-Time Computing Laboratory (RTCL), the University of Michigan, Ann Arbor, MI, USA, in 2017. He is an Associate Professor at the College of Computer Science, Chongqing University, Chongqing, China. He has published more than 20 research papers in important conferences and journals, such as Association for Computing Machinery Ubiquitous Computing (ACM UbiComp), IEEE TRANSACTIONS ON MOBILE COMPUTING (TMC), IEEE TRANSACTIONS ON PARALLEL AND DISTRIBUTED SYSTEMS (TPDS), IEEE TRANSACTIONS ON INTELLIGENT TRANSPORTATION SYSTEMS (T-ITS), and IEEE TRANSACTIONS ON NETWORK SCIENCE AND ENGINEERING (TNSE). His current research interests include crowd-sensing networks and Internet of Things (IoT).
\end{IEEEbiography}

\begin{IEEEbiography}[{\includegraphics[width=1in,height=1.25in,clip,keepaspectratio]{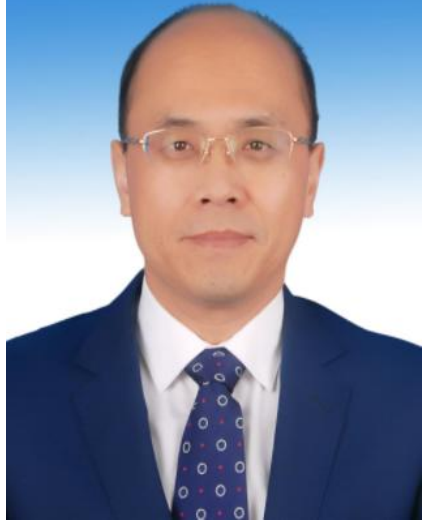}}]{Xiangrong Tong}
received the Ph.D. degree in School of Computer and Information Technology from Beijing Jiaotong University. Currently, he is a Full Professor of Yantai University. His research interests are computer science, intelligent information processing and social networks. He has published more than 50 papers in well known journals and conferences. In addition, he has presided and joined 3 national projects and 3 provincial projects.
\end{IEEEbiography}

\begin{IEEEbiography}[{\includegraphics[width=1in,height=1.25in,clip,keepaspectratio]{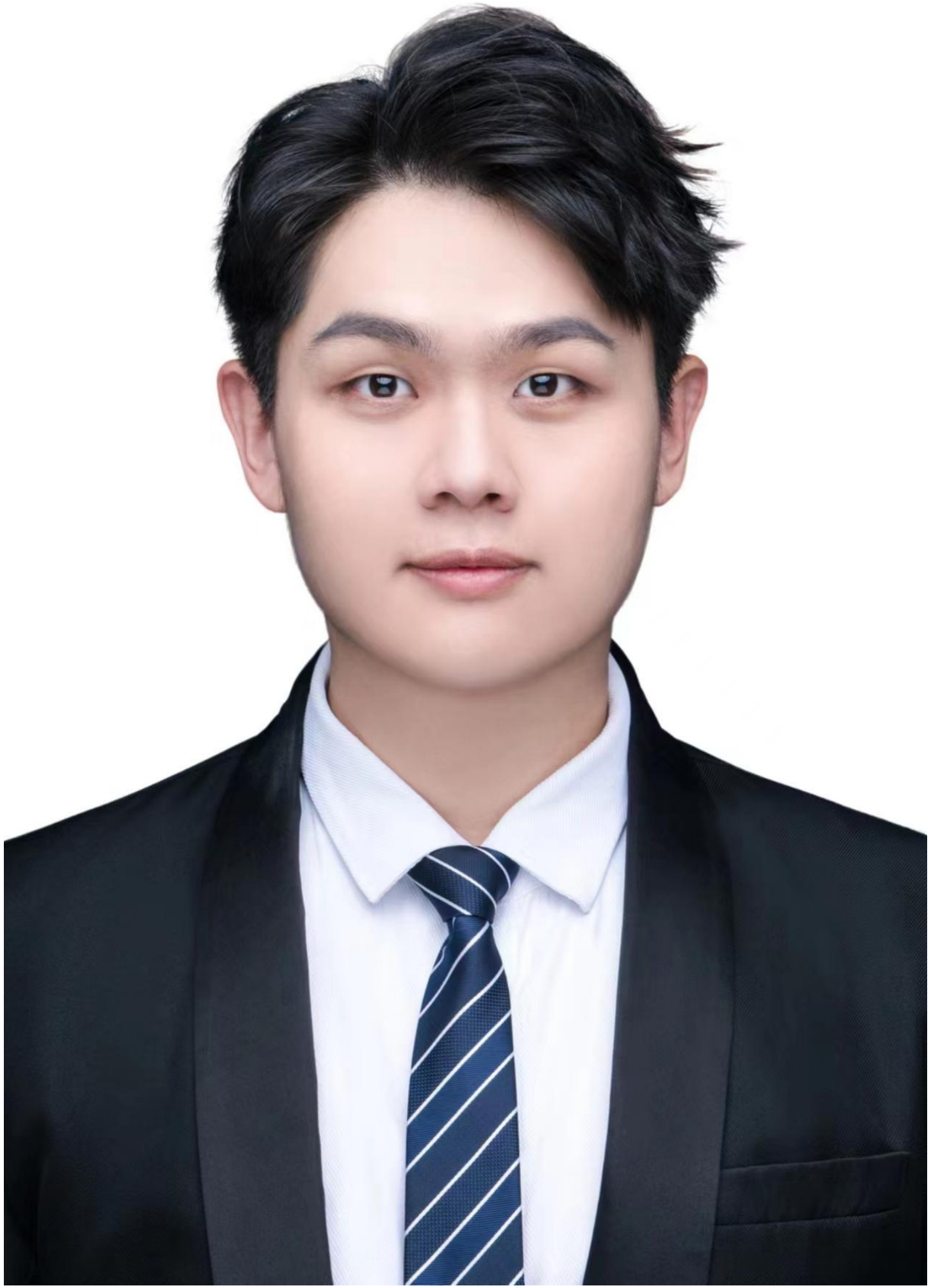}}]{Weilong Wang}
received the B.S. degree in the School of Information Science and Engineering from Shandong Agricultural Engineering College, and the M.S. degree in the School of Computer and Control Engineering from Yantai University. He is currently pursuing the Ph.D. degree with the Department of Computer Science and Engineering,Southeast University. His research interests are mobile crowdsourcing, privacy protection and edge computing.
\end{IEEEbiography}
\vspace{11pt}

\begin{IEEEbiography}[{\includegraphics[width=1in,height=1.25in,clip,keepaspectratio]{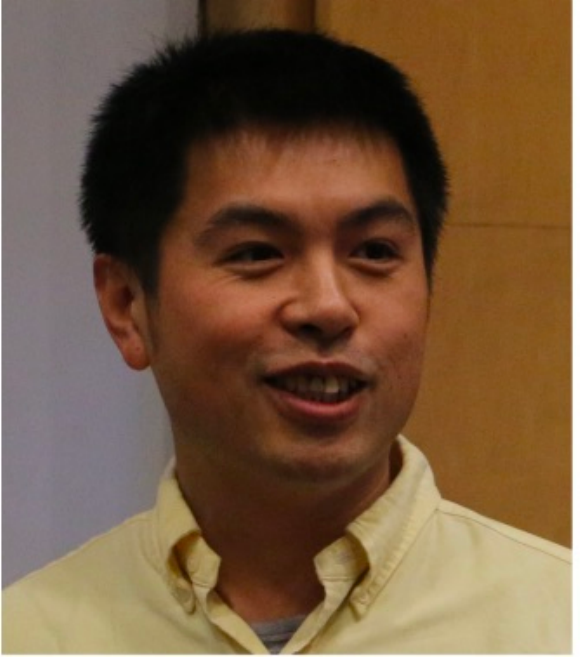}}]{Zhipeng Cai}
(Fellow, IEEE) received his PhD in the Department of Computing Science at University of Alberta, and B.S. degree from Beijing Institute of Technology.  He is currently a Professor in the Department of Computer Science at Georgia State University.  Dr. Cai's research expertise lies in the areas of Resource Management and Scheduling, Privacy, Networking, and Big Data.  His research has received funding from multiple academic and industrial sponsors, including the National Science Foundation and the U.S. Department of State, and has resulted in over 100 publications in top journals and conferences, with more than 14,500 citations, including over 80 IEEE/ACM Transactions papers.  Dr. Cai is an Editor-in-Chief for Wireless Communications and Mobile Computing, and an Associate Editor-in-Chief for Elsevier High-Confidence Computing Journal, as well as an editor for various prestigious journals, such as IEEE Transactions on Knowledge and Data Engineering (TKDE), IEEE Transactions on Vehicular Technology (TVT), IEEE Transactions on Wireless Communications (TWC), and IEEE Transactions on Computational Social Systems (TCSS).  Dr. Cai is the recipient of an NSF CAREER Award and a Fellow of IEEE.
\end{IEEEbiography}

\end{document}